\documentclass[preprint,pre,a4paper]{revtex4-1}
\usepackage{graphicx}
\usepackage{epstopdf}
\usepackage{epsfig}
\usepackage{amssymb,amsmath,stmaryrd,tabularx}
\usepackage{wrapfig}
\usepackage{bm}
\usepackage[center]{subfigure}
\usepackage{enumerate}
\usepackage{hyperref}
\usepackage{forest}
\usepackage[frozencache,newfloat]{minted}
\usepackage{dcolumn}

\usepackage{caption}
\usepackage{float}

\newenvironment{code}{\captionsetup{type=listing}}{}
\SetupFloatingEnvironment{listing}{name=Listing}

\renewcommand{\v}[1]{\ensuremath{\mathbf{#1}}} % for vectors
\let\crossprod=\times
\renewcommand{\times}{\cdot} % times symbol
\newcommand{\gv}[1]{\ensuremath{\mbox{\boldmath$ #1 $}}} 
\newcommand{\diff}{\mathrm{d}}
\newcommand{\pd}[2]{\frac{\partial #1}{\partial #2}}

\newcommand{\grad}[1]{\gv{\nabla} #1} % for gradient
\renewcommand{\div}[1]{\gv{\nabla} \cdot #1} % for divergence
\newcommand{\laplacian}[1]{\grad^2 #1}
\newcommand{\emphy}[1]{\textsc{#1}}
\newcommand{\pdt}[1]{\partial_t #1}
\newcommand{\sym}[1]{\mathrm{sym} #1 }

\newcommand{\wall}{\mathrm{wall}}
\newcommand{\inlet}{\mathrm{in}}
\newcommand{\outlet}{\mathrm{out}}

\newcommand{\jump}[1]{\left[ #1 \right]^+_-}

\newcommand{\pdtau}[1]{\partial_\tau^- #1}
\newcommand{\I}[2]{\left( #1, #2\right)}

\newcommand{\IA}[1]{\int_\Gamma #1 \, \diff \Gamma}

\newcommand{\norm}[1]{\left\lVert #1 \right\rVert}
\newcommand{\kmone}{{k-1}}

\begin{document}
\title{Bernaise: A flexible framework for simulating two-phase electrohydrodynamic flows in complex domains}

\newcommand{\NBI}{Niels Bohr Institute, University of Copenhagen, Blegdamsvej 17, DK-2100 Copenhagen, Denmark.}
\author{Gaute Linga}
\email{linga@nbi.dk}
\affiliation{\NBI}
\author{Asger Bolet}
%\email{asger.bolet@nbi.dk}
\affiliation{\NBI}
\author{Joachim Mathiesen}
%\email{mathies@nbi.dk}
\affiliation{\NBI}

% NOTE:
% All sentences should be kept at one line only.
% This way it is easier to track changes in git.

\begin{abstract}
  \emph{Bernaise} (\emphy{B}inary
  \emphy{E}lect\emphy{r}ohydrody\emphy{na}m\emphy{i}c Solv\emphy{e}r) is a flexible high-level finite element solver of two-phase electrohydrodynamic flow in complex geometries. 
  Two-phase flow with electrolytes is relevant across a broad range of systems and scales, from 'lab-on-a-chip' devices for medical diagnostics to enhanced oil recovery at the reservoir scale.
%  The flow of two immiscible fluids, wherein dissolved ions are transported subject to an electric field, and where the fluids, ions, and fluid-fluid interfaces interact non-linearly with the confining complex boundaries.
%  \emphy{E}lect\emphy{r}ohydrody\emphy{na}m\emphy{i}c Solv\emphy{e}r) is a flexible and robust high-level finite element solver of two-phase electrohydrodynamic flow in complex geometries.
  %Two-phase flow with electrolytes has applications ranging from 'lab-on-a-chip' devices for medical diagnostics to enhanced oil recovery, and 
  %The problem concerns the flow of two immiscible fluids, wherein dissolved ions are transported subject to an electric field, and where the fluids, ions, and fluid-fluid interfaces interact non-linearly with the confining complex boundaries.
%  The two phases may have different properties (such as density, viscosity, permittivity), and the dissolved ions may have different properties (such as diffusivity and solubility) in the two phases.
  For the strongly coupled multi-physics problem, we employ a recently developed thermodynamically consistent model which combines a generalized Nernst--Planck equation for ion transport, the Poisson equation for electrostatics, the Cahn--Hilliard equation for the phase field (describing the interface separating the phases), and the Navier--Stokes equations for fluid flow.
  As an efficient alternative to solving the coupled system of partial differential equations in a monolithic manner, we present a linear, decoupled numerical scheme which sequentially solves the three sets of equations.
  The scheme is validated by comparison to limiting cases where analytical solutions are available, benchmark cases, and by the method of manufactured solution.
  The solver operates on unstructured meshes and is therefore well suited to handle arbitrarily shaped domains and problem set-ups where, e.g., very different resolutions are required in different parts of the domain.
  \emph{Bernaise} is implemented in Python via the FEniCS framework, which effectively utilizes %the linear algebra backend PETSc,
  MPI and domain decomposition, and should therefore be suitable for large-scale/high-performance computing.
  Further, new solvers and problem set-ups can be specified and added with ease to the \emph{Bernaise} framework by experienced Python users.
\end{abstract}

\maketitle

\section{Introduction}
Two-phase flow with electrolytes is encountered in many natural and industrial settings.
Although Lippmann already in the 19th century \cite{lippmann1875,mugele2005} made the observation that an applied electric field changes the wetting behaviour of electrolyte solutions, the phenomenon of electrowetting has remained elusive.
Recent decades have seen an increased theoretical and experimental interest in understanding the basic mechanisms of electrokinetic or electrohydrodynamic flow \cite{ristenpart2009,mugele2009}.
Progress in micro- and nanofluidics \cite{squires2005,schoch2008} has enabled the use electrowetting to control small amounts of fluid with very high precision (see e.g.~the comprehensive reviews by Mugele and coworkers \cite{mugele2005,mugele2010} and \citet{nelson2012} and references therein).
This yields potential applications in, e.g., ``lab-on-chip'' biomedical devices or microelectromechanical systems \cite{pollack2002,srinivasan2004,lee2000}, membranes for harnessing blue energy \cite{siria2017}, energy storage in fluid capacitors, and electronic displays \cite{beni1981,beni1981b,beni1982,hayes2003}.

It is known that electrohydrodynamic phenomena affects transport properties and energy dissipation in geological systems, as a fluid moving in a fluid-saturated porous medium sets up an electric field that counteracts the fluid motion \cite{pride1991,fiorentino2016a,fiorentino2016b}.
Electrowetting may also be an important factor in enhanced oil recovery \cite{hassenkam2011,hilner2015}.
Here, the injection of water of a particular salinity, or ``smart water'' \cite{rezaeidoust2009}, is known to increase the recovery of oil from reservoirs as compared to brine \cite{pedersen2016}.
Further, transport in sub-micrometer scale pores in low-permeability rocks in the Earth's crust may be driven by gradients in the electrochemical potential \cite{plumper2017}, which may have consequences for, e.g., transport of methane-water mixtures in dense rocks.

Hence, a deepened understanding of electrowetting and two-phase electrohydrodynamics would be of both geological and technological importance.
While wetting phenomena (or more generally, two-phase flow) on one hand, and electrohydrodynamics on the other, remain in themselves two mature and active areas of research which both encompass a remarkably rich set of phenomena, this article is concerned with the interface between these fields.
For interested readers, there are several reviews available regarding wetting phenomena \cite{degennes1985,bonn2009,snoeijer2013} and electrohydrodynamics \cite{melcher1969,saville1997,fylladitakis2014}.
% wo-phase flow and related wetting phenomena are encountered in a variety of natural, technological, and industrial applications---from microfluidics to oil recovery \cite{bonn2009}.
Notably, the ``leaky dielectric'' model originally proposed by \citet{taylor1966} (and revisited by \citet{melcher1969}) to describe drop deformation, is arguably the most popular description of electrohydrodynamics, but it does not describe ionic transport and considers all dielectrics to be weak conductors.
In this work, we shall employ a model that does not make such simplifications.
Recently, \citet{schnitzer2015} showed rigorously that models of the latter type reduce to the Taylor--Melcher model in the double limit of small Debye length and strong electric fields.
The simplified model may therefore have advantages in settings where those assumptions are justified, e.g., in simulations on larger scales; while the class of models considered here are more general and expected to be valid down to the smallest scale where the continuum hypothesis still holds.

Experimental and theoretical approaches \cite{zholkovskij2002,monroe2006,monroe2006b} in two-phase electrohydrodynamic flows need to be supplemented with good numerical simulation tools.
This is a challenging task, however: the two phases have different densities, viscosities and permittivities, the ions have different diffusivities and solubilites in the two phases, and moreover, the interface between the phases must be described in a consistent manner.
Hence, much due to the complex physics involved, simulation of two-phase electrohydrodynamic phenomena with ionic transport is still in its infancy.
It has been carried out with success e.g.~in order to understand deformation of droplets due to electric fields \cite{yang2013,yang2014,berry2013}, or for the purpose of controlling microfluidic devices (see e.g.~\cite{zeng2011}).
\citet{lu2007} simulated and performed experiments on droplet dynamics in a Hele-Shaw cell. 
Notably, \citet{walker2009} simulated electrowetting with contact line pinning, and compared to experiments.
In practical applications, such as in environmental remediation or oil recovery, the complex pore geometry is essential and it is therefore of interest to simulate and study electrowetting in such configurations.
However, to our knowledge, there have been few numerical studies of these phenomena in the context of more complex geometries.

In this article, we introduce and describe \emph{Bernaise} (\emphy{B}inary \emphy{E}lect\emphy{r}ohydrody\emphy{na}m\emphy{i}c Solv\emphy{e}r), which is an open-source software/framework for simulating two-phase electrohydrodynamics.
It is suitable for use in complex domains, operating on arbitrary unstructured meshes.
The finite-element solver is written entirely in Python and built on top of the FEniCS framework \cite{logg2012}, which (among other things) effectively uses the PETSc backend for scalability.
FEniCS has in recent years found success in related applications, such as in high-performance simulation of turbulent flow \cite{mortensen2015}, and for single-phase, steady-state electrohydrodynamic flow simulation in nanopores \cite{mitscha-baude2017} and model fractures \cite{bolet2018}.
Since \emph{Bernaise} was inspired by the \emph{Oasis} solver for fluid flow \cite{mortensen2015}, it is similar to the latter in both implementation and use.

In this work, we employ a phase-field model to propagate the interface between the two phases.
Such \emph{diffuse interface} models, as opposed to e.g.~sharp interface models (see for instance \cite{prosperetti2009}), assume that the fluid-fluid interface has a finite size, and have the advantage that no explicit tracking of the interface is necessary.
Hence, using a phase-field model has several advantages in our setting: it takes on a natural formulation using the finite element method; in sub-micrometer scale applications, the diffuse interface and finite interface thickness present in these models might correspond to the physical interface thickness (typically nanometer scale \cite{yang1996}); and the diffuse interface may resolve the moving contact line conundrum \cite{qian2006,snoeijer2013}.
%The latter is due to the interface diffusivity, which may be tuned to experiments or molecular dynamics simulations.
Note that although \emph{ab initio} and molecular dynamics simulation methods are in rapid growth due to the increase in computational power, and do not require explicit tracking of the interface or phenomenological boundary conditions, such methods are restricted to significantly smaller scales than continuum models are.
Nevertheless, they serve as valuable tools for calibration of the continuum methods \cite{qian2003,qian2006,sui2014,ervik2016}.
We note also that sharp-interface methods such as level-set \cite{walker2006,teigen2010} and volume-of-fluid methods \cite{tomar2007,lopez-herrera2011,berry2013} are viable options for simulating electrohydrodynamics, but such methods shall not be considered here.

The use of phase field models to describe multiphase flow has a long history in fluid mechanics \cite{anderson1998}.
Notably, the ``Model H'' of \citet{hohenberg1977}, for two incompressible fluids with matched densities and viscosities, is based on the coupled Navier--Stokes--Cahn--Hilliard system, and was introduced to describe phase transitions of binary fluids or single-phase fluid near the critical point.
\citet{lowengrub1998} later derived a thermodynamically consistent generalization of Model H where densities and viscosities were different in the two phases, however with the numerical difficulty that the velocity field was not divergence free.
To circumvent this issue, \citet{abels2012} developed a thermodynamically consistent and frame invariant phase-field model for two-phase flow, where the velocity field was divergence free, allowing for the use of more efficient numerical methods.
\citet{lu2007} proposed a phase-field model to describe electrohydrodynamics, but was restricted to flow in Hele-Shaw cells, using a Darcy equation to describe the flow between the parallel plates \footnote{Instead of the full Navier--Stokes equations, which would be necessary in the presence of boundaries in the two in-plane dimensions.}.
A phase-field approach to the leaky-dielectric model was presented by \citet{lin2012}.
Using the Onsager variational principle, \citet{campillo-funollet2012} augmented the model of \citet{abels2012} with electrodynamics, i.e.~inclusion of ions, electric fields and forces.
This can be seen as a more physically sound version of the model proposed by \citet{eck2009}, which only contained a single ``net charge'' electrolyte species.
A model for two-phase electrohydrodynamics was derived, with emphasis on contact line pinning, by \citet{nochetto2014}, but this does not appear to be frame-invariant, as the chemical potential depends quadratically on velocity \cite{campillo-funollet2012}.
In this work, we will therefore focus on the model by \citet{campillo-funollet2012}.

There is a vast literature on the discretization and simulations of immiscible two-phase flows including phase-field models (see e.g.~\cite{prosperetti2009,anderson1998}), but here we focus on research which is immediately relevant concerning the discretization and implementation of the model by \citet{campillo-funollet2012}.
\citet{grun2014} discretized the model in Ref.~\cite{abels2012} (without electrohydrodynamics) with a dual mesh formulation, using a finite volume method on the dual mesh for advection terms, and a finite element method for the rest.
Based on the sharp-interface model benchmarks of \citet{hysing2009}, \citet{aland2012} provided benchmarks of bubble dynamics comparing several formulations of phase-field models (without electrodynamics).
Energy-stable numerical schemes for the same case were presented and analyzed in \cite{guillen-gonzalez2014,grun2016}.
\citet{campillo-funollet2012} provided preliminary simulations of the two-phase electrohydrodynamics model in their paper, however with a simplified formulation of the chemical potential of the solutes.
A scheme for the model in \cite{campillo-funollet2012} which decouples the Navier--Stokes equations from the Cahn--Hilliard--Poisson--Nernst--Planck problem, was presented and demonstrated by \citet{metzger2015,metzger2018}.
In the particular case of equal phasic permittivities, the Cahn--Hilliard problem could be decoupled from the Poisson--Nernst--Planck problem.
Recently, a stable finite element approximation of two-phase EHD, with the simplifying assumptions of Stokes flow and no electrolytes, was proposed by \citet{nurnberg2017}.

The main contributions of this article is to give a straightforward description of \emph{Bernaise}, including the necessary background theory, an overview of the implementation, and a demonstration of its ease of use.
Solving the coupled set of equations in a monolithic manner (as is done in Ref.~\cite{campillo-funollet2012} using their in-house \textsc{EconDrop} software) is a computationally expensive task, and we therefore propose a new linear splitting scheme which sequentially solves the phase-field, chemical transport and the fluid flow \emph{subproblems} at each time step.
We demonstrate the validity of the approach and numerical convergence of the proposed scheme by comparing to limiting cases where analytical solutions are available, benchmark solutions, and using the method of manufactured solution.
We demonstrate how the framework can be extended by supplying user-specified problems and solvers.
We believe that due to its flexibility, scalability and open-source licensing, this framework has advantages over software which to our knowledge may have \emph{some} of the same functionality, such as \textsc{EconDrop} (in-house code of Gr\"{u}n and co-workers) and \textsc{Comsol} (proprietary).
Compared to sharp-interface methods, the method employed in the current framework is automatically capable of handling topological changes and contact line motion, and the full three-dimensional (3D) capabilities allows to study more general phenomena than what can be achieved by axisymmetric formulations \cite{berry2013}.
We expect \emph{Bernaise} to be a valuable tool that may facilitate the development of microfluidic devices, as well as a deepened understanding of electrohydrodynamic phenomena in many natural or industrial settings.

The outline of this paper is as follows.
In Sec.~\ref{sec:model}, we introduce the sharp-interface equations describing two-phase electrohydrodynamics; then we present the thermodynamically consistent model of electrohydrodynamics by \citet{campillo-funollet2012}.
In Sec.~\ref{sec:numerics}, we write down the variational form of the model, present the monolithic scheme, and present a linear splitting scheme for solving the full-fledged two-phase electrohydrodynamics.
Sec.~\ref{sec:bernaise} gives a brief presentation of \emph{Bernaise}, and demonstrates its ease use through a minimal example.
Further, we describe how \emph{Bernaise} can be extended with user-specified problems and solvers.
In Sec.~\ref{sec:validation}, we validate the approach as described in the preceding paragraph.
Finally, in Sec.~\ref{sec:application}, we apply the framework to a geologically relevant setting where dynamic electrowetting effects enter, and present full 3D simulations of droplet coalescence and breakup.
Finally, in Sec.~\ref{sec:conclusion} we draw conclusions and point to future work.

We expect the reader to have a basic familiarity with the finite element method, the Python language, and the FEniCS package.
Otherwise, we refer to the tutorial by \citet{langtangen2017}.

\section{Model}
\label{sec:model}
The governing equations of two-phase electrohydrodynamics can be summarized as the coupled system of two-phase flow, chemical transport (diffusion and migration), and electrostatics \cite{campillo-funollet2012}.
We will now describe the sharp-interface equations that the phase-field model should reproduce, and subsequently the phase-field model for electrohydrodynamics.
For the purpose of keeping the notation short, we consider a general electrokinetic scaling of the equations.
The relations between the dimensionless quantities and their physical quantities are elaborated in Appendix \ref{sec:electrokinetic_scaling}.

\subsection{Sharp-interface equations}
In the following, we present each equation of the physical (sharp-interface) model.
With validity down to the nanometer scale, the fluid flow is described by the incompressible Navier--Stokes equations, augmented by some additional force terms due to electrochemistry:
\begin{gather}
  \rho_i \left( \pdt {\v v} + (\v v \cdot \grad ) \v v \right)  - \mu_i \laplacian \v v + \grad p = - \sum_j {c_j} \grad g_{c_j}, \label{eq:sharp_NS1}\\
  \div \v v = 0. \label{eq:sharp_NS2}
\end{gather}
Here, $\rho_i$ is the density of phase $i$, $\v v$ is the velocity field, $\mu_i$ is the dynamic viscosity of phase $i$, $p(\v x, t)$ is the pressure field \footnote{The interpretation of this pressure depends on the formulation of the force on the right hand side of Eq.~\eqref{eq:sharp_NS2}.}, $c_j (\v x, t)$ is the concentration of solute species $j$, and $g_{c_j}$ is the associated electrochemical potential.
%$ \rho_e = \sum_j z_j c_j$ is the total charge density ($z_j$ is the charge if solute species $j$), and finally $V$ is the electric potential.
The form of the right hand side of Eq.\ \eqref{eq:sharp_NS1} is somewhat unconventional (and relies on a specific interpretation of the pressure), but has numerical advantages over other formulations as it avoids, e.g., pressure build-up in the electrical double layers \cite{nielsen2014}.

The transport of the concentration field of species $i$ is governed by the conservative (advection--diffusion--migration) equation:
\begin{equation}
  \pdt c_j + \v v \cdot \grad c_j - \div (K_{ij} c_j \grad g_{c_j}) = 0,
  \label{eq:sharp_conc}
\end{equation}
where $K_{ij}$ is the diffusivity of species $j$ in phase $i$.
The electrochemical potential is in general given by
\begin{equation}
  g_{c_j} (c_j, V) = \alpha'(c_j) + \beta_{ij} + z_j V,
  \label{eq:sharp_chempot}
\end{equation}
where $\alpha'(c) = \partial \alpha / \partial c (c)$, and $\alpha(c)$ is a convex function describing the chemical free energy, $\beta_{ij}$ is a parameter describing the solubility of species $j$ in phase $i$, $z_j$ is the charge if solute species $j$, and $V$ is the electric potential.
Eq.~\eqref{eq:sharp_conc} can be seen as a generalized Nernst--Planck equation.
With an appropriate choice of $\alpha(c)$, Eq.~\eqref{eq:sharp_conc} reduces to the phenomenological Nernst--Planck equation, which has been established for the transport of charged species in dilute solutions under influence of an electric field. %The Nernst--Planck equation reads
%\begin{equation}
%  \pdt c_i + \v v \cdot \grad c_i + \div \left[ D_{ij} \grad c_i + \frac{D_{ij} z_i}{k_{\rm B} T} c_i \grad V \right] = 0,
%  \label{eq:nernst-planck}
%\end{equation}
%which coincides with Eq.~\eqref{eq:sharp_conc} with the choices $g_{c_i} = k_{\rm B} T \log c_i + \textrm{const.}$, $K_{ij} = D_{ij}/(k_{\rm B} T)$.
The latter amounts to a dilute solution, using the ideal gas approximation,
\begin{equation}
  \alpha(c_j) \propto c_j (\ln c_j - 1).
  \label{eq:alphac_nernstplanck}
\end{equation}
With this choice of $\alpha$, the solubility parameter $\beta_{ij}$ can be interpreted as related to a reference concentration $c^{{\rm ref}, i}_j$, through the relation
\begin{equation}
  \beta_{ij} = - \ln c^{{\rm ref}, i}_j.
  \label{eq:beta_ij_interpretation}
\end{equation}
This gives a chemical energy $\mathcal G_j = \alpha(c_j) + \beta_{ij} c_j = c_j ( \ln (c_j/ c^{{\rm ref}, i}_j) - 1)$ which has a minimum at $c_j = c^{{\rm ref}, i}_j$ (see also \cite{linga2018decoupled}).

Since the dynamics of the electric field is much faster than that of charge transport, we can safely assume electrostatic conditions (i.e., neglect magnetic fields).
This amounts to solving the Poisson problem (Gauss' law):
\begin{equation}
  \div ( \varepsilon_i \grad V ) = - \rho_e,
  \label{eq:sharp_poisson}
\end{equation}
Here, $\varepsilon_i$ is the electrical permittivity of phase $i$, and $\rho_e = \sum_j z_j c_j$ is the total charge density.

%$\varepsilon_i = \varepsilon_0 \varepsilon_{\textrm{r}, i}$ is the permittivity of phase $i$, where  $\varepsilon_0$ is the vacuum permitivity and $\varepsilon_{\textrm{r}, i}$ is the relative permeability of phase $i$. 

In the absence of advection, for the case of two symmetric charges, and under certain boundary conditions, Eqs.~\eqref{eq:sharp_conc}--\eqref{eq:sharp_poisson} lead to the simpler Poisson--Boltzmann equation (see Appendix \ref{sec:app_pb}).

\subsubsection{Fluid-fluid interface conditions}
It is necessary to define jump conditions over the interface between the two fluids.
We denote the jump in a physical quantity $\chi$ across the interface by $\jump{\chi}$, and the unit vector $\hat{\v n}_{\rm int}$ normal to the interface.

Firstly, due to incompressibility, the velocity field must be continuous:
\begin{equation}
  \jump{\v v} = 0.
\end{equation}
The electrochemical potential must be continuous across the interface,
\begin{equation}
  \jump{g_{c_j}} = 0.
\end{equation}
Due to conservation of the electrolytes, the flux of ion species $j$ \emph{into} the interface must equal the flux \emph{out of} the interface,
\begin{equation}
  \jump{ K_{ij} c_j \grad g_{c_j} } \cdot \hat{\v n}_{\rm int} = 0,
\end{equation}
and the normal flux of the electric displacement field $\v D = - \epsilon_i \grad V$, and the electric potential, should be continuous (since by assumption, no free charge is located \emph{betweeen} the fluids):
\begin{equation}
  \jump{\epsilon_i \grad V} \cdot \hat{\v n}_{\rm int} = 0, \quad \jump{V} = 0.
\end{equation}
Finally, interfacial stress balance yields the condition
\begin{equation}
  \jump{p} \hat{\v n}_{\rm int} - \jump{ 2 \mu_i \mathcal{D} \v v } \cdot \hat{\v n}_{\rm int} 
  %- \sum_j \jump{ \alpha(c_j) + \beta_{ij} c_j } \cdot \hat{\v n} 
  - \jump{\varepsilon_i \v E \otimes \v E - \frac{1}{2} \varepsilon_i |\v E|^2 \v I} \cdot \hat{\v n}_{\rm int}  = \sigma \kappa \hat{\v n}_{\rm int},
  \label{eq:sharp_jump_stress}
\end{equation}
where $\sigma$ is the surface tension, $\kappa$ is the curvature, and $\v E = - \grad V$ is the electric field.
Moreover, we have defined the shorthand symmetric (vector) gradient,
\begin{equation}
  \mathcal{D} \v v = \sym\left(\grad \v v\right) = \frac{1}{2}\left( \grad \v v + \grad \v v^T\right).
\end{equation}
Further, all gradient terms have been absorbed into the pressure.
Note that Eq.~\eqref{eq:sharp_jump_stress} leads to a modified Young--Laplace law in equilibrium, which include Maxwell stresses.

\subsubsection{Boundary conditions}
There are a range of applicable boundary conditions for two-phase electrohydrodynamics.
Here, we briefly discuss a few viable options.
In the following, we let $\hat{\v n}$ be a unit normal vector pointing out of the domain, and $\hat{\v t}$ be a tangent vector to the boundary.

For the velocity, it is customary to use the no-slip condition $\v u = \v 0$ at the solid boundary.
Alternatively, the Navier slip condition, which may be of use for modelling moving contact lines \cite{sui2014}, could be used:
\newcommand{\rmn}{\textrm{n}}
\newcommand{\rmt}{\textrm{t}}
\begin{equation}
  v_\rmn = 0, \quad \gamma v_\rmt = \mu ( \partial_\rmn v_\rmt + \partial_\rmt v_\rmn ),
\end{equation}
where $\gamma$ is a slip parameter, and the subscripts denote tangential (t) or normal (n) components.
The slip length $\mu/\gamma$ is typically of nanometer scale and dependent on the materials in question.
However, since the implementation of such conditions may become slightly involved, we omit it in the following.

With regards to the electrolytes, it is natural to specify either a prescribed concentration at the boundary, $c_i = c_0$, or a no-flux condition out of the domain,
\begin{equation}
  \hat{\v n} \cdot \left( - \v u c_j + K_{ij} c_j \grad g_{c_j} \right) = 0.
\end{equation}

For the electric potential, it is natural to prescribe either the Dirichlet condition $V=\bar V$, or a prescribed surface charge $\sigma_e(\v x)$, 
\begin{equation}
  \hat{\v n} \cdot \grad V = \frac{\sigma_e}{\epsilon_i}.
\end{equation}

\subsection{Phase-field formulation}
In order to track the interface between the phases, we introduce an order parameter field $\phi$ which attains the values $\pm 1$ respectively in the two phases, and interpolates between the two across a diffuse interface of thickness $\epsilon$.
In the sharp-interface limit $\epsilon \to 0$, the equations should reproduce the correct physics, and reduce to the model above, including the interface conditions.
A thermodynamically consistent phase-field model which
% with a proper modelling of the phase field mobility)
reduces to this formulation was proposed in Ref.~\cite{campillo-funollet2012}:
\begin{gather}
  \begin{split}
    %\pdt (\rho (\phi) \v v) + \div \left( \rho(\phi) \v v \otimes \v v \right) - \div \left[ \mu (\phi) (\grad \v v + \grad \v v ^T) + \v v \otimes \rho'(\phi) M(\phi) \grad g_\phi \right] + \grad p\\
    %= g_\phi \grad \phi + \sum_i g_{c_i} \grad c_i - \rho_e \grad V + \rho(\phi) \v g,
    \pdt (\rho (\phi) \v v) + \div \left( \rho(\phi) \v v \otimes \v v \right) - \div \left[ 2 \mu (\phi) \mathcal D \v v + \v v \otimes \rho'(\phi) M(\phi) \grad g_\phi \right] + \grad p\\
    = - \phi \grad g_\phi - \sum_i {c_i} \grad g_{c_i} %+ \rho(\phi) \v g
    ,
  \end{split} \label{eq:PF_NS1}\\
  \div \v v = 0, \label{eq:PF_NS2}\\
  \pdt \phi + \v v \cdot \grad \phi - \div(M(\phi) \grad g_\phi ) = 0, \label{eq:PF_PF1} \\
  \pdt c_j + \v v \cdot \grad c_j - \div ( K_j (\phi) c_j \grad g_{c_j}) = 0, \label{eq:PF_c} \\
  \div ( \varepsilon (\phi) \grad V ) = - \rho_e. \label{eq:PF_V}
\end{gather}
Here, $\phi$ is the phase field, and it takes the value $\phi=-1$ in phase $i=1$, and the value $\phi=1$ in phase $i=2$.
Eq.~\eqref{eq:PF_PF1} governs the conservative evolution of the phase field, wherein the diffusion term is controlled by the phase field mobility $M(\phi)$.
Here, $\rho$, $\mu$, $\varepsilon$, $K_j$ depend on which phase they are in, and are considered slave variables of the phase field $\phi$.
Across the interface these quantities interpolate between the values in the two phases:
\begin{align}
  \label{eq:rho_intp}
  \rho (\phi) &= \frac{\rho_1 + \rho_2}{2} + \frac{\rho_1 - \rho_2}{2} \phi, \\
  \label{eq:mu_intp}
  \mu (\phi ) &= \frac{\mu_1 + \mu_2}{2} + \frac{\mu_1 - \mu_2}{2} \phi, \\
  \label{eq:eps_intp}
  \varepsilon (\phi) &= \frac{\varepsilon_1 + \varepsilon_2}{2} + \frac{\varepsilon_1 - \varepsilon_2}{2} \phi, \\
  \label{eq:Ki_intp}
  K_j (\phi) &= \frac{K_{1, j} + K_{2, j}}{2} + \frac{K_{1, j} - K_{2, j}}{2} \phi.
\end{align}
These averages are all weighted arithmetically, although other options are available.
For example, \citet{tomar2007} found that, in the case of a level-set method with smoothly interpolated phase properties, using a weighted harmonic mean gave more accurate computation of the electric field.
However, \citet{lopez-herrera2011} found no indication that the harmonic mean was superior when free charges were present, and hence we adopt for simplicity and computational performance the arithmetic mean, although it remains unsettled which mean would yield the most accurate result.

Further, the chemical potential of species $c_j$ is given by
\begin{align}
  g_{c_j}(c_j, \phi) = \alpha'(c_j) + \beta_j (\phi) + z_j V,
  \label{eq:PF_chempot}
\end{align}
where we, for dilute solutions, may model $\alpha(c) = c (\log c - 1) $ to obtain consistency with the standard Nernst--Planck equation.
Further, we use a weighted arithmetic mean for the solubility parameters $\beta_j$:
\begin{equation}
  \beta_j (\phi) = \frac{\beta_{1, j} + \beta_{2, j}}{2} + \frac{\beta_{1,j} - \beta_{2, j}}{2} \phi,
  \label{eq:beta_wam}
\end{equation}
which, under the assumption of dilute solutions and with the interpretation \eqref{eq:beta_ij_interpretation}, corresponds to a weighted geometric mean for the reference concentrations:
\begin{equation}
  c_j^{\rm ref}(\phi) = \left( c^{{\rm ref}, 1}_j \right)^{\frac{1+\phi}{2}} \cdot \left( c^{{\rm ref}, 2}_j \right)^{\frac{1-\phi}{2}} .
\end{equation}
In analogy with $g_{c_j}$ being the chemical potential of species $c_j$, we denote $g_\phi$ as the chemical potential of the phase field $\phi$.
%The chemical potential of the phase field is defined as
It is given by:
\begin{equation}
  % g_\phi = \frac{3 \sigma}{2 \sqrt{2}} \left[ \epsilon^{-1} W'(\phi) - \epsilon \laplacian \phi \right] + ,
  g_\phi = \pd f \phi - \div \pd f {\grad \phi} + \sum_j \beta_j ' (\phi) c_j - \frac 1 2 \varepsilon'(\phi) | \grad V |^2 .
\end{equation}
The free energy functional $f$ of the phase field is defined by
\begin{equation}
  f(\phi, \grad \phi) = \frac{3 \sigma}{2 \sqrt{2}} \left[ \frac{\epsilon}{2} |\grad \phi|^2 + \epsilon^{-1} W(\phi) \right] = \tilde \sigma \left[ \frac{\epsilon}{2} |\grad \phi|^2 + \epsilon^{-1} W(\phi) \right],
\end{equation}
where where $\sigma$ is the surface tension, $\epsilon$ is the interface thickness, and $W(\phi)$ is a double well potential.
Here, we use %, unless otherwise stated, 
 $W(\phi) = (1-\phi^2)^2/4$.
With this free energy, we obtain
\begin{equation}
  g_\phi = \tilde\sigma\epsilon^{-1}W'(\phi) - \tilde\sigma\epsilon\laplacian\phi + \sum_j \beta_j ' (\phi) c_j - \frac 1 2 \varepsilon'(\phi) | \grad V |^2 .
  \label{eq:PF_g}
\end{equation}
We will assume this form throughout.

% Due to the form of Eq.~\eqref{eq:PF_chempot}, we can write
% \begin{equation}
%   g_{c_i} \grad c_i = \alpha'(c_i )\grad c_i + \beta_i (\phi) \grad c_i = \grad ( \alpha(c_i) + \beta_i(\phi) c_i) - \beta_i'(\phi) c_i \grad \phi,
% \end{equation}
% and thereby,
% \begin{align}
%   g_\phi \grad \phi + \sum_i g_{c_i} \grad c_i &= \left( \pd f \phi - \div \pd f {\grad \phi} - \frac 1 2 \varepsilon'(\phi) | \grad V |^2\right) \grad \phi + \grad ( \alpha(c_i) + \beta_i(\phi) c_i) \\
%                                                \begin{split} = \grad \left( f(\phi, \grad \phi) + \sum_i ( \alpha(c_i) + \beta_i(\phi) c_i) \right) - \div \left[ \pd f {\grad \phi} \otimes \grad \phi \right] \\ - \frac 1 2 \grad \varepsilon(\phi) | \grad V |^2.
%                                                  \end{split}
% \end{align}
% The contribution from the first term can be absorbed into the pressure by the (re)definition
% \begin{equation}
%   \tilde p = p - f(\phi, \grad \phi) - \sum_i ( \alpha(c_i) + \beta_i(\phi) c_i ) .
% \end{equation}
% The second and third terms are interface forces.
After some rewriting, exploiting Eq.~\eqref{eq:PF_NS2} and the fact that $\rho'(\phi)$ is constant due to Eq.~\eqref{eq:rho_intp}, Eq.~\eqref{eq:PF_NS1} can be expressed as
\begin{equation}
  \begin{split}
    % \rho(\phi) \pdt \v v + \left( \left(\rho(\phi) \v v - \rho'(\phi) M(\phi) \grad g_\phi  \right) \cdot \grad \right) \v v -
    % \div \left[ 2 \mu(\phi) \sym (\grad \v v) \right] + \grad p \\
    % = - \rho_e \grad V - \div \left[ \pd f {\grad \phi} \otimes \grad \phi \right]  - \frac 1 2 \varepsilon'(\phi) | \grad V |^2 \grad \phi + \rho(\phi) \v g
    \rho(\phi) \pdt \v v + \left( \left(\rho(\phi) \v v - \rho'(\phi) M(\phi) \grad g_\phi  \right) \cdot \grad \right) \v v -
    \div \left[ 2 \mu(\phi) \mathcal D \v v \right] + \grad p \\
    = - \phi \grad g_\phi - \sum_j c_j \grad g_{c_j} %+ \rho(\phi) \v g 
    .
  \end{split}
\end{equation}
% where we have omitted the tilde over $p$. Inserting for $f$, we obtain
% \begin{equation}
%   \begin{split}
%     \rho(\phi) \pdt \v v + \left( \left(\rho(\phi) \v v - \rho'(\phi) M(\phi) \grad g_\phi  \right) \cdot \grad \right) \v v -
%     \div \left[ 2 \mu(\phi) \sym (\grad \v v) \right] + \grad p \\
%     = - \rho_e \grad V - \tilde \sigma \epsilon\div \left[ \grad \phi \otimes \grad \phi \right]  - \frac 1 2 \varepsilon'(\phi) | \grad V |^2 \grad \phi + \rho(\phi) \v g .
%   \end{split}
% \end{equation}

\subsubsection{Phase field mobility}
Given a proper definition of the phase-field mobility $M(\phi)$, the phase-field model should reduce to the sharp-interface model given in the previous section.
As discussed at length in Ref.~\cite{campillo-funollet2012}, the two following ways are viable options:
\begin{subequations}
\begin{align}
  M(\phi) &= \epsilon M_0, \label{eq:pf_mobility_1} \\
  M(\phi) &= M_0 (1-\phi^2)_+. \label{eq:pf_mobility_2}
\end{align}
\end{subequations}
Here $M_0$ is a constant, and $(\cdot)_+ = \max(\cdot, 0)$.
Other formulations of $M$ are possible; some of these will in the limit of vanishing interface width reduce to a sharp-interface model where the interface velocity does not equal the fluid velocity \cite{abels2012,campillo-funollet2012}.

\subsubsection{Boundary conditions}
Some of the interface conditions from the sharp-interface model carry over to the phase field model, but in addition, some new conditions must be specified for the phase field.
Here we give a brief summary.
We assume that the boundary of the domain $\Omega$, $\partial\Omega$, can be divided into an inlet part $\partial\Omega_\inlet$, an outlet part $\partial\Omega_\outlet$, and a wall part $\partial\Omega_\wall$.
We shall primarily discuss the latter here.

For the velocity field, we assume the no-slip condition
\begin{equation}
  \v v (\v x, t) = \v 0 \quad \textrm{for} \quad \v x \in \partial\Omega_\wall.
\end{equation}
Alternatively, a no-flux condition and a slip law could have been used; in particular, a generalized Navier boundary condition (GNBC) has been shown to hold yield a consistent description of the contact line motion \cite{qian2003,qian2006}.
%However, the needed slip length is typically the same order of magnitude as the mean free path of a molecule in the fluid, and hence only relevant in nanofluidics.
%This is orders of magnitude smaller than the settings we consider.
However, to limit the scope, the moving contact line paradox will in this work be overcome by interface diffusion.

With regards to the flow problem, the pressure gauge needs to be fixed.
To this end, the pressure could be fixed somewhere on the boundary, or the pressure nullspace could be removed.

For the concentrations $c_j$, we may use a prescribed concentration, or the no-flux condition
\begin{align}
  \hat{\v n} \cdot \left(K_j(\phi) c_j \grad g_{c_j} \right) = 0 \quad \textrm{on} \quad \partial\Omega_\wall.
\end{align}
For the electric potential, we use either the Dirichlet condition $V=\bar V$ (which is reasonable at either inlet or outlet), or in the presence of charged (or neutral) boundaries, the condition
\begin{equation}
  \hat{\v n} \cdot \grad V = \frac{\sigma_e}{\varepsilon(\phi)} \quad \textrm{on} \quad \partial\Omega_\wall,
\end{equation}
%where $\sigma_e$ is the free surface charge per area.
similar to the sharp-interface condition.
Note that $\sigma_e(\v x)$ is prescribed and can vary over the boundary. 

We assume that the no-flux conditons hold on the phase field chemical potential,
\begin{equation}
  \hat{\v n} \cdot \grad g_\phi = 0 \quad \textrm{on} \quad \partial\Omega_\wall.
\end{equation}
For the phase field itself, a general dynamic wetting boundary condition can be expressed as \cite{carlson2012}:
\begin{equation}
  \epsilon \tau_w \pdt \phi = - \tilde\sigma \epsilon \hat{\v n} \cdot \grad \phi + \tilde\sigma \cos(\theta_e) f_w' (\phi),
  \label{eq:BC_phi_dynamic}
\end{equation}
where $\theta_e$ is the equilibrium contact angle, $\tau_w$ is a relaxation parameter, and $f_w(\phi) = (2+3\phi-\phi^3)/4$ interpolates smoothly between 0 (at $\phi=-1$) and 1 (at $\phi=1$).
In this work, we limit ourselves to studying fixed contact angles, i.e.~considering Eq.~\eqref{eq:BC_phi_dynamic} with $\tau_w = 0$.
For a GNBC, the phase-field boundary condition \eqref{eq:BC_phi_dynamic} must be modelled consistently with the slip condition on the velocity \cite{qian2006}.

%\subsection{Nondimensionalization}
%Binary ElectRohydrydyNAmIc SolvEr

\section{Discretization}
\label{sec:numerics}
For solving the equations of two-phase EHD, i.e.~the model consisting of Eqs.~\eqref{eq:PF_NS1}--\eqref{eq:PF_V}, there are four operations that must be performed:% (either simultaneously or in arbitrary order):
\begin{enumerate}
\item Propagate the phase field $\phi$. \label{pt:splitscheme_phi}
\item Propagate the chemical species concentrations $c_i$. \label{pt:splitscheme_ci}
\item Update the electric potential $V$ \label{pt:splitscheme_V}
\item Propagate the velocity $\v v$ and pressure $p$. \label{pt:splitscheme_vp}
\end{enumerate}
The whole system of equations could in principle be solved simultaneously using implicit Euler discretization in time and e.g.~Newton's method to solve the nonlinear system.
However, in order to simulate larger systems faster, it is preferable to use a splitting scheme to solve for each field sequentially.
One such splitting scheme was outlined in \cite{metzger2015}, based on the energy-stable scheme without electrochemistry as developed by \cite{guillen-gonzalez2014,grun2016}.
However, that scheme did not take into account that the electric permittivities in the two fluids may differ, and when they do, the phase field and the electrochemistry computations become coupled through the electric field \cite{metzger2018}.
We will here discuss two strategies for solving the coupled problem of two-phase electrohydrodynamics.
First, we present the fully monolithic, non-linear scheme, and secondly, we propose a new, fully practical linear operator splitting scheme.
As we are not aware of any splitting schemes that are second-order accurate in time for the case of unmatched densities, we shall constrain our discussion to first-order in time schemes.

In the forthcoming, we will denote the inner product of any two scalar, vector, or tensor fields $\mathcal A, \mathcal B$ by $\I{\mathcal A}{\mathcal B}$.
Further, we consider a discrete time step $\tau$, and denote the (first-order) discrete time derivative by
\begin{equation}
  \pdtau \mathcal A^k = \frac{\mathcal A ^k - \mathcal A ^\kmone}{\tau}.
\end{equation}
The equations are discretized on the domain $\Omega \subset \mathbb R^d$, $d=2,3$, with the no-slip boundary $\Gamma$.
Since we do not consider explicitly in- and outlet boundary conditions in this work, we will omit this possible part of the domain for the sake of brevity.

We define the following finite element subspaces:
\begin{align}
  \v V_h &= (V_h)^d \quad \textrm{where} \quad V_h = \left\{ v \in H^1 (\Omega) \right\} \quad \textrm{for velocity}, \\
  P_h &= \left\{ p \in L_0^2 (\Omega) \right\} \quad \textrm{for pressure}, \\
  \Phi_h       &= \left\{ \phi \in H^1 (\Omega) \right\} \quad \textrm{for phase field}, \\
  G_h &= \left\{ g \in H^1(\Omega) \right\} \quad \textrm{for phase field chemical potential}, \\
  C_h &= \left\{ c \in H^1 (\Omega) \right\} \quad \textrm{for concentrations}, \\
  U_h &= \left\{ V \in H^1 (\Omega) \right\} \quad \textrm{for the electrostatic potential}.
\end{align}

\subsection{Monolithic scheme}
Here we give the fully implicit scheme that follows from a na\"{i}ve implicit Euler discretizion of the model \eqref{eq:PF_NS1}--\eqref{eq:PF_V}, and supplemented by Eq.~\eqref{eq:PF_g}. %eq:PF_NS2eq:PF_PF1eq:PF_ceq:PF_V.

Assume that  $(\v v^\kmone, p^\kmone, \phi^\kmone, g_\phi^\kmone, c_1^\kmone, \ldots, c_M^\kmone, V^\kmone)$  is given.
The scheme can then be summarized by the following.
Find $(\v v^{k}, p^k, \phi^k, g_\phi^k, c_1^k, \ldots, c_N^k, V^k) \in \v V_h \crossprod P_h \crossprod \Phi_h \crossprod G_h \crossprod (C_h)^N \crossprod \mathcal U_h$ such that
\begin{subequations}
  \label{eq:scheme_monolithic}
\begin{multline}
  \I{\rho^k \pdtau{ \v v^k }}{\v u}
  + \I{\left( \v m^k \cdot \grad \right) \v v^k }{\v u}
  % - \I{ \div \left[ 2 \mu^k \mathcal D \v v^k \right]}{\v u}
  + \I{ 2 \mu^k \mathcal D \v v^k }{\mathcal D \v u}
  % + \I{\grad p^k}{\v u} \\
  - \I{p^k}{\div \v u} \\
  = - \I{\phi^k \grad g_\phi^k}{\v u}
  - \sum_j \I{c_j^k \grad g_{c_j}^k}{\v u} %+ \I{\rho^k \v g}{\v u} 
  ,
\end{multline}
\begin{equation}
  \I{\div \v v^k}{q} = 0,
\end{equation}
\begin{equation}
  \I{\pdtau \phi^k}{\psi}
  - \I{\v v^k \phi^k}{\grad \psi}
  % - \I{\div(M^k \grad g_\phi^k )}{\psi}
  + \I{M^k \grad g_\phi^k }{\grad \psi}
  = 0,
\end{equation}
\begin{multline}
  \I{g_\phi^k}{g_\psi} =
  \I{\tilde\sigma\epsilon^{-1}W'(\phi^k)}{g_\psi}
  % - \I{\tilde\sigma\epsilon\laplacian \phi^k}{h}
  %% - \IA{\tilde\sigma\epsilon \hat{\v n} \grad \phi^k h} \tilde\sigma \cos(\theta_e) f_w' (\phi)
  - \tilde\sigma \cos(\theta_e) \IA{ f_w' (\phi^k) g_\psi}
  + \I{\tilde\sigma\epsilon\grad \phi^k}{\grad g_\psi} \\
  + \sum_j \I{ \beta_j ' c_j^k}{g_\psi}
  - \I{\frac 1 2 \varepsilon' | \grad V^k |^2}{g_\psi} ,
\end{multline}
\begin{equation}
  \I{\pdtau c_j^k}{b_j}
  - \I{\v v^k c_j^k}{\grad b_j}
  % - \I{\div (  K_j^k c_j^k \grad g_{c_j}^k)}{b_j}
  + \I{ K_j^k c_j^k \grad g_{c_j}^k}{\grad b_j}
  = 0,
\end{equation}
\begin{equation}
  % \I{\div ( \varepsilon ^k \grad V^k )}{U} = - \I{\rho_e^k}{U}
  \I{\varepsilon ^k \grad V^k }{\grad U} = \I{\rho_e^k}{U} + \IA{ \sigma_e  U},
\end{equation}
\end{subequations}
for all test functions $(\v u, q, \psi, g_\psi, b_1, \ldots, b_N, U) \in \v V_h \crossprod P_h \crossprod \Phi_h \crossprod G_h \crossprod (C_h)^N \crossprod \mathcal U_h$. 
Here we have used
\begin{equation}
  \v m^k = \rho^k \v v^k - \rho' M^k \grad g_\phi^k
\end{equation}
and the shorthands
\begin{gather*}
  \rho^k = \rho(\phi^k), \quad
  \mu^k = \mu(\phi^k), \quad
  M^k = M(\phi^k), \quad
  \varepsilon^k = \varepsilon (\phi^k), \\
  K_j^k = K_j (\phi^k), \quad \textrm{and} \quad
  \rho_e^k = \rho_e ( \{ c_j^k \} ). 
\end{gather*}
Note that Eqs.~\eqref{eq:scheme_monolithic} constitute a fully coupled non-linear system and the equations must thus be solved simultaneously, preferably using a Newton method.
This results in a large system matrix which must be assembled and solved iteratively, and for which there are in general no suitable preconditioners available.
On the other hand, the scheme is fully implicit and hence expected to be fairly robust with regards to e.g.~time step size.
There are in general several options for constructing the linearized variational form to be used in a Newton scheme. %, but in this work we shall resort to the solver automatically generated by FEniCS.

\subsection{A linear splitting scheme}
Now, we introduce a linear operator splitting scheme.
This scheme splits between the processes of phase-field transport, chemical transport under an electric field, and hydrodynamic flow, such that the equations governing each of these processes are solved separately.

\paragraph{Phase field step}
Find $(\phi^k, g_\phi^k) \in \Phi_h \crossprod G_h$ such that
\begin{subequations}
\begin{equation}
  % \int_\Omega \frac{\phi^{k+1}-\phi^k}{\Delta t} \psi \, \diff V + \int_\Omega \v v^k \cdot \grad \phi^{k+1} \psi \, \diff V + \int_\Omega M(\phi^k) \grad g_\phi^{k+1} \cdot \grad \psi \, \diff V %- \int_{\partial\Omega} M(\phi^k) \grad g_\phi^{k+1} \psi  \cdot \, \diff \v S
  % = 0
  \I{ \pdtau \phi^k}{ \psi } - \I{ \v v^\kmone \phi^k }{ \grad \psi } + \I{ M^\kmone \grad g_\phi^k }{ \grad \psi }
  = 0
\end{equation}
\begin{multline}
  % \int_\Omega g_\phi^{k+1} h \, \diff V = \tilde\sigma\epsilon^{-1} \int_\Omega \overline{W'}(\phi^{k+1}, \phi^{k}) h \, \diff V + \tilde\sigma\epsilon \int_\Omega \grad \phi^{k+1} \cdot \grad h \, \diff V \\
  % % - \tilde\sigma\epsilon \int_{\partial \Omega} \v n \cdot \grad \phi^{k+1} h \, \diff A
  % + \int_{\partial \Omega}  \left( \epsilon \tau_w \frac{\phi^{k+1}-\phi^k}{\Delta t}  - \tilde\sigma\cos(\theta_e) \overline{f_w'}(\phi^{k+1}, \phi^{k}) \right) h \, \diff A \\
  % + \sum_i \beta_i' \int_\Omega  ' c_i^k h \, \diff V - \frac 1 2 \varepsilon'  \int_\Omega | \grad V^k |^2 h \, \diff V
  \I{ g_\phi^{k} }{g_\psi} = \tilde\sigma\epsilon^{-1} \I{ \overline{W'}(\phi^{k}, \phi^{\kmone}) }{g_\psi} + \tilde\sigma\epsilon \I{ \grad \phi^{k} }{ \grad g_\psi } \\
  - \tilde\sigma \cos(\theta_e) \IA{ \overline{f_w'}(\phi^{k}, \phi^{\kmone}) \, g_\psi }
  + \sum_j \beta_j'  \I{ c_j^\kmone }{g_\psi} - \frac 1 2 \varepsilon'  \I { | \grad V^\kmone |^2}{g_\psi},
\end{multline}
\end{subequations}
for all test functions $(\psi, g_\psi) \in \Phi_h \crossprod G_h$.
Here, $\overline{W'}(\phi^{k}, \phi^{\kmone})$ is a linearization of $W'(\phi^{k})$ around $\phi^\kmone$:
\begin{equation}
  \overline{W'}(\phi^{k}, \phi^{\kmone}) = W'(\phi^\kmone) + W''(\phi^\kmone) (\phi^{k}-\phi^\kmone).
\end{equation}
We have also used the discretization of Eq.~\ref{eq:BC_phi_dynamic}
\begin{equation}
  % \tilde\sigma\epsilon \v n \cdot \grad \phi^{k+1} = - \epsilon \tau_w \frac{\phi^{k+1}-\phi^k}{\Delta t}  + \tilde\sigma\cos(\theta_e) \overline{f_w'}(\phi^{k+1}, \phi^{k}),
  \tilde\sigma\epsilon \v n \cdot \grad \phi^{k} = \tilde\sigma\cos(\theta_e) \overline{f_w'}(\phi^{k}, \phi^{\kmone}),
\end{equation}
where we have used the linearization
\begin{equation}
  \overline{f_w'}(\phi^{k}, \phi^{\kmone}) = f_w'(\phi^\kmone) + f_w''(\phi^\kmone)(\phi^{k}-\phi^\kmone).
\end{equation}

\paragraph{Electrochemistry step}
Find $(c_1, \ldots, c_N, V) \in (C_h)^N \crossprod U_h$ such that
\begin{subequations}
\begin{equation}
  \I{\pdtau c_j^k }{ b_j } - \I{ \v v^\kmone c_j^{k}}{ \grad b_j } 
  + \I{ \bar{\v J}_{c_j}^k}{\grad b_i } = 0
\end{equation}
\begin{equation}
  \I{ \varepsilon^{k} \grad V^{k} }{ \grad U }
  % - \int_{\partial\Omega} \varepsilon(\phi^{k+1}) \v n \cdot \grad V^{k+1}  U \, \diff A
  + \IA{ \sigma_e  U }
  + \I{  \rho_e^{k}}{ U } = 0
\end{equation}
\end{subequations}
for all test functions $(b_1, \ldots, b_N, U) \in (C_h)^N \crossprod U_h$.
Here $\bar{\v J}_{c_j}^k$ is a linear approximation of the diffusive chemical flux $\v J_{c_j} = K_j(\phi) c_j \grad g_{c_j}$.
For conciseness, we here constrain our analysis to ideal chemical solutions, i.e.\ we assume a common chemical energy function on the form $\alpha({c}) = c (\ln c - 1)$.
To this end, we approximate the flux by:
\begin{equation}
  \bar{\v J}_{c_j}^k = K_j^k (\grad c_i^{k} + c_i^k \beta_i' \grad\phi^k  + z_i c_i^{\kmone} \grad V^{k} ).
\end{equation}

\paragraph{Fluid flow step}
Find $(\v v^k, p^k) \in \v V_h \crossprod P_h$ such that 
\begin{subequations}
\begin{multline}
  \I{ \rho^\kmone \pdtau \v v^k }{\v u}
  + \I{ \left( \bar{\v m}^\kmone \cdot \grad \right) \v v^{k} }{ \v u } \\
  %+ \frac{1}{2} \I{ \left(\pdtau \rho^k + \div \bar{\v m}^\kmone \right) \v v^{k}}{\v u}
  + \frac{1}{2} \I{ \v v^{k} \pdtau \rho^k }{\v u}
  - \frac{1}{2} \I{ \bar{\v m}^\kmone }{ \grad (\v v^{k} \cdot \v u)}
  + \I{ 2\mu^{k} \mathcal D \v v^{k}}{ \mathcal D \v u }
  - \I{ p^{k} }{\div \v u } \\
  %+ \int_{\partial\Omega_\inlet} p_\inlet \v n \cdot \v u \, \diff A \\
%  + \int_{\partial\Omega_\outlet} p_\outlet \v n \cdot \v u \, \diff A \\
  = %- \int_\Omega \rho_e (\{ c_i^{k+1} \}) \grad V^{k+1} \cdot \v u \, \diff V
  %+ \tilde\sigma\epsilon \int_\Omega ( \grad \phi^{k+1} \otimes \grad \phi^{k+1} ) : \grad \v u \, \diff V \\
  % - \int_{\partial\Omega} \left[ \tilde\sigma \epsilon \v n \cdot \grad\phi \right] \v u \cdot \grad\phi \, \diff A\\
  %- \int_{\partial\Omega} \left[ - \epsilon \tau_w \frac{\phi^{k+1}-\phi^k}{\Delta t}  + \tilde\sigma\cos(\theta_e) \overline{f_w'}(\phi^{k+1}, \phi^{k}) \right] \v u \cdot \grad\phi^{k+1} \, \diff A\\
  %- \frac{1}{2} \varepsilon' \int_\Omega |\grad V^{k+1}|^2 \grad\phi^{k+1} \cdot \v u \, \diff V
  - \I{\phi^{k} \grad g_\phi^k}{\v u}
  - \sum_j \I{c_j^{k} \grad g_{c_j}^k}{\v u}
  %+ \I{ \rho^k \v g }{ \v u }
  \label{eq:splitscheme_NS1}
\end{multline}
\begin{equation}
  \I{ q }{ \div \v v^k } = 0
\end{equation}
\end{subequations}
for all test functions $(\v u, q) \in \v V_h \crossprod P_h$.
Here, we have used the following approximation of the advective momentum:
\begin{align}
  \bar{\v m}^\kmone = \rho^\kmone \v v^\kmone - \rho' M^k \grad g_\phi^{k}.
\end{align}
Note that the terms in \eqref{eq:splitscheme_NS1} involving $ \pdtau \rho^k + \div \bar{\v m}^\kmone $, which is a discrete approximation of $\pdt \rho + \div \v m = 0$, is included to satisfy a discrete energy dissipation law \cite{shen2015} (i.e., to improve stability).
This step requires solving for the velocity and pressure in a coupled manner. 
This has the advantage that it yields accurate computation of the pressure, but the drawback that it is computationally challenging to precondition and solve, related to the Babuska--Brezzi (BB) condition (see e.g.\ \cite{brenner2007}).
Alternatively, it might be worthwhile to further split the fluid flow step into the following three substeps, at the cost of some lost accuracy \cite{langtangen2002}.
\begin{subequations}
\begin{itemize}
  \item Tentative velocity step: Find $\tilde{\v v}^k \in \v V_h$ such that for all $\v u \in \v V_h$,
\begin{multline}
  \left(\rho^{k-1} \frac{\tilde{\v v}^{k}-\v v^{k-1}}{\tau}, \v u\right) 
  + \left( (\bar{\v m}^{k-1} \cdot \grad ) \tilde{\v v}^{k} , \v u \right)
  + \left( 2 \mu^k \mathcal D \tilde{\v v}^{k}, \mathcal D \v u \right)
  - \left(p^{k-1}, \div \v u \right) \\
  + \frac{1}{2} \left(  \tilde{\v v}^k \pdtau \rho^k, \v u \right) 
  - \frac{1}{2} \left( \bar{\v m}^{k-1} ,  \grad (\tilde{\v v}^k \cdot \v u) \right)
  = 
  - \I{\phi^{k} \grad g_\phi^k}{\v u}
  - \sum_i \left( c_i^{k-1} \grad g_i^k, \v u \right),
  \label{eq:split_u_1}
\end{multline}
with the Dirichlet boundary condition $\tilde{\v v}^k = \v 0$ on $\Gamma$.
\item Pressure correction step:
Find $p^k \in P_h$ such that for all $q \in P_h$, we have
\begin{align}
  \left( \frac{1}{\rho_0} \grad (p^k - p^{k-1}) , \grad q \right) = - \frac{1}{\tau} \left( \div \tilde{\v v}^k, q \right).
  \label{eq:split_p}
\end{align}
%where the artificial Neumann condition $\v n \cdot \grad (p^k - p^{k-1}) = 0$ has been enforced.
%Note that this induces an $O(\tau)$ error.
\item Velocity correction step:
Then, find $\v v^k \in \v V_h$ such that for all $\v u \in \v V_h$,
\begin{align}
  \left( \rho^{k} \frac{\v v^k -\tilde{\v v}^k}{\tau}, \v u \right) = \left( p^k - p^{k-1} , \div \v u \right),
  \label{eq:split_u_2}
\end{align}
which we solve by explicitly imposing the Dirichlet boundary condition $\v u^k = \v 0$ on $\Gamma$. %, which supresses the error from the Neumann condition above.
\end{itemize}
\end{subequations}
Eqs.\ \eqref{eq:split_u_1}, \eqref{eq:split_p}, and \eqref{eq:split_u_2} should be solved sequentially, and constitutes a variant of a projection scheme, i.e., a fractional-step approach to the fluid flow equations \cite{chorin1967,chorin1968,langtangen2002,guermond2006,shen2015}.
We will in this paper refer to the coupled solution of the fluid flow equations, unless stated otherwise.
Specifically, the fractional-step fluid flow scheme will only be demonstrated in the full 3D simulations in Sec.\ \ref{sec:charged_droplets_3d}.
% By a redefinition of the pressure, we may write \eqref{eq:splitscheme_NS1} as
% \begin{multline}
%   \I{ \rho^\kmone \pdtau \v v^k }{\v u}
%   + \I{ \left( \bar{\v m}^\kmone \cdot \grad \right) \v v^{k} }{ \v u } \\
%   + \frac{1}{2} \I{ \left(\pdtau \rho^k  \right) \v v^{k}}{\v u}
%   - \frac{1}{2} \I{ \bar{\v m}^\kmone }{ \grad (\v v^{k} \cdot \v u)}
%   + \I{ 2\mu^{k} \mathcal D \v v^{k}}{ \mathcal D \v u }
%   - \I{ p^{k} }{\div \v u } \\
%   = - \I{ \rho_e^{k} \grad V^{k} }{ \v u }
%   + \tilde\sigma\epsilon \I{ \grad \phi^{k} \otimes \grad \phi^{k} }{ \grad \v u } \\
%   - \tilde\sigma\cos(\theta_e) \IA{ \overline{f_w'}(\phi^{k}, \phi^\kmone) \v u \cdot \grad\phi^{k} }%\\
%   - \frac{1}{2} \varepsilon' \I{ |\grad V^{k}|^2 \grad\phi^{k} }{ \v u }
%   %+ \I{ \rho^k \v g }{ \v u }
% \end{multline}

The scheme presented above consists in sequentially solving three decoupled subproblems (or five decoupled subproblems for the fractional-step fluid flow alternative).
The subproblems are all linear, and hence attainable for specialized linear solvers which could improve the efficiency.
We note that the splitting introduces an error of order $\tau$, i.e.~the same as the scheme itself.
Moreover, our scheme does not preserve the same energy dissipation law on the discrete level, that the original model does on the continuous level.
We are currently not aware of any scheme for two-phase electrohydrodynamics with this property, apart from the fully implicit scheme presented in the previous section.

\section{Bernaise}
\label{sec:bernaise}
We have now introduced the governing equations and two strategies for solving them.
Now, we will introduce the Bernaise package, and describe an implementation of a generic simulation problem and a generic solver in this framework.
For a complete description of the software, we refer to the online Git repository \cite{bernaise2018}.
The work presented herein refers to version 1.0 of \emph{Bernaise}, which is compatible with version 2017.2.0 of FEniCS \cite{logg2012}.

\subsection{Python package}
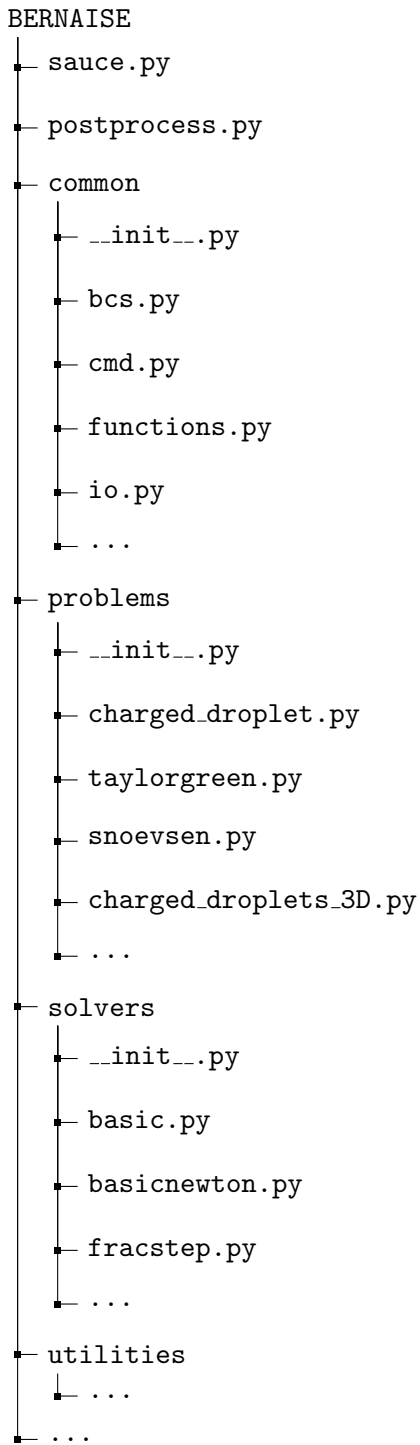
\begin{figure}[H]
  \begin{forest}
    for tree={
      font=\ttfamily,
      grow'=0,
      child anchor=west,
      parent anchor=south,
      anchor=west,
      calign=first,
      edge path={
        \noexpand\path [draw, \forestoption{edge}]
        (!u.south west) +(7.5pt,0) |- node[fill,inner sep=1.25pt] {} (.child anchor)\forestoption{edge label};
      },
      before typesetting nodes={
        if n=1
        {insert before={[,phantom]}}
        {}
      },
      fit=band,
      before computing xy={l=15pt},
    }
    [BERNAISE
    % [README.md]
      [sauce.py]
      [postprocess.py]
      [common
        [\_\_init\_\_.py]
        [bcs.py]
        [cmd.py]
        [functions.py]
        [io.py]
        [...]
      ]
      [problems
        [\_\_init\_\_.py]
        [charged\_droplet.py]
        [taylorgreen.py]
        [snoevsen.py]
        [charged\_droplets\_3D.py]
        [...]
      ]
      [solvers
        [\_\_init\_\_.py]
        [basic.py]
        [basicnewton.py]
        [fracstep.py]
        [...]
      ]
      [utilities
        [...]
      ]
      [...]
    ]
  \end{forest}
  \caption{\label{fig:dirtree}
    Part of the directory structure of Bernaise.}
\end{figure}

\emph{Bernaise} is designed as a Python package, and the main structure of the package is shown in Fig.~\ref{fig:dirtree}.
The package contains two main submodules, \texttt{problems} and \texttt{solvers}.
As suggested by the name, the \texttt{problems} submodule contains scripts where problem-specific geometries (or meshes), physical parameters, boundary conditions, initial states, etc., are specified.
We will in Sec.~\ref{subsec:problems} dive into the constituents of a problem script.
The \texttt{solvers} submodule, on the other hand, contains scripts that are implementations of the numerical schemes required to solve the governing equations.
Two notable examples that are implemented in \emph{Bernaise} are the monolithic scheme (implemented as \texttt{basicnewton}) and the linear splitting scheme (implemented as \texttt{basic}).
We shall in Sec.~\ref{subsec:solvers} describe the building blocks of such a solver.
Further, a default solver compatible with a given problem is specified in the problem, but this setting can---along with most other settings specified in a problem---be overridden by providing an additional keyword to the main script call (see below).
Note that \emph{not all} solvers are compatible with \emph{all} problems, and vice versa.

A simulation is typically run from a terminal, pointing to the \emph{Bernaise} directory, using the command
\begin{minted}[fontsize=\small]{bash}
>> python sauce.py problem=charged_droplet
\end{minted}
where \texttt{charged\_droplet} may be exchanged with another problem script of choice; albeit we will use \texttt{charged\_droplet} as a pedagogical example in the forthcoming.
The main script \texttt{sauce.py} fetches a \texttt{problem} and connects it with the \texttt{solver}.
It sets up the finite element problem with all the given parameters, initializes the finite element fields with the specified initial state, and solves it with the specified boundary condition at each time step, until the specified (physical) simulation time \texttt{T} is exceeded.
Any parameter in the problem can be overridden by specifying an additional keyword from the command line; for example, the simulation time can be set to 1000 by running the command:

\begin{minted}[fontsize=\small]{bash}
>> python sauce.py problem=charged_droplet T=1000
\end{minted}

After every given interval of steps, specified by the parameter \texttt{checkpoint\_interval}, a checkpoint is stored, including all fields, and all problem parameters at the time of writing to file.
The checkpoint can be loaded, and the simulation can be continued, by running the command:
\begin{minted}[fontsize=\small]{bash}
>> python sauce.py problem=charged_droplet \
   restart_folder=results_charged_droplet/1/Checkpoint/
\end{minted}
where the \texttt{restart\_folder} points to an appropriate checkpoint folder.
Here, the problem parameters stored within the checkpoint have precedence over the default parameters given in the \texttt{problem} script.
Further, any parameters specified by command line keywords have precedence over the checkpoint parameters.

The role of the main module \texttt{sauce.py} is to allocate the required variables to run a simulation, to import routines from the specified problem and solver, to iterate the solver in time, and to output and store data at appropriate times.
Hence, the main module works as a general interface to problems and solvers.
This is enabled by overloading a series of functions, such that problem- and solver-specific functions are defined within the problem and solver, respectively.
The structure of \texttt{sauce.py} is by choice similar to the \texttt{NSfracStep.py} script in the \emph{Oasis} solver \cite{mortensen2015}; both in order to appeal to overlapping user bases, and to keep the code readable and consistent with and similar to common FEniCS examples.
However, an additional layer of abstraction in e.g.~setting up functions and function spaces is necessary in order to handle a flexible number of subproblems and subspaces, depending on e.g.~whether phase field, electrochemistry or flow is disabled, or whether we are running with a monolithic or operator splitting scheme.
To keep the \emph{Bernaise} code as readable and easily maintainable as possible, we have consciously avoided uneccessary abstraction.
Only the boundary conditions (found in \texttt{common/bcs.py} are implemented as classes.

\subsection{The problems submodule}
\label{subsec:problems}
The basic user typically interacts with \emph{Bernaise} by implementing a \emph{problem} to be solved.
This is accessible to \emph{Bernaise} when put in the subfolder \texttt{problems}.
The implementation consists in overloading a certain set of functions; all of which are listed in the \texttt{problems/\_\_init\_\_.py} file in the problems folder.
The mandatory functions that must be overloaded for each problem are:
\begin{itemize}
\item \texttt{mesh}: defines the geometry.
  Equivalent to the \texttt{mesh} function in \emph{Oasis} \cite{mortensen2015}.
\item \texttt{problem}: sets up all parameters to be overloaded, including defining solutes and types of finite elements.
  The default parameters are defined in the \texttt{problems/\_\_init\_\_.py} file.
\item \texttt{initialize}: initializes all fields.
\item \texttt{create\_bcs}: sets all subdomains, and defines boundary conditions (including pointwise boundary condtions, such as pressure pinning).
  The boundary conditions are more thoroughly explained below.
\end{itemize}
Further, there are functions that \emph{may} be overloaded.
\begin{itemize}
\item \texttt{constrained\_domain}: set if the boundary is to be considered periodic.
\item \texttt{pf\_mobility}: phase field mobility function; cf.\ \eqref{eq:pf_mobility_1} and \eqref{eq:pf_mobility_2}.
\item \texttt{start\_hook}: hook called before the temporal loop.
\item \texttt{tstep\_hook}: hook called at each time step in the loop.
\item \texttt{end\_hook}: hook called at the end of the program.
\item \texttt{rhs\_source}: explicit source terms to be added to the right hand side of given fields; used e.g.~in the method of manufactured solution.
\end{itemize}
Note here the use of three \emph{hooks} that are called during the course of a simulation.
These are useful for outputting certain quantities during a simulation, e.g.~the flux through a cross section, or total charge in the domain.
The \texttt{start\_hook} could also be used to call a steady-state solver to initialize the system closer to equilibrium, e.g.~a solver that solves only the electrochemistry subproblem such that we do not have to resolve the very fast time scale of the initial charge equilibration.

In Listing \ref{code:problems}, we show an implementation of the \texttt{problems} function, which sets the necessary parameters that are required for the \texttt{charged\_droplet} case to run.
Here, the \texttt{solutes} array (which defines the solutes), contains only one species, but it can in principle contain arbitrarily many.
\begin{code}
\begin{minted}[frame=single,fontsize=\small]{python}
def problem():
    info_cyan("Charged droplet in an electric field.")

    # Define solutes
    # Format: name, valency, diffusivity in phase 1, diffusivity in phase 2,
    #         solubility energy in phase 1, solubility energy in phase 2
    solutes = [["c_p",  1, 1e-5, 1e-3, 4., 1.]]

    # Default parameters to be loaded unless starting from checkpoint.
    parameters = dict(
        solver="basic",                    # Solver to be used.
        folder="results_charged_droplet",  # Folder to store results in.
        dt=0.08,                           # Timestep
        t_0=0.,                            # Starting time
        T=8.,                              # Total simulation time
        grid_spacing=1./32,                # Mesh size
        interface_thickness=0.03,          # Extent of diffuse interface
        solutes=solutes,                   # Array of solutes defined above
        Lx=2.,                             # Length of domain along x
        Ly=1.,                             # Length of domain along y
        rad_init=0.25,                     # Initial droplet radius
        V_left=10.,                        # Potential at left side
        V_right=0.,                        # Potential at right side
        surface_tension=5.,                # Surface tension
        concentration_init=10.,            # Initial (total) concentration
        pf_mobility_coeff=0.00002,         # Phase field mobility coeff. (M_0)
        density=[200., 100.],              # Density in phase 1, phase 2
        viscosity=[10., 1.],               # Viscosity in phase 1, phase 2
        permittivity=[1., 1.]              # Permittivity in phase 1, phase 2
    )
    return parameters
\end{minted}
\captionof{listing}{The \texttt{problems} function for the \texttt{charged\_droplet} case.}
\label{code:problems}
\end{code}

In Listing \ref{code:initialize}, we show the code for the initialization stage.
Here, \texttt{initial\_pf} and \texttt{initial\_c} are functions defined locally inside the \texttt{charged\_droplet.py} problem script, that set the initial distributions of the phase field and the concentration field, respectively.
Here, it should be noted how the (boolean) parameters \texttt{enable\_PF}, \texttt{enable\_EC} and \texttt{enable\_NS} allow to switch on or off either the phase field, the electrochemistry or the hydrodynamics, respectively.
\begin{code}
\begin{minted}[frame=single,fontsize=\small]{python}
def initialize(Lx, Ly, rad_init, interface_thickness, solutes,
               concentration_init, restart_folder, field_to_subspace,
               enable_NS, enable_PF, enable_EC, **namespace):
    """ Create the initial state. """
    w_init_field = dict()
    if not restart_folder:
        x0, y0, rad0, c0 = Lx/4, Ly/2, rad_init, concentration_init
        # Initialize phase field
        if enable_PF:
            w_init_field["phi"] = initial_pf(
                x0, y0, rad0, interface_thickness,
                field_to_subspace["phi"].collapse())

        # Initialize electrochemistry
        if enable_EC:
            w_init_field[solutes[0][0]] = initial_c(
                x0, y0, rad0/3., c0, interface_thickness,
                field_to_subspace[solutes[0][0]].collapse())

    return w_init_field
\end{minted}
\captionof{listing}{The \texttt{initialize} function for the \texttt{charged\_droplet} case.}
\label{code:initialize}
\end{code}

\subsection{The solvers submodule}
\label{subsec:solvers}
Advanced users may develop solvers that can be placed in the \texttt{solvers} subdirectory.
In the same way as with the \texttt{problems} submodule, a solver implementation constists of overloading a range of functions which are defined in \texttt{solvers/\_\_init\_\_.py}.
\begin{itemize}
\item \texttt{get\_subproblems}: Returns a dictionary (\texttt{dict}) of the subproblems which the solver splits the problem into.
  This dictionary has points to the name of the fields and the elements (specified in problem) which the subspace is made up of.
\item \texttt{setup}: Sets up the FEniCS solvers for each subproblem.
\item \texttt{solve}: Defines the routines for solving the finite element problems, which are called at every time step.
\item \texttt{update}: Defines the routines for assigning updated values to fields, which are called at the end of every time step.
\end{itemize}
The module \texttt{solvers/basicnewton.py} implements the monolithic scheme, while the module \texttt{solvers/basic.py} implements the segregated solver \footnote{The latter also contains an equilibrium solver for the quiescent electrochemistry problem, mainly to be used for initialization purposes.}.
The problem is split up into the subproblems corresponding to whether we have a monolothic or segragated solver in the function \texttt{get\_subproblems}.
Within the \texttt{setup} function, the variational forms are defined, and the solver routines are initialized.
The latter are eventually called in the \texttt{solve} routine at every time step.
Note that the element types are defined within the problem, and that the solvers in general can be applied for higher-order spatial accuracy without further ado.
The task of \texttt{get\_subproblems} is simply to link the subproblem to the element specification.

In Listing \ref{code:get_subproblems}, we show how the \texttt{get\_subproblems} function is implemented in the \texttt{basic} solver.
As can be readily seen, the function formally splits the problem into the three subproblems \texttt{NS}, \texttt{PF}, and \texttt{EC}.
\begin{code}
\begin{minted}[frame=single,fontsize=\small]{python}
def get_subproblems(solutes, enable_NS, enable_PF, enable_EC, **namespace):
    """ Returns dict of subproblems the solver splits the problem into. """
    subproblems = dict()
    if enable_NS:
        subproblems["NS"] = [dict(name="u", element="u"),
                             dict(name="p", element="p")]
    if enable_PF:
        subproblems["PF"] = [dict(name="phi", element="phi"),
                             dict(name="g", element="g")]
    if enable_EC:
        subproblems["EC"] = ([dict(name=solute[0], element="c")
                              for solute in solutes]
                             + [dict(name="V", element="V")])
    return subproblems
\end{minted}
\captionof{listing}{The \texttt{get\_subproblems} subroutine of the \texttt{basic} solver.}
\label{code:get_subproblems}
\end{code}
The other functions (such as \texttt{setup}) are somewhat more involved, but can be found at the Git repository \cite{bernaise2018}.

Note that the implementations of the solvers presented above are sought to be short and humanly readable, and therefore quite straightforwardly %and na\"{i}vely
implemented.
There are several ways to improve the efficiency (and hence scalability) of a solver, at the cost of lost intuitiveness \cite{mortensen2015}.
%In the next section, we shall consider a more optimized implementation of a solver.

\subsection{Boundary conditions}
\label{subsec:bernaise_bcs}
Boundary conditions are among the few components of \emph{Bernaise} which are implemented as classes.
Physical boundary conditons may consist of a combination of Dirichlet and Neumann (or Robin) conditions, and the latter must be incorporated into the variational form.
The boundary conditions are specified in the specific problem script, while the variational form is set up in the solver. 
To promote code reuse, keeping the physical boundary conditions accessible from the problems side, and simultaneously independent of the solver, the various boundary conditions are stored as classes in a separate module.
The boundaries themselves should be set by the user within the \texttt{problem}.
By importing various boundary condition classes from \texttt{common/bcs.py}, the boundary conditions can be inferred at user-specified boundaries.

Within the \texttt{bcs} module, the base class \texttt{GenericBC} is defined.
The boolean member functions \texttt{is\_dbc} and \texttt{is\_nbc} specifies, respectively, whether the concrete boundary conditions impose a Dirichlet and Neumann condition, and both return false by default.
The base class is inherited by various concrete boundary conditon classes, and by overloading these two member functions, the member functions \texttt{dbc} or \texttt{nbc} are respectively called at appropriate times in the code.
There is a hierarchy of boundary conditions which inherit from each other.
Some of the boundary conditions currently implemented in \emph{Bernaise} are:
\begin{itemize}
\item \texttt{GenericBC}: Base class for all boundary conditions.
  \begin{itemize}
  \item \texttt{Fixed}: Dirichlet condition, applicable for all fields.
    \begin{itemize}
    \item \texttt{NoSlip}: The no-slip condition---a pure Dirichlet condition with the value $\v 0$, applicable for velocity.
    \item \texttt{Pressure}: Constant pressure boundary condition---adds a Neumann condition to the velocity, i.e.~a boundary term in the variational form.
    \end{itemize}
  \item \texttt{Charged}: A charged boundary---a Neumann conditon intended for use with the electric potential $V$.
  \item \texttt{Open}: An open boundary---a Neumann condition is applied.
  \end{itemize}
\end{itemize}
We note that when a no-flux condition is to be applied, no specific boundary condition class needs to be supplied, since the boundary term in the variational form then disappears (in particular when considering conservative PDEs).

As an example, we show in Listing \ref{code:create_bcs} the \texttt{create\_bcs} function within the \texttt{charged\_droplet} case.
Here, the boundaries \texttt{Wall}, \texttt{Left}, etc., are defined in the standard Dolfin way as instances of a \texttt{SubDomain} class.
\begin{code}
\begin{minted}[frame=single,fontsize=\small]{python}
def create_bcs(field_to_subspace, Lx, Ly, solutes, V_left, V_right,
               enable_NS, enable_PF, enable_EC,
               **namespace):
    """ The boundary conditions are defined in terms of field. """

    boundaries = dict(
        wall=[Wall(Lx)],
        left=[Left()],
        right=[Right(Lx)]
    )

    noslip = Fixed((0., 0.))

    bcs = dict()
    bcs_pointwise = dict()

    bcs["wall"] = dict()
    bcs["left"] = dict()
    bcs["right"] = dict()

    if enable_NS:
        bcs["wall"]["u"] = noslip
        bcs["left"]["u"] = noslip
        bcs["right"]["u"] = noslip
        bcs_pointwise["p"] = (0., "x[0] < DOLFIN_EPS && x[1] < DOLFIN_EPS")

    if enable_EC:
        bcs["left"]["V"] = Fixed(V_left)
        bcs["right"]["V"] = Fixed(V_right)

    return boundaries, bcs, bcs_pointwise
\end{minted}
\captionof{listing}{The \texttt{create\_bcs} function within the \texttt{charged\_droplet} case.}
\label{code:create_bcs}
\end{code}

\subsection{Post-processing}
\label{subsec:postprocessing}
An additional module provided in \emph{Bernaise} is the post-processing module.
It operates with methods analogously to how the main \emph{Bernaise} script operates with problems.
The base script \texttt{postprocess.py} pulls in the required method and analyses or operates on a specified folder.
The methods are located in the folder \texttt{analysis\_scripts/} and new methods can be implemented by users by adding scripts to this folder.

To exemplify its usage, we consider a method to analyse the temporal development of the energy.
This is done by navigating to the root folder and calling
\begin{minted}[fontsize=\small]{bash}
>> python postprocess.py method=energy_in_time folder=results_charged_droplet/1/
\end{minted}
where we assume that the output of the simulation, we want to analyse, is found in the folder \texttt{results\_charged\_droplet/1/}.
The analysis method \texttt{energy\_in\_time} above can, of course, be exchanged with another method of choice.
A list of available methods can be produced by supplying the help argument from a terminal call:
\begin{minted}[fontsize=\small]{bash}
>> python postprocess.py -h
\end{minted}

Similar to the \texttt{problems} submodule, the methods are implemented by overloading a set of routines, where default routines are found in \texttt{analysis\_scripts/\_\_init\_\_.py}.
The routines required to implement an analysis method are the following:
\begin{itemize}
\item \texttt{description}: routine called when a question mark is added to the end of the method name during a call from the terminal, meant to obtain a description of the method without having to inspect the code.
\item \texttt{method}: the routine that performs the desired analysis.
\end{itemize}
The implementation hinges on the \texttt{TimeSeries} class (located in \texttt{utilities/TimeSeries.py}), which efficiently imports the XDMF/HDF5 data files and the parameter files produced by a \emph{Bernaise} simulation.
Several plotting routines are implemented in \texttt{utilities/plot.py}, and these are extensively used in various analysis methods.

\section{Validation}
\label{sec:validation}
With the aim of using \emph{Bernaise} for quantitative purposes, it is essential to establish that the schemes presented in the above converges to the correct solution---in two senses:
\begin{itemize}
\item The numerical schemes should converge to the correct solution of the phase-field model.
\item The solution of the phase-field model should converge to the correct sharp-interface equations \footnote{Obviously, when the physical interface thickness may be resolved by the phase field, the sharp-interface assumption might be less sensible than the diffuse. Hence, in such cases this point might be too crude.}.
\end{itemize}
Unless otherwise stated, we mean by convergence that the error in all fields $\chi$ should behave like,
\begin{gather}
  %\norm{\chi-\chi_e}_h \sim h^{k_h} \quad \textrm{in space,}\\
  %\norm{\chi-\chi_e}_h \sim \tau^{k_\tau} \quad \textrm{in time},
  \norm{\chi-\chi_e}_h \sim C_h h^{k_h} + C_\tau \tau^{k_\tau} \label{eq:convergence_def}
\end{gather}
where $\norm{\cdot}_h$ is an $L^2$ norm, $\chi$ is the simulated field, $\chi_e$ is the exact solution, $h$ is the mesh size, $\tau$ is the time step, $k_h$ is the order of spatial convergence, $k_\tau$ is the order of temporal convergence ($k_\tau=1$ in this work), and $C_h$ and $C_\tau$ are constants.

In the following, we present convergence test in three cases.
Firstly, in the limiting case of a stable bulk intrusion without electrochemistry, an analytical solution is available to test against.
Secondly, using the method of manufactured solution, convergence of the full two-phase EHD problem to an augmented Taylor--Green vortex is shown.
Thirdly, we show convergence towards a highly resolved reference solution for an electrically driven charged droplet.

We note that the aim of \emph{Bernaise} is to solve coupled multi-physics problems, and while the solvers may contain subtle errors, they may be negligible for many applications, and dominant only in limiting cases.
In addition to testing the whole, coupled multi-physics problem of two-phase EHD, a proper testing should also consider simplified settings where fewer physical mechanisms are involved simultaneously.
A brief discussion of testing and such reduced models is given in Appendix \ref{sec:testing}.
In this section, we show the convergence of the schemes in a few relevant cases, which we believe represent the efficacy of our approach.
Tests of simplified-physics problems are found in the GitHub repository \cite{bernaise2018}.

\subsection{Stable bulk intrusion}
\begin{figure}[htb]
  \includegraphics[width=0.9\textwidth]{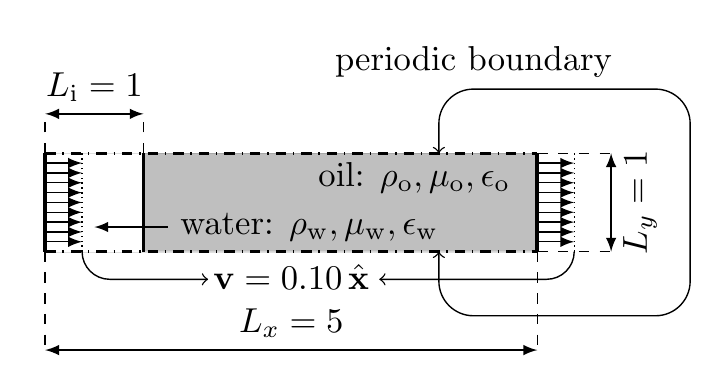}
  \caption{\label{fig:schematic_intrusion}
    Schematic set-up of the stable bulk flow intrusion test case.
    Here, the `water' (subscript w) displaces the `oil' (subscript o). 
    At the left and right boundaries, a constant velocity is prescribed.
  }
\end{figure}

A case where an analytic solution is available, is the stable intrusion of one fluid into another, in the absence of electrolytes and electric fields.
A schematic view of the initial set-up is shown in Fig.~\ref{fig:schematic_intrusion}.
A constant velocity $\v v = v_0 \hat{\v x}$ is applied at both the left and right sides of the reservoir, and periodic boundary conditions are imposed at the perpendicular direction.
We shall here consider the convergence to the solution of the phase-field equation, i.e.~retaining a finite interface thickness $\epsilon$.
This effectively one-dimensional problem is implemented in \texttt{problems/intrusion\_bulk.py}.

Due to the Galilean invariance, we expect the velocity field to be uniformly equal to the inlet and outlet velocities, i.e.~$\v v(\v x, t) = v_0 \hat{\v x}$.
The exact analytical solution for the phase field is given by
\begin{equation}
  \phi(\v x, t) = \tanh\left(\frac{x-x_0-v_0 t }{\sqrt{2} \epsilon}\right),
\end{equation}
for which we shall consider the error norm.
Note that the only parameters this analytical solution depends on are the initial position of the interface $x_0$, the injection velocity $v_0$, and the interface width $\epsilon$.
%Further, it should be noted that this solves the phase field equations \eqref{eq:PF_NS1}--\eqref{eq:PF_NS2} exactly only when the densities are equal in the two phases.
We consider the parameters $\rho_1=\rho_2=1000$, $\mu_1=100$, $\mu_2=1$, $\sigma=2.45$, $\epsilon=0.03$, $M (\phi) = M_0 = 2\cdot 10^{-5}$, $x_0=1$, $L_x=5$, $L_y=1$ and $v_0=0.1$.

\begin{figure}[htb]
  \includegraphics[width=0.59\columnwidth]{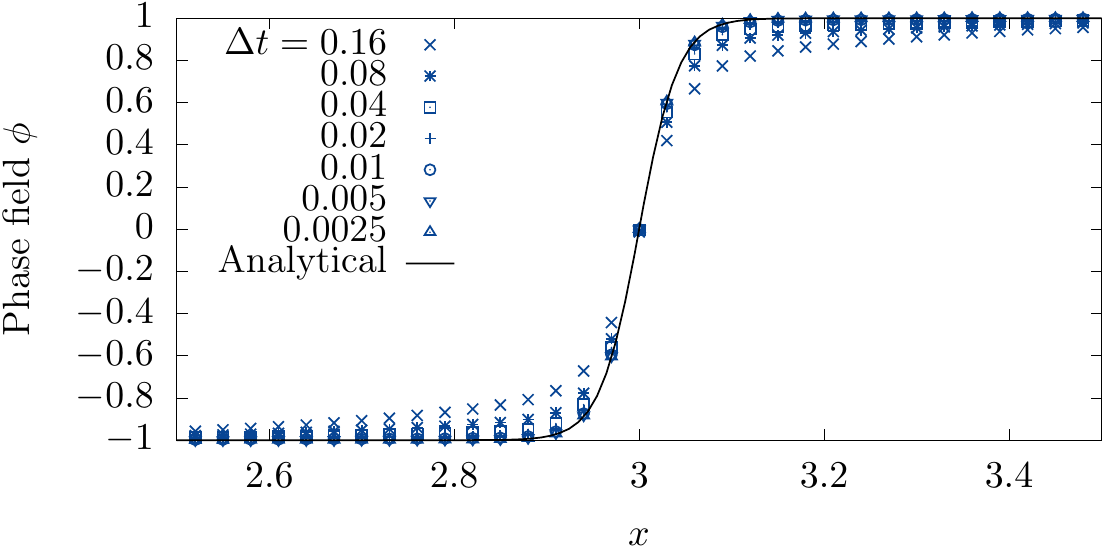}~
  \includegraphics[width=0.4\columnwidth]{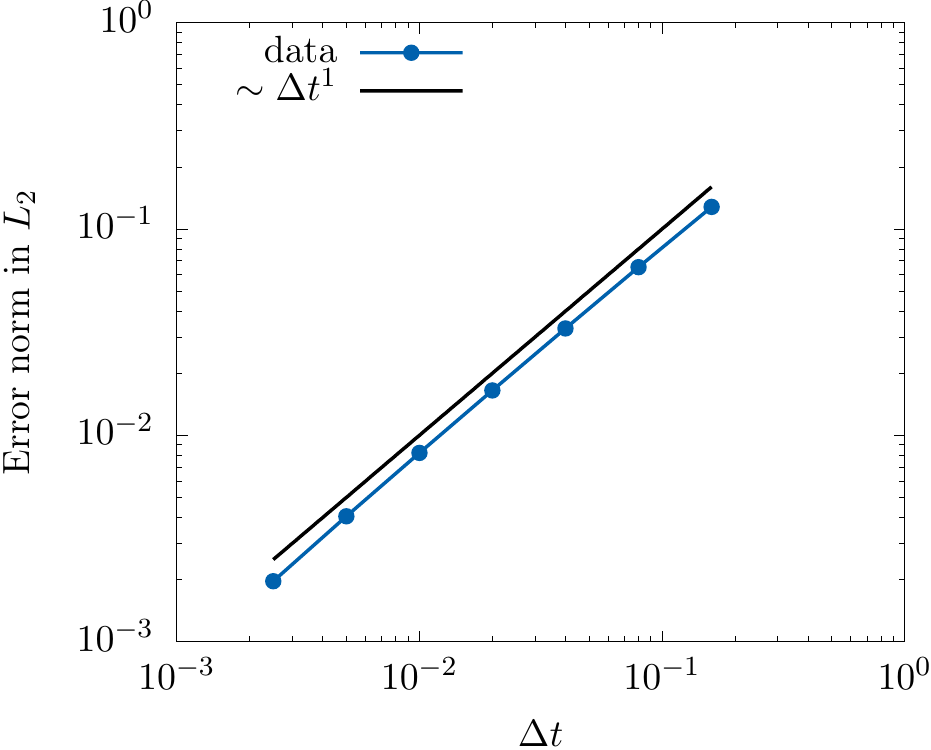}
  \caption{\label{fig:convergence_intrusion_bulk_dt}
    Convergence in time for the case of stable intrusion.
    The mesh size is held fixed at $h=0.0039$.
    Left: We show the phase field interpolated at equidistant points along the centerline for increasing temporal resolution.
    The solid black line is the analytical solution.
    Right: The integrated $L^2$ norm of the phase field plotted against time step.
    The solid black line shows the theoretical convergence order of the scheme ($\sim \tau$).
    As can be seen from the figure, it displays close to ideal scaling.
  }
\end{figure}

Fig.~\ref{fig:convergence_intrusion_bulk_dt} shows the convergence to the analytical solution with regards to temporal resolution.
The order of convergence is consistent with the order of the scheme, indicating that the scheme is appreciable at least in the lack of electrostatic interactions.

\begin{figure}[htb]
  \includegraphics[width=0.59\columnwidth]{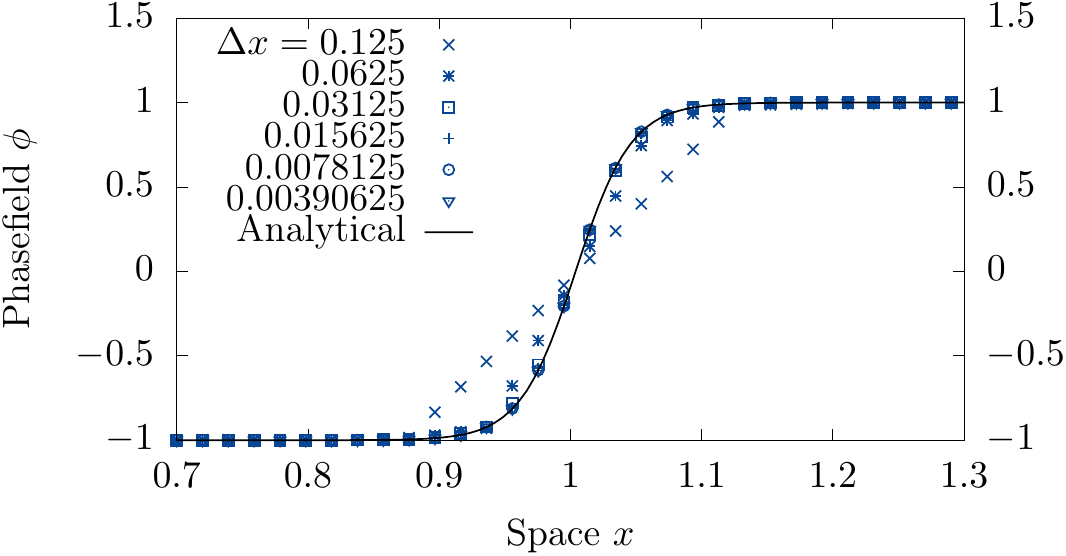}~
  \includegraphics[width=0.4\columnwidth]{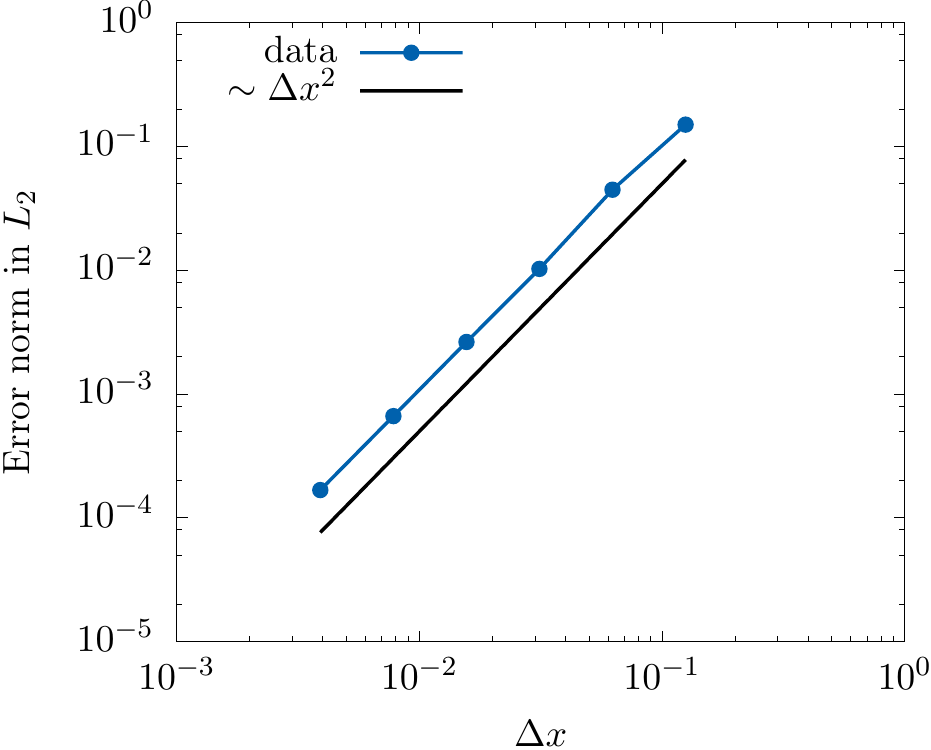}
  \caption{\label{fig:convergence_intrusion_bulk_dx}
    Convergence in space for the case of stable intrusion.
    The time step is held fixed at $\tau = 0.0025$.
    Left: Phase field interpolated at equidistant points along the centerline for increasing spatial resolution.
    Right: The $L^2$ norm of the phase field is plotted against mesh resolution.
    The solid black line shows the theoretical convergence order ($\sim h^2$).
  }
\end{figure}

Fig.~\ref{fig:convergence_intrusion_bulk_dx} shows the convergence of the phase field with regards to the spatial resolution.
The scheme is seen to converge at the theoretical rate, $\sim h^2$.

\subsection{Method of manufactured solution: a two-phase electrohydrodynamic Taylor--Green vortex}
Having established convergence in the practically one-dimensional case, we now consider a slightly more involved setting where we use the method of manufactured solution to obtain a quasi-analytical test case.

The Taylor--Green vortex is a standard benchmark problem in computational fluid dynamics because it stands out as one of the few cases where exact analytical solutions to the Navier--Stokes equations are available.
However, in the case of two-phase electrohydrodynamics, the Navier--Stokes equations couple to both the electrochemical and the phase field subproblems.
In Ref.~\cite{linga2018decoupled} the authors augmented the Taylor--Green vortex with electrohydrodynamics, and in this work we supplement the latter with a phase field and non-matching densities of the two phases.

We consider the full set of equations on the domain $\Omega=[0, 2\pi] \crossprod [0, 2\pi]$, where all quantities may differ in the two phases. 
The two ionic species have opposite valency $\pm z$.
The fields are given by
\begin{subequations}
  \begin{align}
    \v u &= U(t) (\hat{\v x} \cos x \sin y - \hat{\v y} \sin x \cos y ), \label{eq:tg_ref_u} \\
    p    &= -\sum_{mn} \mathcal{P}_{mn}(t) \cos (2mx) \cos (2ny), \\
    \phi &= \Phi(t) \cos x \cos y,\\
    c_\pm &= c_0 (1 \pm \cos x \cos y \, C(t)), \\
    V    &= \frac{z c_0 C(t)}{\varepsilon} \cos x \cos y . \label{eq:tg_ref_V}
  \end{align}
\end{subequations}
Here, the time-dependent coefficients are given by
\begin{align}
  U(t) &= U_0 \exp\left(- \frac{2 \bar{\mu}}{\bar{\rho}} t \right), \\
  C(t) &= C_0 \exp\left(-2 \bar D \left( 1 + \frac{c_0}{\bar{\varepsilon}} \right) t \right), \\
  \Phi(t) &= \Phi_0 \exp\left(-2 M \tilde{\sigma} \left(2 \epsilon-\frac{1}{\epsilon}\right)t \right), 
\end{align}
where $U_0, C_0$ and $\Phi_0$ are scalars, and
% \begin{align}
%  \mathcal P _{mn} = \begin{cases}
%    \frac{\mathcal Q_1 (t)+\mathcal Q_2(t) + 3\mathcal Q_3 (t)}{4} \quad \textrm{for} \quad (m,n) \in \{ (0,1), (1,0) \}, \\
%    \frac{\mathcal Q_2(t) + 4\mathcal Q_3 (t)}{4} \quad \textrm{for} \quad (m,n) \in \{ (1,1) \}, \\
%    \frac{3\mathcal Q_3 (t)}{16} \quad \textrm{for} \quad (m,n) \in \{ (0,2), (2,0) \}, \\
%    \frac{\mathcal Q_3 (t)}{4} \quad \textrm{for} \quad (m,n) \in \{ (1,2), (2,1) \}, \\
%    \frac{\mathcal Q_3 (t)}{16} \quad \textrm{for} \quad (m,n) \in \{ (2,2) \}, \\
%    0 \quad \textrm{otherwise.}
%  \end{cases}
% \end{align} 
\begin{align}
  \mathcal P _{mn} = \begin{cases}
    \mathcal Q_1 (t)+\mathcal Q_2(t) \quad \textrm{for} \quad (m,n) \in \{ (0,1), (1,0) \}, \\
    \mathcal Q_2(t) \quad \textrm{for} \quad (m,n) \in \{ (1,1) \}, \\
    0 \quad \textrm{otherwise.}
 \end{cases}
\end{align} 
where
% \begin{equation}
%   \mathcal Q_1 = \rho U_0^2(t), \quad
%   \mathcal Q_2 = \frac{z^2 c_0^2 C^2(t)}{\epsilon} + \frac{\sigma \Phi^2(t)}{2} (2\epsilon - \epsilon^{-1}), \quad
%   \mathcal Q_3 = \frac{3 \sigma \Phi^4(t)}{16 \epsilon}.
% \end{equation}
\begin{equation}
  \mathcal Q_1 = \frac{1}{4} \rho U_0^2(t), \quad \textrm{and} \quad
  \mathcal Q_2 = \frac{z^2 c_0^2 C^2(t)}{4 \epsilon}.
\end{equation}
Further, a bar indicates the arithmetic average over the value in the two phases, i.e.~$\bar{\chi} = (\chi_1 + \chi_2)/2$ for any quantity $\chi$, and $\bar D = (\bar D_+ + \bar D_-)/2 = (D_{+,1} + D_{+,2} + D_{-,1} + D_{-,2})/4$ is the arithmetic average over all diffusivities.
The time-dependent boundary conditions are set by prescribing the reference solutions at the boundary of $\Omega$ for all fields given in \eqref{eq:tg_ref_u}--\eqref{eq:tg_ref_V}, except the pressure $p$, which is set (to the reference value) only at the corner point $(x, y) = (0, 0)$.
The method of manufactured solution now consists in augmenting the conservation equations \eqref{eq:PF_NS1}, \eqref{eq:PF_PF1}, \eqref{eq:PF_c} and \eqref{eq:PF_V} by appropriate source terms, such that the reference solution \eqref{eq:tg_ref_u}--\eqref{eq:tg_ref_V} solves the system exactly.
These source terms were computed in Python using the \emph{Sympy} package, and are rather involved algebraic expressions.
The expressions are therefore omitted here, but can be found as a utility script in the \emph{Bernaise} package.
Note that in the special case of single-phase flow without electrodynamics, i.e.~$\phi \equiv 1$ and $z = 0$, we retrieve the classic Taylor--Green flow (with a passive tracer concentration field), where all artificial source terms vanish.

We consider now the convergence towards the manufactured solution.
We let the grid size $\Delta h \in [2\pi/256, 2\pi/16]$ and the time step $\tau \in [0.0001, 0.01]$, and evaluate the solution at the final time $T=0.1$.
The parameters for two phases used the simulation are given in Table \ref{tab:tg_phasic_parameters}, while the non-phase specific parameters are given in Table \ref{tab:tg_parameters}.
Note that in order to test all parts of the implementation, all parameters are kept roughly in the same order of magnitude.
When all the physical processes are included, the manufactured solution becomes an increasingly bad approximation and thus the resulting source terms become large.
Thus, in order to avoid numerical instabilities, it was necessary to evaluate the error at a relatively short final time $T$.
However, it should be enough to locate errors in most parts of the code.

\begin{table}
\caption{\label{tab:tg_phasic_parameters} Phasic parameters used in the Taylor--Green simulations.}
\begin{ruledtabular}
\begin{tabular}{l l l l}
Parameter          & Symbol        & Value in phase  1 & Value in phase 2 \\
\hline
Density            & $\rho$        & $3$ & $1$ \\
Viscosity          & $\mu$         & $3$ & $5$ \\
Permittivity       & $\varepsilon$ & $3$ & $4$ \\
Cation diffusivity & $D_+$         & $3$ & $1$ \\
Anion diffusivity  & $D_-$         & $4$ & $2$ \\
Cation solubility  & $\beta_+$     & $2$ & $-2$ \\
Anion solubility   & $\beta_-$     & $1$ & $-1$ \\
\end{tabular}
\end{ruledtabular}
\end{table}

\begin{table}
\caption{\label{tab:tg_parameters} Non-phase-specific parameters used in the Taylor--Green simulations.}
\begin{ruledtabular}
\begin{tabular}{l l l l}
Parameter                & Symbol        & Value \\
\hline
Surface tension          & $\sigma$      & $0.1$ \\
Interface thickness      & $\epsilon$    & $1/\sqrt{2}$ \\
Phase field mobility     & $M$           & $1$ \\
Initial velocity         & $U_0$         & $1$ \\
Initial concentration    & $c_0$         & $1$ \\
Initial phase field      & $\Phi_0$      & $1$ \\
Initial conc.\ deviation & $C_0$         & $0.5$ \\
\end{tabular}
\end{ruledtabular}
\end{table}

\begin{figure}[htb]
  \includegraphics[width=0.5\columnwidth]{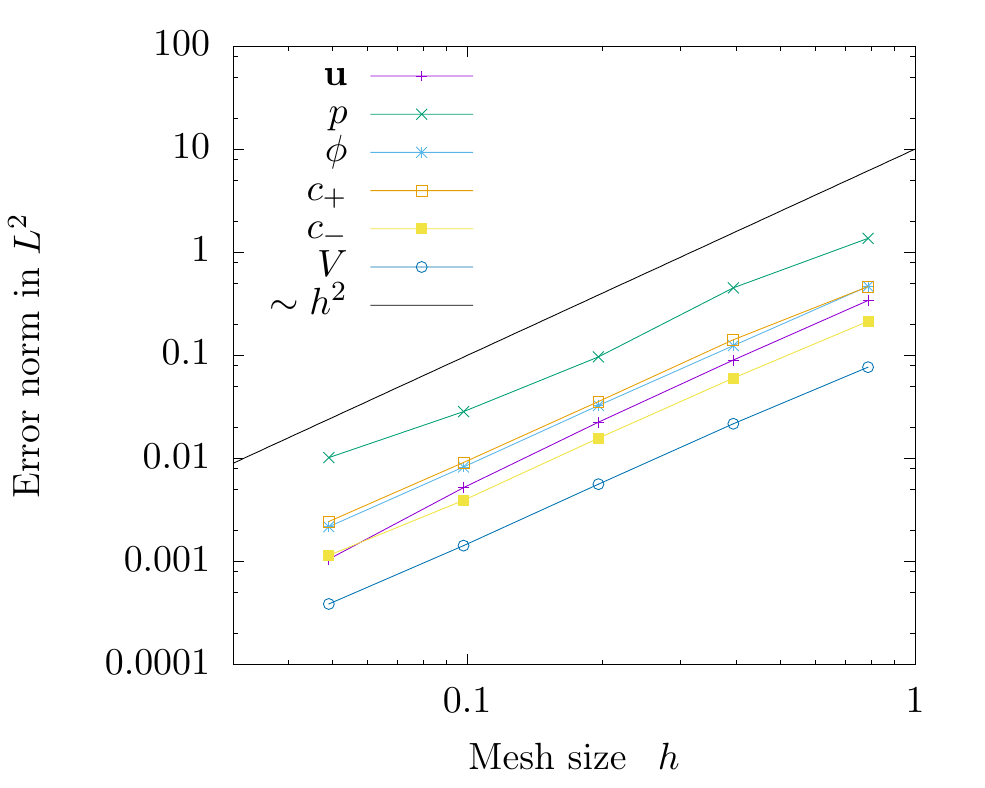}
  \caption{\label{fig:tg_convergence_space}
    Convergence in space for the two-phase electrohydrodynamic Taylor--Green manufactured solution.
    The solid black line shows the theoretical convergence rate based on the order of the finite elements chosen ($\sim h^2$).
    All fields display close to ideal convergence.
  }
\end{figure}

We plot the $L^2$ errors of all the fields as a function of the grid size $h$ in Fig.~\ref{fig:tg_convergence_space}.
In these simulations, we used a small time step $\tau=0.0001$ to rule out the contribution of time discretization to the error, cf.~Eq.~\eqref{eq:convergence_def}.
It is clear that the spatial convergence is close to ideal for all fields, indicating that the scheme approaches the correct solution.
The pressure $p$ displays slightly worse convergence and higher error norm than the other fields, which may be due to the pointwise way of enforcing the pressure boundary condition (all other fields have Dirichlet conditions on the entire boundary).

\begin{figure}[htb]
  \includegraphics[width=0.5\columnwidth]{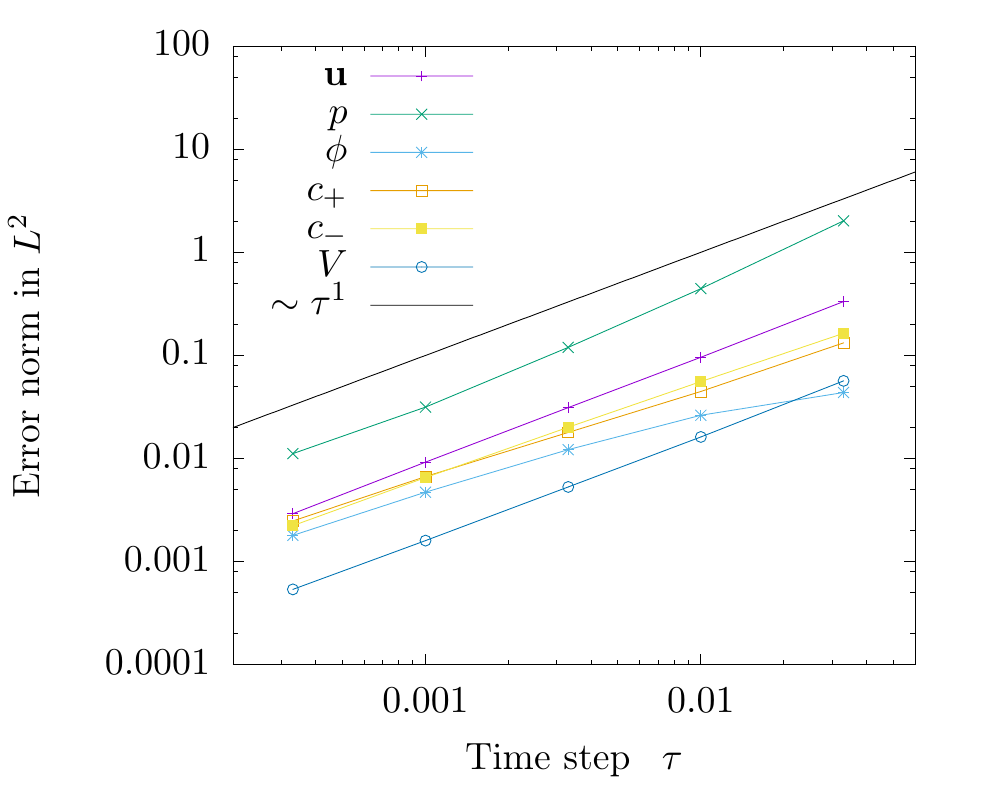}
  \caption{\label{fig:tg_convergence_time}
    Convergence in time for the two-phase electrohydrodynamic Taylor--Green manufactured solution.
    The solid black line shows the theoretical convergence rate of the scheme ($\sim \tau^1$).
    All fields display close to ideal convergence.
  }
\end{figure}

In Fig.~\ref{fig:tg_convergence_time}, we plot the $L^2$ errors of the same fields as in Fig.~\ref{fig:tg_convergence_space}, but as a function of the time step $\tau$.
In the simulations plotted here, we used a fine grid resolution with $h=2\pi/256$ to rule out the contribution of spatial discretization to the error, cf.~Eq.~\eqref{eq:convergence_def}.
Clearly, first order convergence is achieved for sufficient refinement, for all fields including the pressure.

\subsection{Droplet motion driven by an electric field}
\begin{figure}[htb]
  \includegraphics[width=0.9\columnwidth]{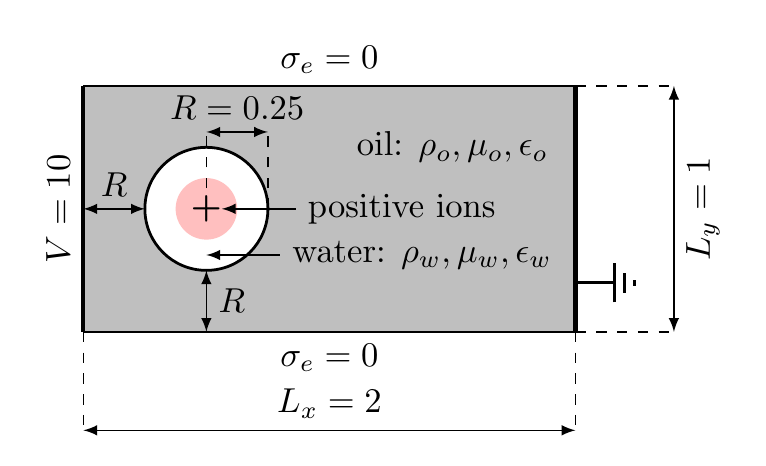}
  \caption{\label{fig:schematic_droplet}
    Schematic set-up of the test case of droplet motion driven by an electric field.
    The `water' droplet contains positive ions and is driven by the electric field set up between the high potential on the left wall and the grounded right wall.
  }
\end{figure}

We now consider a charged droplet moving due to an imposed electric field; a problem for which there is no analytical solution available.
However, by comparing to a highly resolved numerical solution, convergence for the fully coupled two-phase electrohydrodynamic problem can be verified.
This problem has already been partly presented in the above, and is implemented in \texttt{problems/charged\_droplet.py}.
A sketch showing the initial state is shown in Fig.~\ref{fig:schematic_droplet}.
We consider an initially circular droplet, where a positive charge concentration is initiated as a Gaussian distribution, with variance $\delta_c^2$, in the middle of the droplet.
In this set-up, we consider only a single, positive species.
The total amount of solute, i.e.~integrated concentration, is $C_0 = \int_\Omega c_0 \, \diff A$. 
The left wall of the reservoir is kept at a positive potential, $V=\Delta V$, while the right wall is grounded, $V=0$.
The top and bottom walls are assumed to be perfectly insulating, i.e.~a no-flux condition is applied on concentration fields and electric fields, and a no-slip condition is applied on the velocity.
The fluid surrounding the droplet is neutral, and its parameters are chosen such that the solute is only very weakly soluble in the surrounding fluid, and the diffusivity here is very low here to prevent leakage.
The droplet is accelerated  by the electric field towards the right, before it is slowed down due to viscous effects upon approaching the wall.

With regards to reproducing the sharp-interface equations, we consider now the case of reducing the interface thickness $\epsilon \to 0$.
To this end, we keep the ratio $h/\tau$ between mesh size and time step fixed, and further we keep the interface thickness $\epsilon$ proportional to $h$.
The latter spans roughly 3-4 elements.
Since the interface thickness $\epsilon$ changes, an important parameter in the phase-field model changes, which couples back to the equations, and thus the L$_2$ norm does not necessarily constitute a proper convergence measure.
We therefore resort to using the \emph{picture norm} or contour of the droplet as a measure, i.e.~the zero-level set of the phase field $\phi=0$.
In particular, we will consider two observables: circumference and the center of mass (along $x$) of the droplet, as a function of resolution.
A similar approach was taken for the case of phase-field models without electrodynamics by \citet{aland2012} who compared their benchmarks to sharp interface results by \citet{hysing2009}.

The resolutions used in our simulations are given in Table \ref{tab:charged_droplet_res}.
In order not to have to adjust the phase field mobility when refining, whilst still expecting to retrieve the sharp-interface model in the limit $\epsilon \to 0$, we choose the phase field mobility given by \eqref{eq:pf_mobility_2}.
All parameters for the phasic quantities are given in Table \ref{tab:charged_droplet_phase_params}, while the remaining parameters are given in Table \ref{tab:charged_droplet_params}.
From these parameters, using the unit scaling adopted in this paper, we find an approximate Debye length $\lambda_{D} = \sqrt{{\varepsilon}/(2 z^2 c_R)} \simeq \sqrt{1/(2 \cdot 10)} \simeq 0.2$ (see Section \ref{sec:app_net_charge} in the Appendix for this expression), since we can approximate the order of magnitude of $c_R < C/(\pi R^2) = 10/(\pi \cdot 0.25^2)$ for a moderate screening.
\begin{table}
\caption{\label{tab:charged_droplet_res}
  Numerical parameters that vary with resolution in the charged droplet simulations:
  Mesh size $h$, time step $\tau$, and interface thickness $\epsilon$.}
\begin{ruledtabular}
\begin{tabular}{l l l}
  $h$                & $\tau$        & $\epsilon$ \\ 
  \hline
  0.04               & 0.04          & 0.06 \\ % lizard/5
  0.02               & 0.02          & 0.03 \\ % lizard/4
  0.01               & 0.01          & 0.015 \\ % duke/4
  0.005              & 0.005         & 0.0075 \\ % lizard/2
  0.0025             & 0.0025        & 0.00375 \\ % lizard/3
\end{tabular}
\end{ruledtabular}
\end{table}
\begin{table}
\caption{\label{tab:charged_droplet_phase_params}
  Numerical parameters for the phases that are common for all charged droplet simulations.}
\begin{ruledtabular}
\begin{tabular}{l l l l}
  Parameter     & Symbol          & Value, phase 1 & Value, phase 2\\ 
  \hline
  Density       & $\rho$        & 200.0          & 100.0 \\
  Permittivity  & $\varepsilon$ & 1.0            & 1.0 \\
  Diffusivity   & $D$           & $1 \cdot 10^{-5}$ ($\simeq 0$) & 0.001 \\
  Solubility    & $\beta$       & 4.0            & 1.0 \\
  Viscosity     & $\mu$         & 10.0           & 1.0 \\
\end{tabular}
\end{ruledtabular}
\end{table}
\begin{table}
\caption{\label{tab:charged_droplet_params}
  Numerical parameters not specific to phase for the charged droplet simulations.}
\begin{ruledtabular}
\begin{tabular}{l l l}
  Parameter              & Symbol          & Value \\ 
  \hline
  Potential difference   & $\Delta V$      & 10.0  \\
  Integrated concentration & $C_0$ & 10.0  \\
  Phase field mobility coeff. & $ M_0  $   & $1.5 \cdot 10^{-5}$ \\
  Initial droplet radius & $R$           & 0.25 \\
  Initial conc.~std.~dev. & $\delta_c$     & 0.0833 \\
  Surface tension        & $\sigma$        & 5.0 \\
  Length in $x$-direction & $L_x$          & 2.0 \\
  Length in $y$-direction & $L_y$          & 1.0 \\
\end{tabular}
\end{ruledtabular}
\end{table}

\begin{figure}[htb]
  \includegraphics[width=0.6\columnwidth]{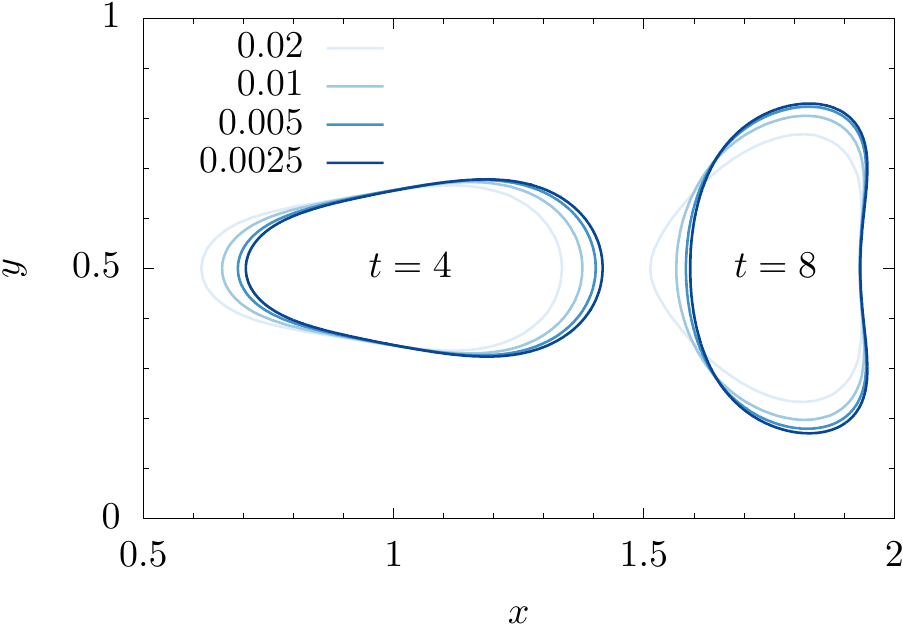}
  \caption{\label{fig:droplet_shape_comparison}
    Shape comparison of electrically driven charged droplet at two time instances.
    The effect of the four resolutions given in Table \ref{tab:charged_droplet_res} is shown.
    The legend shown in the figure refers to both spatial ($h$) and temporal resolution ($\tau$).
  }
\end{figure}

In Fig.~\ref{fig:droplet_shape_comparison}, we show the contour of the driven droplet at two time instances $t=4$ and $t=8$, and compare increasing resolution (simultaneously in space, time and interface thickness).
Qualitatively inspecting the contours by eye, the droplet shapes seem to converge to a well defined shape with increasing resolution at both time instances.
However, qualitive comparison is clearly not enough to assess the convergence.
As in Refs.~\cite{aland2012,hysing2009}, we define three observables:
\begin{itemize}
\item Center of mass: We consider the center of mass of the dispersed phase (phase 2, i.e.~$\phi < 0$),
\begin{equation} 
  x_{\rm CM} = \frac{\int_{\phi < 0} x \, \diff A }{ \int_{\phi < 0} \diff A},
\end{equation}
where we approximate the integral over the droplet (phase 2) by $\int_{\phi < 0} (\cdot) \, \diff A = \int_\Omega (1-\phi)(\cdot)/2 \, \diff A$.
\item Drift velocity: Similarly as above, the velocity at which the droplet is driven is measured by
\begin{equation}
  \mathcal{V} = \frac{\int_{\phi < 0} \v u \cdot \hat{\v x} \, \diff A }{ \int_{\phi < 0} \diff A}.
\end{equation}
\item Circularity: Defined as the ratio of the circumference of the area-equivalent circle to the droplet circumference,
\begin{equation}
  \mathcal C = \frac{2 \sqrt{ \pi \int_{\phi < 0} \, \diff A }}{\ell}.
\end{equation}.
\end{itemize}
The circumference $\ell$ and the integrals are computed by the post-processing method \texttt{geometry\_in\_time} which is built into \emph{Bernaise}.

\begin{figure}[htb]
  \includegraphics[width=0.75\columnwidth]{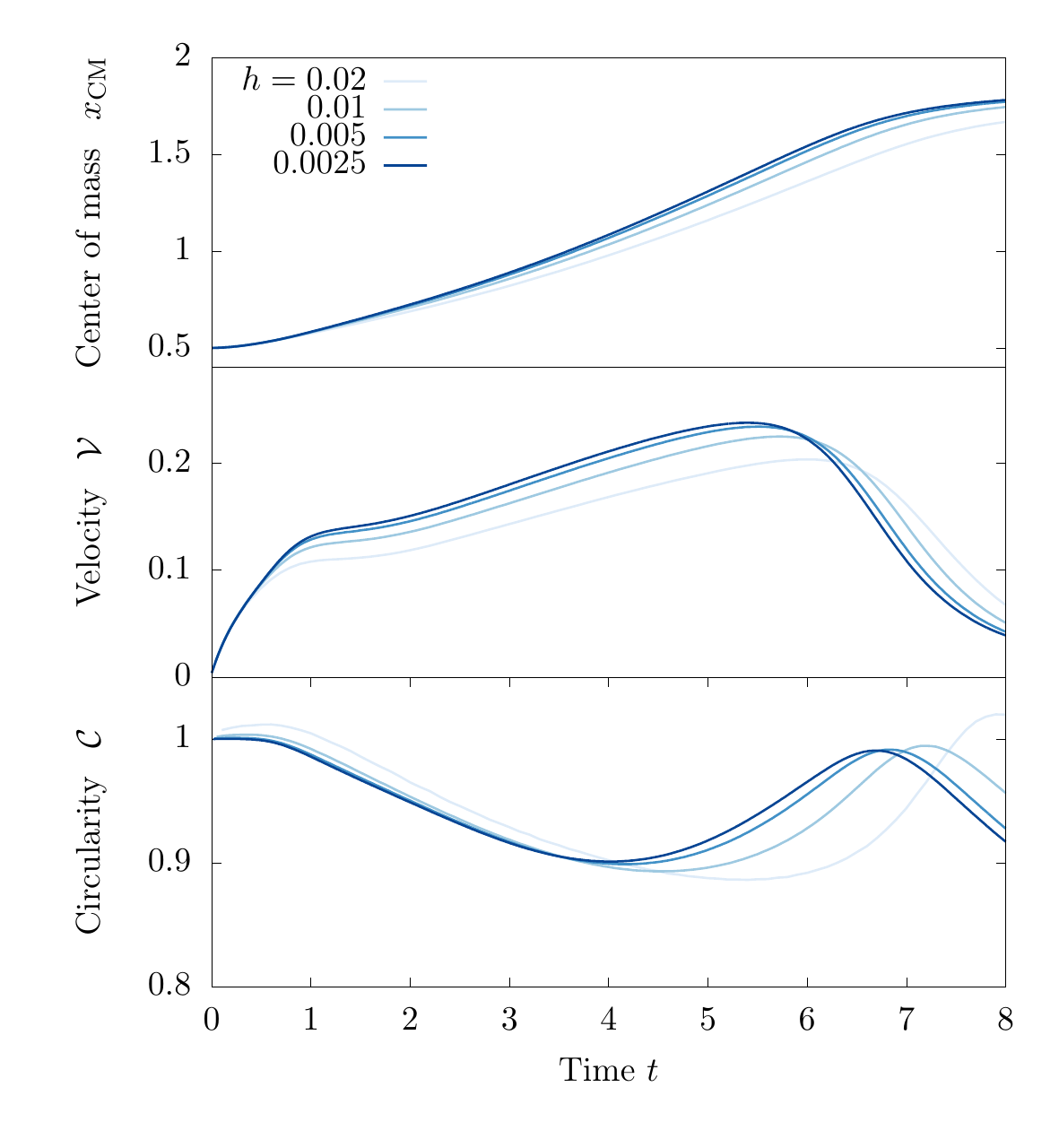}
  \caption{\label{fig:droplet_time_data_comparison}
    Observable quantities as a function of time.
    Increasing resolutions (spatial and temporal) are compared. 
  }
\end{figure}

Fig.~\ref{fig:droplet_time_data_comparison} shows the three quantities as a function of time for increasing resolution.
(Here we have omitted the coarsest resolution $h=0.04$ for visual clarity.)
The curves seem to converge towards well-defined trajectories with resolution.
For a more quantitative comparison, we define the time-integrated error norm,
\begin{equation}
  \norm{e}_p = \left({\frac{\int_0^T |q_{\rm ref}(t) - q (t)|^p \, \diff t}{\int_0^T |q_{\rm ref}(t) |^p \, \diff t}} \right)^{1/p}
  \label{eq:error_time}
\end{equation}
for a given quantity $q$.
We can compute an empirical convergence rate of this norm,
\begin{equation}
  k_{p,i} = \frac{\log\left( {\norm{e}_p (h_{i+1})} / {\norm{e}_p (h_i)} \right)}{\log\left( h_{i+1}/h_i\right)}
\end{equation}
for two successive resolutions ($h_{i+1} > h_i$).
Here we shall consider the $L^2$ error norm in time, i.e.~$p=2$, and in practice we compute the integrals in time by cubic spline interpolation of measurement points saved at every 5 time steps.
There is no exact solution, or reference high-resolution sharp-interface solution available for this set-up.
However, if we now assume that the finest resolution \emph{is} the exact solution, and use this as the reference field in Eq.~\eqref{eq:error_time}, we can compute error norms and convergence rates.
These values are reported in Table \ref{tab:charged_droplet_convergence}.
\begin{table}
\caption{\label{tab:charged_droplet_convergence}
  Mesh size $h$, error norm $\norm{e}_2$, and empirical convergence rate $k_2$ for increasing grid refinement, assuming the solution for the finest resolution to be exact.}
\begin{ruledtabular}
\begin{tabular}{l l l}
  $h$            & $\norm{e}_2$   & $k_2$ \\ 
  \hline
  Center of mass \\
  \hline
  0.04  & 0.1798 & \\
  0.02  & 0.0955 & 0.9129 \\
  0.01  & 0.0410 & 1.2186 \\
  0.005 & 0.0126 & 1.7033 \\
  \hline
  Drift velocity \\
  \hline
  0.04  & 0.3427 & \\
  0.02  & 0.2067 & 0.7293 \\
  0.01  & 0.1032 & 1.0025 \\
  0.005 & 0.0341 & 1.5932 \\
  \hline
  Circularity \\
  \hline
  0.04  & 0.0891 & \\
  0.02  & 0.0423 & 1.0757 \\
  0.01  & 0.0205 & 1.0467 \\
  0.005 & 0.0060 & 1.7612 \\
\end{tabular}
\end{ruledtabular}
\end{table}
The computed convergence rates increase for all three observables and reach 1.6--1.7 with increasing resolution, indicating also quantitatively a convergence that is in agreement with the anticipated convergence rate.
Considering Eq.~\eqref{eq:convergence_def}, from the temporal discretization, we expect $k_2 \simeq 1$, and from the spatial $k_2 \simeq 2$.
Depending on which term contributes most to the error, we will measure either of these rates.
The values measured here indicate that both terms may be comparable in magnitude; however if we instead of using directly the finest solution as reference, extrapolated the trajectories further, we would presumptively have achieved lower convergence rates.
This might indicate that the convergence error is eventually dominated by the temporal discretization, cf.\ Eq.\ \eqref{eq:convergence_def}.

\section{Applications}
\label{sec:application}

\subsection{Oil extrusion from a dead-end pore}
\begin{figure}[htb]
  \centering
  \includegraphics[width=0.9\textwidth]{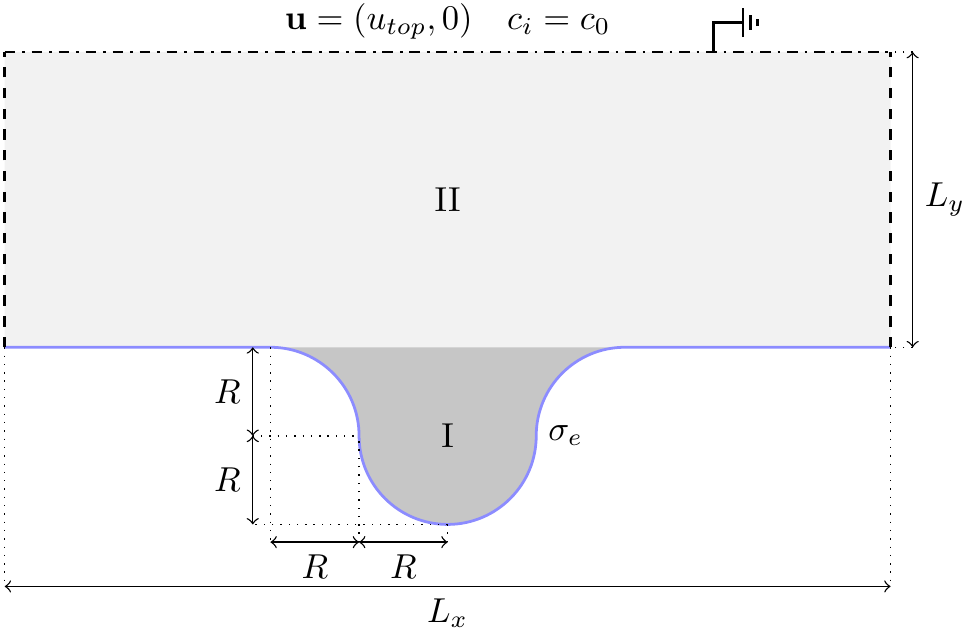}
  \caption[Dead-end pore geometry]
  {A schematic depiction of the ``dead-end pore'' geometry, with the appropriate boundary conditions for the problem and specified initial conditions for the phase field. 
    The geometry is specified by the two lengths $L_x$, $L_y$, and the radius $R$ used to define the dead-end pore in the center of the channel by a circle and a circular smoothed inlet.
    The roman numerals indicate the phase, along with the tone of gray.
    The darker phase is the oil-like phase ($\mathrm{I}$), and the lighter one is the water-like phase ($\mathrm{II}$). 
  }
  \label{fig:snovsen_setup}
\end{figure}

Here, we present a demonstration of the method in a potential geophysical application.
We consider a shear flow of one phase (``water'') over a dead-end pore which is initially filled with a second phase (``oil'').
The water phase contains initially a uniform concentration of positive and negative ions, $c_\pm |_{t=0} = c_0$, and the water--oil interface is modelled to be impermeable.
The simulation of the dead-end pore is carried out to preliminarily assess the hypothesis that electrowetting could be responsible for the increased expelling of oil in low-salinity enhanced oil recovery.
The problem set-up is schematically shown in Fig.~\ref{fig:snovsen_setup}.
The phasic parameters used in the simulations are given in Table \ref{tab:snovsen_phasic_parameters}, and the remaining parameters are given in Table \ref{tab:snovsen_parameters}.
This problem is implemented in the file \texttt{problems/snoevsen.py}.

\begin{table}
  \caption{
    \label{tab:snovsen_phasic_parameters}
    Phasic parameters for the simulations of shear flow over a dead-end pore.
    The subscript $\pm$ indicates the value for both the positive and negative ions.
  }
  \begin{ruledtabular}
    \begin{tabular}{l c r r}
      Parameter       & Symbol        & \text{Value in phase 1} & \text{Value in phase 2} \\
      \hline
      Viscosity       & $\mu$       & 1.0    & 1.0 \\
      Density         & $\rho$      & 10.0   & 10.0 \\
      Permittivity    & $\epsilon$  & 1.0    & 1.0 \\
      Solution energy & $\beta_\pm$ & 4      & 1 \\
      Ion mobility    & $D_\pm$     & 0.0001 & 0.01 \\
    \end{tabular}
  \end{ruledtabular}
\end{table}

\begin{table}[htb]
  \caption{
    \label{tab:snovsen_parameters}
    Simulation parameters for the simulations of shear flow over a dead-end pore.
  }
  \begin{ruledtabular}
    \begin{tabular}{l c r}
      Parameter            & Symbol        & Value \\
      \hline
      Length               & $L_x$         & 3.0 \\
      Height               & $L_y$         & 1.0 \\
      Total simulation time & $T$          & 20 \\
      Radius               & $R$           & 0.3 \\
      Time step            & $\tau$        & 0.01 \\
      Resolution           & $h$           & 1/120 \\
      Interface thickness  & $\varepsilon$ & 0.02 \\
      Phase field mobility & $M_0$         & $2.5\cdot10^{-6}$ \\
      Surface tension      & $\sigma$      & 2.45 \\
      Surface charge       & $\sigma_e$    & $\{-10, 0 \}$ \\
      Reference concentration & $c_{0}$    & 2 \\
      Shear velocity       & $u_{\rm top}$  & 0.2 \\
    \end{tabular}
  \end{ruledtabular}
\end{table}

\begin{figure}[h]
  % \centering
  \subfigure[\ $t=3.0$, $\sigma_e=0$]{
    \includegraphics[width=0.45\textwidth]{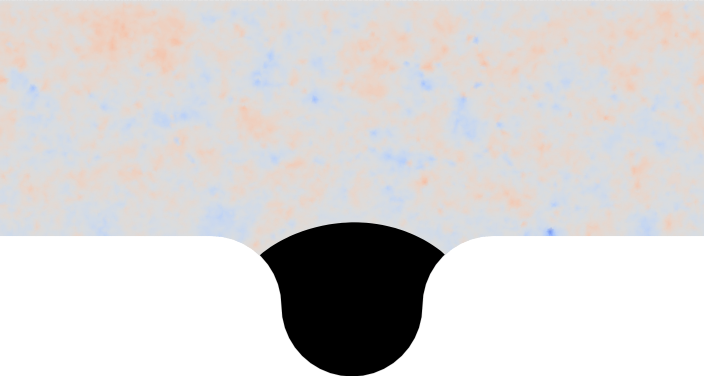}
    \label{fig:snc1}
  }
  \subfigure[\ $t=3.0$, $\sigma_e=-10$]{
    \includegraphics[width=0.45\textwidth]{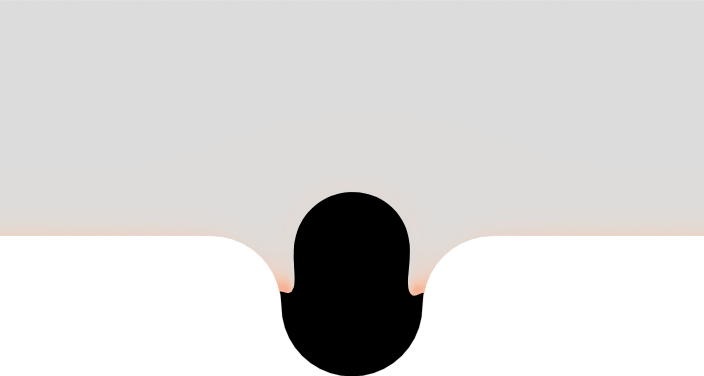}
    \label{fig:sc1}
  }
  \subfigure[\ $t=6.0$, $\sigma_e=0$]{
    \includegraphics[width=0.45\textwidth]{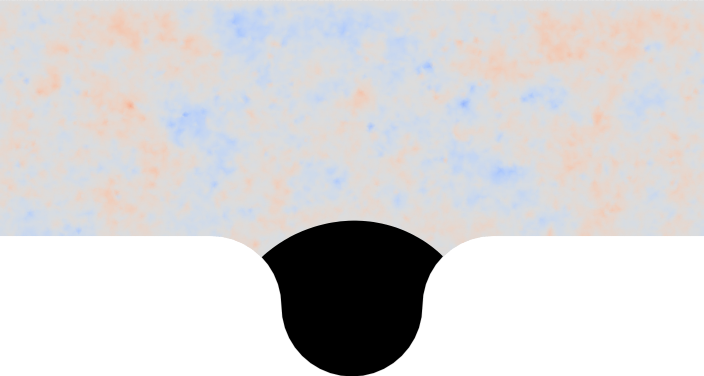}
    \label{fig:snc2}
  }
  \subfigure[\ $t=6.0$, $\sigma_e=-10$]{
    \includegraphics[width=0.45\textwidth]{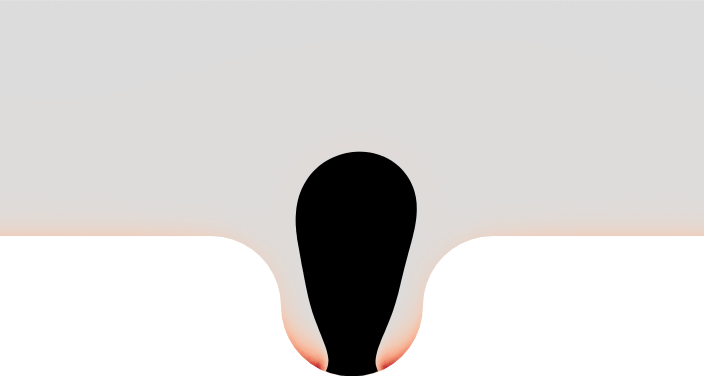}
    \label{fig:sc2}
  }
  \subfigure[\ $t=9.0$, $\sigma_e=0$]{
    \includegraphics[width=0.45\textwidth]{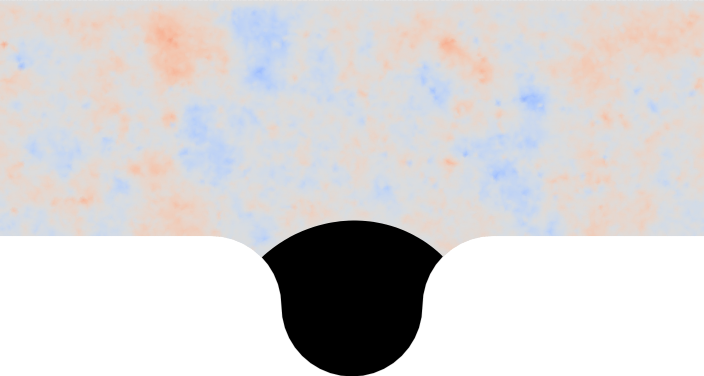}
    \label{fig:snc3}
  }
  \subfigure[\ $t=9.0$, $\sigma_e=-10$]{
    \includegraphics[width=0.45\textwidth]{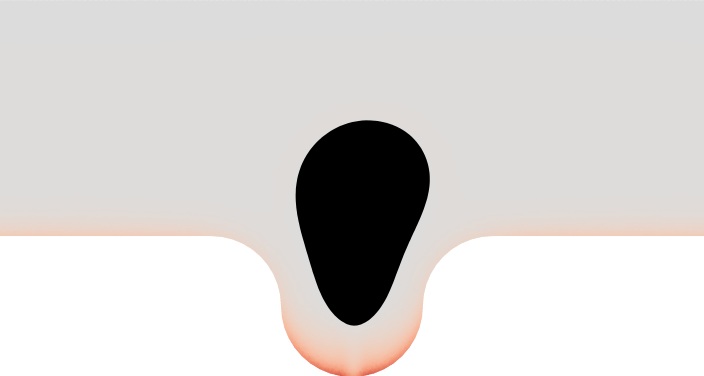}
    \label{fig:sc3}
  }
  \caption{
    Oil released from a dead-end pore.
    We show instantaneous snapshots from the simulations of the dead-end pore under a shear flow.
    The black phase is the oil phase, which does not contain solutes, and the other phase is the water phase, which contains monovalent positive and negative ions.
    The color in the lighter phase indicates the local net charge, red meaning positive charge, blue negative charge, and gray neutral charge.
    The color scale is relative to the maximum deviation from neutral charge for an entire simulation; therefore the neutral simulations display numerical noise (which is of the order of machine precision).
    In the left column the surface charge is zero, and in the right column, a uniform surface charge density $\sigma_e = -10$ is set. 
    % The first row is for $t=300 dt$, the second is for $t=600 dt$, and finally the third row is $t=900 dt$.
    The rows show snapshots at different times $t$.
  }
  \label{fig:snovsen_sim}
\end{figure}

To investigate the effect of including electrostatic interactions, we show in Fig.~\ref{fig:snovsen_sim} instantaneous snapshots of simulations with and without surface charge at different times. 
The left column, Figs.~\ref{fig:snc1}, \ref{fig:snc2}, and \ref{fig:snc3}, shows the results for vanishing surface charge, and the right column, Figs.~\ref{fig:sc1}, \ref{fig:sc2}, and \ref{fig:sc3}, shows the results for a surface charge of $\sigma_e=-10$.

For the uncharged case, the frames that are shown are almost indistinguishable.
In fact, the main difference is the numerical noise of the total charge, which is due to roundoff errors of machine precision.
The initial dynamics of the oil plug interface, which is to equilibrate with the neutral contact angle and the shear flow, mainly happens before the first frame presented; compare Figs.~\ref{fig:snovsen_setup} and \ref{fig:snc1}.

A markedly different behavior is displayed in the right column, Figs.~\ref{fig:sc1}, \ref{fig:sc2}, and \ref{fig:sc3}, where a uniform surface charge density is enforced the walls at the simulation start, $t=0$.
Here, we see first that two tongues are intruding on both sides of the droplet, which push the droplet out into the center of the dead-end pore. 
The process is continued, as shown in the second frame, and finalized, as shown in the third frame, with the complete release of the droplet as the two tongues meet at the bottom of the dead-end pore, cutting the final contact point.

With these simulations, we have demonstrated the effects when a surface charge couples to hydrodynamics.
This has lead to the observation that oil phase, on a larger scale than the Debye length, behaves like it is completely dewetting even when we locally enforce a neutral contact angle.

\subsection{3D simulations of droplet coalescence and breakup in an electric field}
\label{sec:charged_droplets_3d}
Finally, to demonstrate the ability of \emph{Bernaise} to simulate 3D configurations, we present simulations of two oppositely charged droplets that coalesce.
In order to achieve this efficiently, a fully iterative solver was implemented.
The solver consists of a fractional step version of the \texttt{basic} solver, in the sense that within the fluid flow step, it splits between the velocity and pressure computations, as shown in Eqs.\ \eqref{eq:split_u_1}, \eqref{eq:split_p}, and \eqref{eq:split_u_2}.
The splitting introduces a weak compressibility which suffices to stabilize the problem \cite{langtangen2002} (with respect to the BB condition) and thus we can use P$_1$ finite elements also for the velocity.
The combination of fewer degrees of freedom and the applicability of iterative linear solvers imparts significant speed-up compared to coupled solvers, which is of paramount importance for 3D simulations.
This yields advantages over solvers which rely on a mixed-element formulation of the hydrodynamic subproblem \cite{metzger2018}.
The detailed analysis of the fractional step solver will be published in a separate paper, but the implementation can be found in \texttt{solvers/fracstep.py}.
For solving the linear systems iteratively, we use an algebraic multigrid (AMG) preconditioner and a generalized minimal residual (GMRES) linear solver for the electrochemical and the pressure correction step; Jacobi preconditioner (Jacobi) and a stabilized bi-conjugate gradient method (BiCGStab) for the velocity prediction, and Jacobi and GMRES for the velocity correction.
For the phase field we use Jacobi and a conjugate gradient method.

To prevent leakage of ions out of the two coalescing droplets, a weighted geometric mean was used for the diffusivities:
\begin{equation}
  K_j (\phi) = K_{j,1}^{\frac{1+\phi}{2}} \cdot K_{j,2}^{\frac{1-\phi}{2}},
\end{equation}
instead of the arithmetic mean \eqref{eq:Ki_intp} used in most of the article.

We consider a setup of two initially spherical droplets in a domain $\Omega = [0, L_x] \crossprod [0, L_y] \crossprod [0, L_z]$.
The droplets are centered at $(L_x/2, L_y/2, (L_z \pm L_x)/2)$ and have a radius $R$.
The lower droplet (along the $z$-axis) is initialized with a Gaussian concentration distribution of negative ions ($z_- = -1$), whereas the upper droplet is initialized with positive ions ($z_+ = 1$).
The average concentration of the respective ion species within each droplet is $c_0$, such that the total charge in the system is zero, and the initial spread (standard deviation) of the Gaussian distribution is $R/3$.
A potential $V_0$ is set on the top plane at $z = L_z$ and the bottom plane at $z=0$ is taken to be grounded.
We assume no-slip and no-flux conditions on all boundaries, except for the electrostatic potential $V$ at the top and bottom planes, and the fluid is taken to be in a quiescent state at the initial time $t=0$.
The phasic parameters used in the simulations are given in Table \ref{tab:charged_droplets_3d_phasic_parameters}, and the remaining parameters are given in Table \ref{tab:charged_droplets_3d_parameters}.
The problem is implemented in the file \texttt{problems/charged\_droplets\_3D.py}.
\begin{table}
  \caption{
    \label{tab:charged_droplets_3d_phasic_parameters}
    Phasic parameters for the simulations of droplet coalescence and breakup in an electric field.
    The subscript $\pm$ indicates the value for both the positive and negative ions.
  }
  \begin{ruledtabular}
    \begin{tabular}{l c r r}
      Parameter       & Symbol        & \text{Value in phase 1} & \text{Value in phase 2} \\
      \hline
      Viscosity       & $\mu$       & 1.0    & 0.5 \\
      Density         & $\rho$      & 500.0   & 50.0 \\
      Permittivity    & $\epsilon$  & 1.0    & 2.0 \\
      Solution energy & $\beta_\pm$ & 2      & 0 \\
      Ion mobility    & $D_\pm$     & 0.0001 & 0.1 \\
    \end{tabular}
  \end{ruledtabular}
\end{table}

\begin{table}[htb]
  \caption{
    \label{tab:charged_droplets_3d_parameters}
    Simulation parameters for the simulations of droplet coalescence and breakup in an electric field.
  }
  \begin{ruledtabular}
    \begin{tabular}{l c r}
      Parameter            & Symbol        & Value \\
      \hline
      Length along $x$     & $L_x$         & 1.0 \\
      Length along $y$     & $L_y$         & 1.0 \\
      Height               & $L_z$         & 2.0 \\
      Total simulation time & $T$          & 20 \\
      Initial radius       & $R$           & 0.2 \\
      Time step            & $\tau$        & 0.005 \\
      Resolution           & $h$           & 1/64 \\
      Interface thickness  & $\varepsilon$ & 0.01 \\
      Phase field mobility & $M_0$         & $1\cdot10^{-5}$ \\
      Surface tension      & $\sigma$      & 2.0 \\
      Initial avg.\ concentration & $c_{0}$  & 20.0 \\
    \end{tabular}
  \end{ruledtabular}
\end{table}

\begin{figure}[htb]
  \includegraphics[width=0.95\columnwidth]{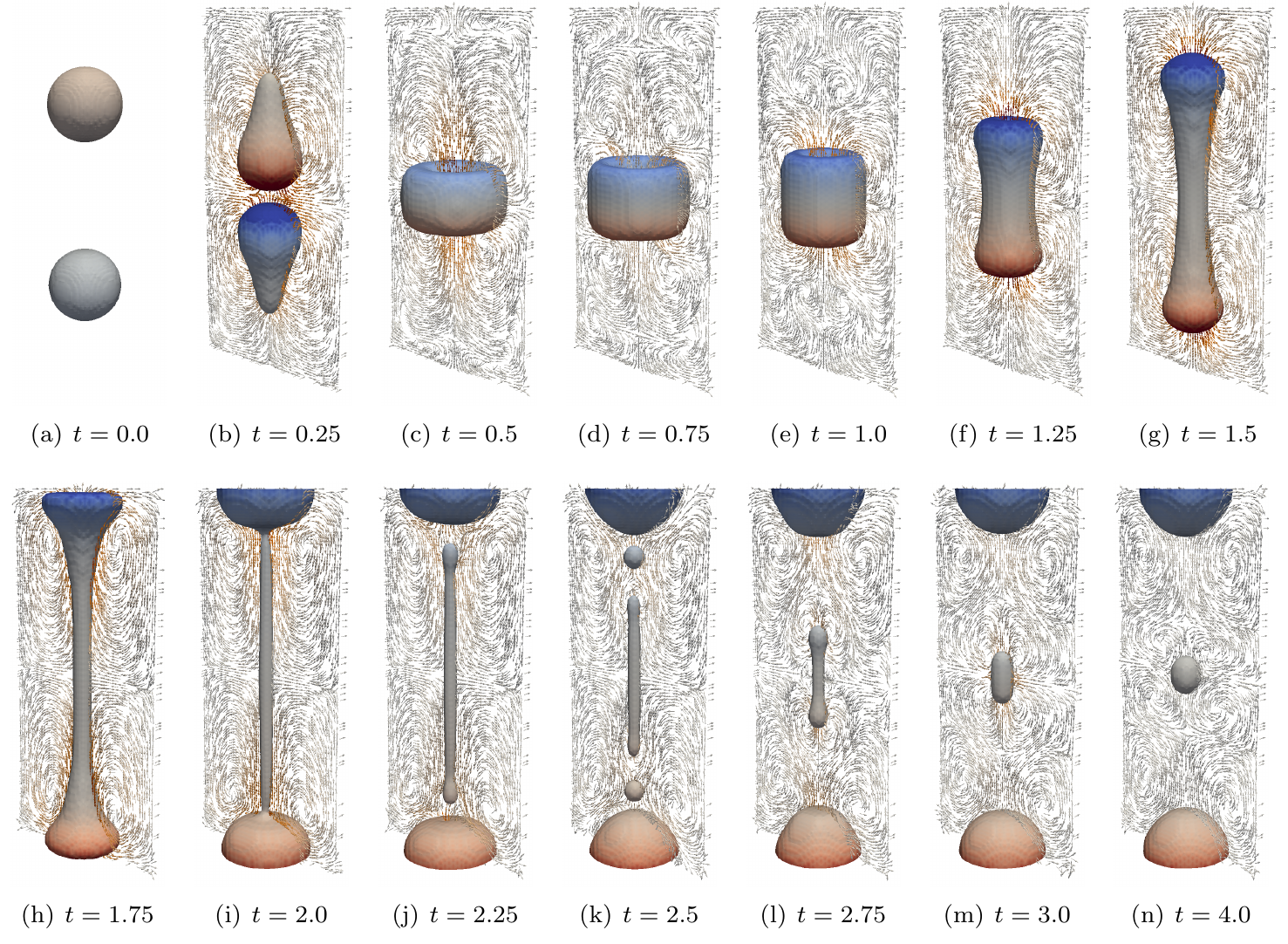}
  \caption{\label{fig:charged_droplets_3d_snapshots}
    Snapshots from the simulations of droplet coalescence and subsequent breakup in an electric field.
    The phase boundary shows the $\phi=0$ isosurface of the phase field.
    The coloring indicates charge: red is positive and blue is negative.
    The color bar goes from -20 (deep blue) to 20 (deep red).
    The quivers show the velocity field in the $x=0.5$ plane (color indicates intensity).
  }
\end{figure}

Fig.\ \ref{fig:charged_droplets_3d_snapshots} shows snapshots from the simulations at several instances of time.
As seen from the figure, the droplets are set in motion towards each other by the electric field and collide with each other.
Subsequently, the unified droplet is stretched, until it touches both electrodes.
The middle part then breaks off, and as it is unstable, it further emits droplets that are  released to two two sides.
Finally, two spherical caps form at each electrode, and a neutral drop is left in the middle, due to the initial symmetry.
Similar behaviour has been observed in axisymmetric simulations (e.g.\ \cite{pillai2015}).

\begin{figure}[htb]
  \includegraphics[width=0.5\columnwidth]{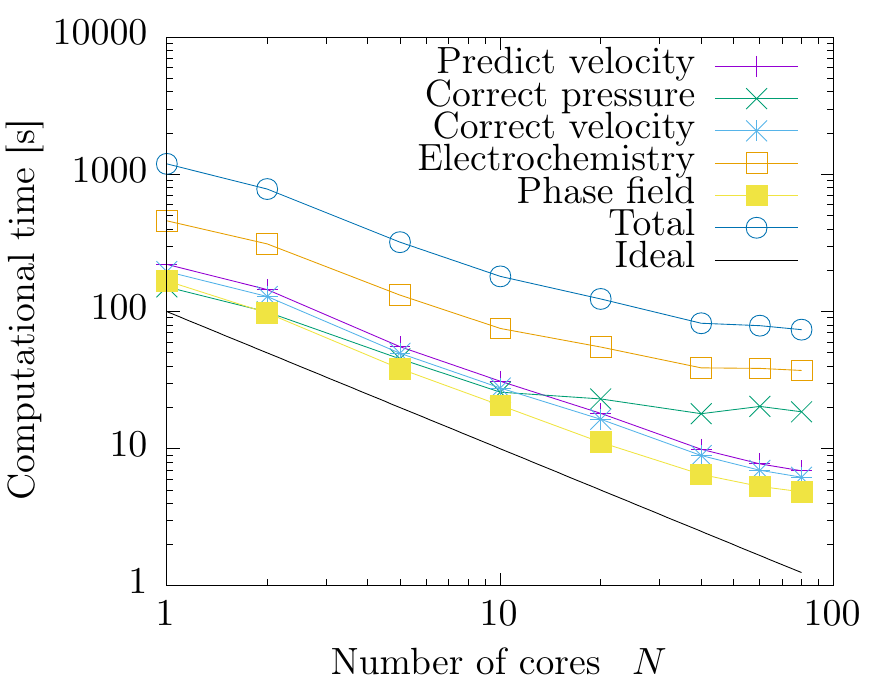}
  \caption{\label{fig:scaling}
    Strong scaling test.
    We show computational time per timestep versus number of processor cores for the coalescence and breakup of droplets in 3D.
    The results are averaged over the 10 first timesteps for simulations with $128\crossprod 128\crossprod 256 = 4 194 304$ degrees of freedom, with a time step $\tau = 0.02$.
  }
\end{figure}

We finally carry out a strong scaling test of the linear iterative solver on a single in-house server with 80 dedicated cores.
The results of average computational time per time step (averaged over 10 time steps) versus number of cores are shown in Fig.\ \ref{fig:scaling}.
We show here the amount of time spent per time step for all substeps in order to illuminate where most of the computational resources are spent.
As can be seen, a significant portion of the computational time is spent on the electrochemical substep.
Overall, the solver displays sublinear scaling with the number of cores, but the results are promising given that neither the solver nor the FEniCS install (a standard PPA install of FEniCS 2017.2.0 on Ubuntu 16.04 server) are fully optimized.
Much could be gained by improving the two steps where solving a Poisson equation is involved; in particular it seems possible that more specifically tailored preconditioners than the straightforward AMG preconditioning could impart speedup.
However, we stress that the division of labour between the steps is highly problem-dependent, and in particular, the electrochemical subproblem is susceptible to how far into the non-linear regime we are (see e.g., \cite{bolet2018}).

\section{Discussion and conclusion}
\label{sec:conclusion}
We have in this work presented \emph{Bernaise}, a flexible open-source framework for simulating two-phase electrohydrodynamics in complex geometries using a phase-field model.
The solver is written in its entirety in Python, and is built on top of the FEniCS/DOLFIN framework \cite{logg2012,logg2010} for solving partial differential equations using the finite element method on unstructured meshes.
FEniCS in turn interfaces to, e.g., scalable state-of-the art linear solvers through its PETSc backend \cite{petsc2017}.
We have proposed a linear operator-splitting scheme to solve the coupled non-linear equations of two-phase electrohydrodynamics.
In contrast to solving the equations directly in a monolithic manner, the scheme sequentially solves the Cahn--Hilliard equation for the phase field describing the interface, the Poisson--Nernst--Planck equations for the electrochemistry (solute transport and electrostatics), and the Navier--Stokes equations for the hydrodynamics, at each time step.
Implementation of new solvers and problems has been demonstrated through representative examples.
Validation of the implementation was carried out by three means: (1) By comparison to analytic solutions in limiting cases where such are available, (2) by the method of manufactured solution through an augmented Taylor--Green vortex, and (3) through convergence to a highly resolved solution of a new two-phase electrohydrodynamics benchmark problem of an electrically driven droplet.
Finally, we have presented applications of the framework in non-trivial settings.
Firstly, to test the applicability of the code in a complicated geometry, and to illuminate the effects of dynamic electrowetting, we simulated a shear flow of water containing an electrolyte over a dead-end pore initially filled with oil.
This problem is relevant from a geophysical standpoint, and exemplifies the potential of the method to simulate the dynamics of the interaction between two-phase flow and electric double layers.
Secondly, the ability of the framework to simulate three-dimensional configurations was demonstrated using a fully iterative version of the operator-splitting scheme, by simulating the coalescence and subsequent breakup of two oppositely charged droplets in an electric field.
%Finally, we have tested the parallel scalability of the linear scheme implementation and compared it to its monolithic counterpart.
The parallel scalability of the latter solver was tested on in-house computing facilities.
The results presented herein underpin our aim that \emph{Bernaise} can become a valuable tool both within the micro- and nanofluidics community and within geophysical simulation.

There are several possible avenues for further development and use of \emph{Bernaise}.
With regards to computational effort, the linear operator-splitting scheme constitutes a major computational improvemnt over a corresponding monolithic scheme.
For the resulting smaller and simpler subproblems, more specialized linear solvers and preconditioners can be used.
However, the implementation of the schemes are still not fully optimized, as in many cases it is not strictly necessary to reassemble entire system matrices (multiple times) at every time step.
Using ideas e.g.~from Ref.~\cite{mortensen2015} on how to effectively preassemble system matrices in FEniCS, one could achieve an implementation that is to a larger extent dominated by the backend linear solvers.
However, as the phase field is updated at every time step, there may be less to gain in performance than what was the case in the latter reference.

With regard to solving the Navier--Stokes equations, the solvers considered herein either rely on a coupled solution of the (the \texttt{basic} and \texttt{basicnewton} solvers) or a fractional step approach that splits between the computations of velocity and pressure (the \texttt{fracstep} solver that was considered in Sec.\ \ref{sec:charged_droplets_3d}).
Using direct linear solvers, the coupled solvers yield accurate prediction of the pressure and can be expected to be more robust.
However, direct solvers have numerical disadvantages when it comes to scalability, and Krylov solvers require specifically tailored preconditioners to achieve robust convergence.
An avenue for further research is to refine the \texttt{fracstep} solver and develop decoupled energy-stable schemes for this problem, which seems possible by building on literature on similar systems \cite{guillen-gonzalez2014,metzger2015,metzger2018,shen2015,grun2016,linga2018decoupled}.
%This could further impart efficient simulations of three-dimensional effects.
The implementation of such enhanced schemes in \emph{Bernaise} is straighforward, as demonstrated in this paper.

A clear enhancement of \emph{Bernaise} would be adaptivity, both in time and space.
Adaptivity in time should be implemented such that time step is variable and controlled by the globally largest propagation velocity (in any field), and a Courant number of choice.
Adaptivity in space is presently only supported as a one-way operation.
Adaptive mesh refinement is already used in the mesh initialization phase in many of the implemented problems.
However, mesh coarsening has currently limited support in FEniCS and to the authors' knowledge there are no concrete plans of adding support for this.
Hence, \emph{Bernaise} lacks an adaptive mesh functionality, but this could be implemented in an \emph{ad hoc} manner with some code restructuring.

In this article, we have not considered any \emph{direct} dependence of the contact angle (i.e.~the surface energies) on an applied electric field.
%This remains an open area of research, and a good starting point may be the critical review of wetting boundary conditions for phase-field models by \citet{huang2015}.
However, the contact angle on scales below the Debye length is generally thought to be unaffected, albeit on scales larger than the insulator thickness, an apparent contact angle forms \cite{mugele2007,linga2018controlling}.
Using the full two-phase electrohydrodynamic model presented herein, effective contact angle dependencies upon the zeta potential could be measured and used in simulations of more macroscopic models; i.e.~models admissible on scales where the electrical double layers are not fully resolved.
This would result in a modified contact angle energy that would be enforced as a boundary condition in a phase field model \cite{huang2015}.

Physically, several extensions of the model could be included in the simulation framework.
Surfactants may influence the dynamics of droplets and interfaces, and could be included as in e.g.~the model by \citet{teigen2011}.
The model in its current form further assumes that we are concerned with dilute solutions (i.e., ideal gas law for the concentration), and hence more complicated electrochemistry could to some extent be incorporated into the chemical free energy $\alpha(c)$.

Finally, the requirement of the electrical double layer to be well-resolved constitutes the main constraint for upscaling of the current method.
Thus, for simulation of two-phase electrohydrodynamic flow on larger scales, if ionic transport need not be accounted for, it would only require minor modifications of the code to run the somewhat simpler Taylor--Melcher leaky dielectric model, e.g.\ in the formulation by \citet{lin2012}, within the current framework.

\begin{acknowledgments}
This project has received funding from the European Union's Horizon 2020 research and innovation program through a Marie Curie initial training networks under grant agreement no.\ 642976 (NanoHeal), and from the Villum Fonden through the grant ``Earth Patterns.''
\end{acknowledgments}

\appendix
\section{Electrokinetic scaling of the equations}
\label{sec:electrokinetic_scaling}
For completeness, we show here, as in a companion paper \cite{linga2018controlling}, how the dimensionless variable scaling assumed in this paper arises from the equations formulated in physical (e.g., SI) units.
The scaling results in equations that are easier to work with, but that need to be scaled back to physical units in order to be e.g., compared to experiments.

For concreteness, we consider the standard Nernst--Planck equation (i.e., dilute solutions) for solute transport, which in physical units can be written as
\begin{equation}
  \pd {c_j} {t} + \v u \cdot \grad c_j = \div \left( D_j \grad c_j - \frac{z_j q_e c_j}{k_{\rm B} T} \v E \right),
  \label{eq:cont_NP}
\end{equation}
where $k_{\rm B}$ is Boltzmann's constant, $T$ is the temperature, and $q_e$ is the elementary charge.
The Poisson equation is in physical units given by
\begin{equation}
  \div \left( \epsilon_0 \epsilon_{\rm r} \v E \right) = \rho_e,
  \label{eq:cont_P}
\end{equation}
where the net charge is given by $\rho_e = q_e \sum_j c_j$.
The Navier--Stokes equations are given by the usual
\begin{gather}
  \rho \left( \pdt \v u + \v u \cdot \grad \v u \right) - \mu \grad^2 \v u + \grad p = - \rho_e \grad V, \label{eq:cont_NS1}\\
  \div \v u = 0. \label{eq:cont_NS2}
\end{gather}
Continuity of the normal stress across the interface between the phases can be formulated as
\begin{equation}
  \Big[ 2 \mu \mathcal D \v u  - p' \v I + 
  \sigma \kappa \v I + \epsilon_0 \epsilon_{\rm r} \v E \otimes \v E - \frac{1}{2} \epsilon_0 \epsilon_{\rm r} \v E^2 \v I \Big] \cdot \hat{\v n}_{\rm int} = \v 0,
  \label{eq:cont_interface}
\end{equation}
where $p'$ is a pressure which has absorbed some extra gradient terms.
We introduce now dimensionless versions of all physical variables, and indicate the dimensionless versions by a tilde.
Further, all reference values are marked with an asterisk.
Hence, we let $\tilde t = t / t^*$, $\tilde \rho = \rho/\rho^*$, $\tilde {\v u} = \v u / u^*$, $\tilde p = p/p^*$, $\tilde\mu = \mu/\mu^*$, $\tilde c_j = c_j / c^*$, $\tilde V = V/V^*$, $\tilde D_\pm = D_\pm / D^*$, $\tilde \epsilon = \epsilon_{\rm r} / \epsilon^*$, and $\tilde \sigma = \sigma / \sigma^*$.
All spatial dimensions are scaled by a reference linear size $R^*$, such that $\tilde {\v x} = {\v x} / R^*$.
The electrostatic potential $V$ is scaled by a thermal voltage,
\begin{equation}
  V^* = V_T = \frac{k_{\rm B} T}{q_e}.
\end{equation}
The other reference values are given by \cite{linga2018controlling}
\begin{gather}
  t^* = \frac{R^*}{u^*}, \quad
  \rho^* = \frac{q_e c^* V_T}{(u^*)^2}, \quad
  D^* = u^* R^*, \quad
  p^* = q_e c^* V_T, \\
  \mu^* = \frac{q_e c^* V_T R^*}{u^*}, \quad
  \epsilon^* = \frac{ q_e c^* (R^*)^2}{\epsilon_0 V_T}, \quad
  \sigma^* = q_e c^* V_T R^* .
\end{gather}
This constitutes an invertible set of relations between the physical and dimensionless variables.
In particular, adopting the dimensionless variables and subsequently dropping the tildes, results in the set of equations \eqref{eq:cont_NP} to  \eqref{eq:cont_interface} with $q_e = k_B T = 1$ and $\epsilon_0 \epsilon_{\rm r} \to \epsilon$.
This is essentially the scaling adopted in this paper.

\section{Poisson--Boltzmann equation for two phases}
\label{sec:app_pb}
Here, we derive a generalized Poisson--Boltzmann equation for the case of two phases, valid in equilibrium.
We are here considering the steady state of the sharp interface equations.
Considering Eq.~\eqref{eq:sharp_conc} with $\partial_t = 0$ and $\v v = \v 0$, taking the inner product of it with $g_{c_j}$, and integrating over the domain $\Omega$, we obtain
\begin{equation}
  \int_\Omega K_{ij} c_j |\grad g_j|^2 \, \diff \Omega = \int_{\partial\Gamma} K_{ij} c_j g_{c_j} \hat{\v n} \cdot \grad g_{c_j} \, \diff \Gamma = 0,
\end{equation}
where the last equality holds, since at equilibrium the fluxes must vanish at the boundary (and hence also in the bulk).
Since $c_j$ is positive, $g_{c_j}$ may not vary.
Hence, the electrochemical potential associated with electrolyte $j$ must satisfy:
\begin{equation}
  g_{c_j} = \alpha'(c_j) + \beta_{ij} + z_j V = C_j,
  \label{eq:pb_g}
\end{equation}
where $C_j$ is a constant.
We assume that one of the two phases is connected to a reservoir far away, such that here $\beta_{ij} = \beta_R$, $c_j = c_R$ and $V=V_R$.
Evaluating Eq.~\eqref{eq:pb_g} at the reservoir, we have
\begin{equation}
  C_j = \alpha'(c_R) + \beta_R + z_j V_R.
  \label{eq:pb_g_res}
\end{equation}
By defining $\chi(\cdot)$ as the inverse function of $\alpha'(\cdot)$, we may combine Eqs.~\eqref{eq:pb_g} and \eqref{eq:pb_g_res} and invert with respect to $c_j$:
\begin{equation}
  c_j = \chi \left( \alpha'(c_R) + \beta_R - \beta_{ij} - z_j (V - V_R) \right).
  \label{eq:pb_c}
\end{equation}
Hence, by Eq.~\eqref{eq:sharp_poisson}, we obtain a closed equation for $V$:
\begin{equation}
  \laplacian V = - \varepsilon_i^{-1} \sum_j z_j \chi \left( \alpha'(c_R) + \beta_R - \beta_{ij} - z_j (V - V_R) \right),
  \label{eq:pb_V}
\end{equation}
with the above boundary conditions at the reservoir.
The interface condition between the phases is $\jump{ V } = 0,$
i.e.~continuity in $V$, and the boundary condition at the reservoir is $V=V_R$.
Next, we consider some special cases of this equation.

\subsection{Standard Poisson--Boltzmann}
\label{subsec:standard_pb}
With two symmetric electrolytes, $j \in \{ \pm \}$, $z_\pm = \pm z$, $\beta_{ij} = \beta_i$, $V_R=V$ and the ideal gas chemical potential, we have that $\alpha'(c) = \ln c$, $\chi (a) = e^a$, and we obtain from Eq.~\eqref{eq:pb_V}:
\begin{equation}
  \laplacian V = \frac{2 z c_R}{\varepsilon_i} e^{\beta_R - \beta_{i}} \sinh \left(zV\right) = \frac{\sinh \left(zV\right)}{\lambda_{D,i}^{2} z},
  \label{eq:standard_pb}
\end{equation}
where we have defined a phase-dependent Debye length $\lambda_{i, D} = \sqrt{{\varepsilon_i}e^{-\beta_R+\beta_{i}}/(2 z^2 c_R)}$.
Now, Eq.~\eqref{eq:pb_c} yields that the concentration is retrieved by
\begin{equation}
  c_\pm = c_R e^{\beta_R - \beta_{i} \mp z V}.
  \label{eq:standard_pb_c_pm}
\end{equation}

\subsubsection{Linearized}
When $|zV| \ll 1$, we may expand Eq.~\eqref{eq:standard_pb} to the first order to obtain the linearized Poisson--Boltzmann equation:
\begin{equation}
  \laplacian V = \frac{ V }{\lambda_{D,i}^{2}}.
  \label{eq:standard_pb_linearized}
\end{equation}
In principle, we can also expand Eq.~\eqref{eq:standard_pb_c_pm}:
\begin{equation}
  c_\pm = c_R e^{\beta_R - \beta_{i}} \left( 1 \mp z V \right),
\end{equation}
so that the total charge density is given by
\begin{equation}
  \rho_e = - 2 z^2 c_R e^{\beta_R - \beta_{i}} V .
  \label{eq:pb_rhoe_linearized}
\end{equation}

\subsection{Net charge}
\label{sec:app_net_charge}
Now we consider the single ``net charge'' model which was proposed in Ref.~\cite{eck2009} and used in the simulations of Ref.~\cite{campillo-funollet2012}.
(Note that these papers redefined the diffusivity to absorb the net charge, effectively~$K_{ij} c_j \to K$; but this does not have consequences in the forthcoming.)
Here, we have only one species $c_1 = \rho_e$ with charge $z$, and $\alpha'(c) = \lambda c$, such that $\chi(a) = \lambda^{-1} a$.
We consider the reservoir to be neutrally charged.
Further, $V_R = 0$, for simplicity.
Eq.~\eqref{eq:pb_V} yields
\begin{equation}
  %\laplacian V = \frac{z}{\varepsilon_i \lambda} \left( - \frac{\lambda c_R  + \beta_R - \beta_{i}}{z} + V \right),
  \laplacian V = \frac{z^2}{\varepsilon_i \lambda} \left( V - \frac{\beta_R - \beta_{i}}{z} \right),
  \label{eq:pb_V_netcharge}
\end{equation}
and Eq.~\eqref{eq:pb_c} becomes
\begin{equation}
  % c = c_R + \lambda^{-1}\left(\beta_R - \beta_{i} - z V \right).
  \rho_e = z\lambda^{-1}\left(\beta_R - \beta_{i} - z V \right).
  \label{eq:pb_c_netcharge}
\end{equation}

Note that in the case of single-phase flow, Eq.~\eqref{eq:pb_V_netcharge} becomes
Eq.~\eqref{eq:pb_V} yields
\begin{align}
  \laplacian V = \frac{z^2}{\varepsilon \lambda} V = \lambda_D^{-2} V,
\end{align}
which is the linearized Poisson--Boltzmann equation (see Sec.~\ref{subsec:standard_pb}), where we have identified a Debye length $\lambda_D = \sqrt{\varepsilon \lambda} / z$.
Eq.~\eqref{eq:pb_c_netcharge} becomes:
\begin{equation}
  \rho_e = - \lambda^{-1} z^2 V .
\end{equation}
Comparison to \eqref{eq:pb_rhoe_linearized} leads us to identify $2c_R = \lambda^{-1}$, which yields $\lambda_{D} = \sqrt{{\varepsilon}/(2 z^2 c_R)}$ in compliance with the definition in Sec.~\ref{subsec:standard_pb}.
Note that even though the equilibrium solution complies with the linearized, equilibrium Nernst--Planck equation, the dynamics, particularly with two phases, may differ significantly.

\subsubsection{Simple case}
It is interesting to investigate this equation for a single phase in a finite 1-D geometry, $x \in [0, L]$.
% We may write
% \begin{equation}
%   \dd V x = b (V-a) = \dd {(V-a)} x,
% \end{equation}
% where
% \begin{equation}
%   a = \frac{\lambda c_R}{z} \quad \textrm{and} \quad b = \frac{z}{\varepsilon_i \lambda}.
% \end{equation}
We assume the boundary conditions $\diff V / \diff x |_{x=0}= - \sigma_e / \varepsilon$ and $V|_{x=L} = 0$.
% The general solution is:
% \begin{equation}
%   V = a + A e^{kx} + B e^{-kx},
% \end{equation}
% Further,
% \begin{equation}
%   \d V x = k (A e^{kx} - B e^{-kx} ).
% \end{equation}
%Enforcing the boundary conditions, we obtain:
The solution is
\begin{equation}
  %V = \frac{\lambda c_R}{z} \left(1 - \frac{\cosh kx}{\cosh kL} \right) - \frac{\sigma_e}{\varepsilon k} \,\cosh kx \left( \tanh kx - \tanh k L \right) ,
  V = - \frac{\sigma_e}{\varepsilon k} \left( \sinh kx - \tanh k L \, \cosh kx \right) ,
\end{equation}
%where $k = \sqrt{b}$.
where $k = z / \sqrt{{\varepsilon \lambda}}$.
Thus,
\begin{equation}
  %c =  c_R \, \frac{\cosh kx}{\cosh kL} + \frac{z \sigma_e}{\varepsilon k \lambda} \,\cosh kx \left( \tanh kx - \tanh k L \right) .
  \rho_e =  \frac{z^2 \sigma_e}{\varepsilon k \lambda} \left( \sinh kx - \tanh k L \,\cosh kx \right) .
\end{equation}
Hence this form of the chemical potential yields an exact solution.
Note however, that there is in principle no mechanism controlling the sign of $\rho_e$, which is natural given that it should here signify a net charge.

The total charge is
\begin{align}
  Q &= \int_0^L \rho_e \, \diff x =  \frac{z^2 \sigma_e}{\varepsilon k \lambda} \int_0^L \left( \sinh kx - \tanh k L \cosh kx \right) \diff x \\
    &= \frac{z^2 \sigma_e}{\varepsilon k^2 \lambda} \left[ \cosh kx - \tanh k L \sinh kx \right]_0^L \\
%    &= - \sigma_e \left[ 1 + \frac{\sinh^2 kL - \cosh^2 kL}{\cosh kL} \right]
    &= - \sigma_e \left[ 1 - \frac{1}{\cosh kL} \right],
\end{align}
which approaches the (negative) applied surface charge in the limit of infinite domain, $L\to \infty$, as it should.

\section{Some considerations on testing and applicability of the framework}
\label{sec:testing}
To simplify the complexity of the problem, the scheme can be reduced to describe settings where fewer physical mechanisms are present simultaneously.
\begin{itemize}
\item The very simplest is \emph{pure single-phase flow}, containing only point \ref{pt:splitscheme_vp} from the list in Sec.~\ref{sec:numerics}.
\item Slightly more demanding is \emph{single-phase flow with transport of a tracer dye} (in the absence of electric charges and fields), using points \ref{pt:splitscheme_ci} and \ref{pt:splitscheme_vp}.
\item More demanding, \emph{pure two-phase flow} (with unmatched densities and viscosities), where only points \ref{pt:splitscheme_phi} and \ref{pt:splitscheme_vp} from the list above enter.
\item In the absence of electric charges and external electric potential, two-phase flow with passive transport of a tracer dye can be modelled, i.e.~using the points \ref{pt:splitscheme_phi}, \ref{pt:splitscheme_ci} and \ref{pt:splitscheme_vp}.
\item Time-dependent single-phase EHD can be modelled using points \ref{pt:splitscheme_ci}, \ref{pt:splitscheme_V} and \ref{pt:splitscheme_vp}.
\item There is also a subtle case where all concentrations $c_i = 0$, but the electric field acts as a force on the interface of the two-phase flow only due to the jump in permittivity $\varepsilon$.
  This includes points \ref{pt:splitscheme_phi}, \ref{pt:splitscheme_V} and \ref{pt:splitscheme_vp}.
\item Finally, the full-fledged two-phase EHD includes all the points in the list in Sec.~\ref{sec:numerics}.
\end{itemize}
Rigorous testing the solver should therefore follow these steps of increasing complexity.

\bibliographystyle{apsrev4-1}
\bibliography{references}

 \newcommand{\acis}{Adv. Colloid Interface Sci.}
  \newcommand{\appphyslett}{Appl. Phys. Lett.} \newcommand{\arfm}{Ann. Rev.
  Fluid Mech.} \newcommand{\commcompphys}{Commun. Comput. Phys.}
  \newcommand{\cmame}{Comput. Methods Appl. Mech. Eng.}
  \newcommand{\cpc}{Comput. Phys. Commun.} \newcommand{\csurfa}{Colloids Surf.
  A} \newcommand{\ejam}{Eur. J. Appl. Math.} \newcommand{\energyfuels}{Energy
  Fuels} \newcommand{\faradaydiscuss}{Faraday Discuss.}
  \newcommand{\gji}{Geophys. J. Int.} \newcommand{\ieeetps}{IEEE Trans. Plasma
  Sci.} \newcommand{\ijhmt}{Int. J. Heat Mass Transf.} \newcommand{\ijms}{Int.
  J. Mol. Sci.} \newcommand{\ijmf}{Int. J. Multiph. Flow}
  \newcommand{\ijnmf}{Int. J. Numer. Methods Fluids} \newcommand{\jap}{J. Appl.
  Phys.} \newcommand{\jast}{J. Adhes. Sci. Technol.} \newcommand{\jchemphys}{J.
  Chem. Phys.} \newcommand{\jcis}{J. Colloid Interface Sci.}
  \newcommand{\jcompm}{J. Comp. Math.} \newcommand{\jcompp}{J. Comput. Phys.}
  \newcommand{\jfm}{J. Fluid Mech.} \newcommand{\jmems}{J. Microelectromech.
  Syst.} \newcommand{\jpc}{J. Phys. Chem.} \newcommand{\jpcm}{J. Phys. Condens.
  Matter} \newcommand{\jpcb}{J. Phys. Chem. B} \newcommand{\labchip}{Lab Chip}
  \newcommand{\mmmas}{Math. Models Methods Appl. Sci.}
  \newcommand{\mman}{ESAIM: Math. Model Numer. Anal.}
  \newcommand{\mathcomput}{Math. Comput.} \newcommand{\natgeo}{Nat. Geosci.}
  \newcommand{\natrevchem}{Nat. Rev. Chem.} \newcommand{\nmpde}{Numer. Methods
  Partial Differ. Equ.} \newcommand{\physfluids}{Phys. Fluids}
  \newcommand{\physreve}{Phys. Rev. E} \newcommand{\physrevfluids}{Phys. Rev.
  Fluids} \newcommand{\physrevlett}{Phys. Rev. Lett.} \newcommand{\prsa}{Proc.
  Royal Soc. A} \newcommand{\revmodphys}{Rev. Mod. Phys.}
  \newcommand{\siap}{SIAM J. Appl. Math.} \newcommand{\sinum}{SIAM J. Numer.
  Anal.} \newcommand{\transpporousmedia}{Transp. Porous Media}
\begin{thebibliography}{93}%
\makeatletter
\providecommand \@ifxundefined [1]{%
 \@ifx{#1\undefined}
}%
\providecommand \@ifnum [1]{%
 \ifnum #1\expandafter \@firstoftwo
 \else \expandafter \@secondoftwo
 \fi
}%
\providecommand \@ifx [1]{%
 \ifx #1\expandafter \@firstoftwo
 \else \expandafter \@secondoftwo
 \fi
}%
\providecommand \natexlab [1]{#1}%
\providecommand \enquote  [1]{``#1''}%
\providecommand \bibnamefont  [1]{#1}%
\providecommand \bibfnamefont [1]{#1}%
\providecommand \citenamefont [1]{#1}%
\providecommand \href@noop [0]{\@secondoftwo}%
\providecommand \href [0]{\begingroup \@sanitize@url \@href}%
\providecommand \@href[1]{\@@startlink{#1}\@@href}%
\providecommand \@@href[1]{\endgroup#1\@@endlink}%
\providecommand \@sanitize@url [0]{\catcode `\\12\catcode `\$12\catcode
  `\&12\catcode `\#12\catcode `\^12\catcode `\_12\catcode `\%12\relax}%
\providecommand \@@startlink[1]{}%
\providecommand \@@endlink[0]{}%
\providecommand \url  [0]{\begingroup\@sanitize@url \@url }%
\providecommand \@url [1]{\endgroup\@href {#1}{\urlprefix }}%
\providecommand \urlprefix  [0]{URL }%
\providecommand \Eprint [0]{\href }%
\providecommand \doibase [0]{http://dx.doi.org/}%
\providecommand \selectlanguage [0]{\@gobble}%
\providecommand \bibinfo  [0]{\@secondoftwo}%
\providecommand \bibfield  [0]{\@secondoftwo}%
\providecommand \translation [1]{[#1]}%
\providecommand \BibitemOpen [0]{}%
\providecommand \bibitemStop [0]{}%
\providecommand \bibitemNoStop [0]{.\EOS\space}%
\providecommand \EOS [0]{\spacefactor3000\relax}%
\providecommand \BibitemShut  [1]{\csname bibitem#1\endcsname}%
\let\auto@bib@innerbib\@empty
%</preamble>
\bibitem [{\citenamefont {Lippmann}(1875)}]{lippmann1875}%
  \BibitemOpen
  \bibfield  {author} {\bibinfo {author} {\bibfnamefont {G.}~\bibnamefont
  {Lippmann}},\ }\emph {\bibinfo {title} {Relations entre les
  ph{\'e}nom{\`e}nes {\'e}lectriques et capillaires}},\ \href@noop {} {Ph.D.
  thesis},\ \bibinfo  {school} {Sorbonne} (\bibinfo {year} {1875})\BibitemShut
  {NoStop}%
\bibitem [{\citenamefont {Mugele}\ and\ \citenamefont
  {Baret}(2005)}]{mugele2005}%
  \BibitemOpen
  \bibfield  {author} {\bibinfo {author} {\bibfnamefont {F.}~\bibnamefont
  {Mugele}}\ and\ \bibinfo {author} {\bibfnamefont {J.-C.}\ \bibnamefont
  {Baret}},\ }\href {\doibase 10.1088/0953-8984/17/28/R01} {\bibfield
  {journal} {\bibinfo  {journal} {\jpcm}\ }\textbf {\bibinfo {volume} {17}},\
  \bibinfo {pages} {R705} (\bibinfo {year} {2005})}\BibitemShut {NoStop}%
\bibitem [{\citenamefont {Ristenpart}\ \emph {et~al.}(2009)\citenamefont
  {Ristenpart}, \citenamefont {Bird}, \citenamefont {Belmonte}, \citenamefont
  {Dollar},\ and\ \citenamefont {Stone}}]{ristenpart2009}%
  \BibitemOpen
  \bibfield  {author} {\bibinfo {author} {\bibfnamefont {W.}~\bibnamefont
  {Ristenpart}}, \bibinfo {author} {\bibfnamefont {J.}~\bibnamefont {Bird}},
  \bibinfo {author} {\bibfnamefont {A.}~\bibnamefont {Belmonte}}, \bibinfo
  {author} {\bibfnamefont {F.}~\bibnamefont {Dollar}}, \ and\ \bibinfo {author}
  {\bibfnamefont {H.}~\bibnamefont {Stone}},\ }\href {\doibase
  10.1038/nature08294} {\bibfield  {journal} {\bibinfo  {journal} {Nature}\
  }\textbf {\bibinfo {volume} {461}},\ \bibinfo {pages} {377} (\bibinfo {year}
  {2009})}\BibitemShut {NoStop}%
\bibitem [{\citenamefont {Mugele}(2009)}]{mugele2009}%
  \BibitemOpen
  \bibfield  {author} {\bibinfo {author} {\bibfnamefont {F.}~\bibnamefont
  {Mugele}},\ }\href@noop {} {\bibfield  {journal} {\bibinfo  {journal}
  {Nature}\ }\textbf {\bibinfo {volume} {461}},\ \bibinfo {pages} {356}
  (\bibinfo {year} {2009})}\BibitemShut {NoStop}%
\bibitem [{\citenamefont {Squires}\ and\ \citenamefont
  {Quake}(2005)}]{squires2005}%
  \BibitemOpen
  \bibfield  {author} {\bibinfo {author} {\bibfnamefont {T.~M.}\ \bibnamefont
  {Squires}}\ and\ \bibinfo {author} {\bibfnamefont {S.~R.}\ \bibnamefont
  {Quake}},\ }\href {\doibase 10.1103/RevModPhys.77.977} {\bibfield  {journal}
  {\bibinfo  {journal} {\revmodphys}\ }\textbf {\bibinfo {volume} {77}},\
  \bibinfo {pages} {977} (\bibinfo {year} {2005})}\BibitemShut {NoStop}%
\bibitem [{\citenamefont {Schoch}\ \emph {et~al.}(2008)\citenamefont {Schoch},
  \citenamefont {Han},\ and\ \citenamefont {Renaud}}]{schoch2008}%
  \BibitemOpen
  \bibfield  {author} {\bibinfo {author} {\bibfnamefont {R.~B.}\ \bibnamefont
  {Schoch}}, \bibinfo {author} {\bibfnamefont {J.}~\bibnamefont {Han}}, \ and\
  \bibinfo {author} {\bibfnamefont {P.}~\bibnamefont {Renaud}},\ }\href
  {\doibase 10.1103/RevModPhys.80.839} {\bibfield  {journal} {\bibinfo
  {journal} {\revmodphys}\ }\textbf {\bibinfo {volume} {80}},\ \bibinfo {pages}
  {839} (\bibinfo {year} {2008})}\BibitemShut {NoStop}%
\bibitem [{\citenamefont {Mugele}\ \emph {et~al.}(2010)\citenamefont {Mugele},
  \citenamefont {Duits},\ and\ \citenamefont {Van~den Ende}}]{mugele2010}%
  \BibitemOpen
  \bibfield  {author} {\bibinfo {author} {\bibfnamefont {F.}~\bibnamefont
  {Mugele}}, \bibinfo {author} {\bibfnamefont {M.}~\bibnamefont {Duits}}, \
  and\ \bibinfo {author} {\bibfnamefont {D.}~\bibnamefont {Van~den Ende}},\
  }\href {\doibase 10.1016/j.cis.2009.11.002} {\bibfield  {journal} {\bibinfo
  {journal} {\acis}\ }\textbf {\bibinfo {volume} {161}},\ \bibinfo {pages}
  {115} (\bibinfo {year} {2010})}\BibitemShut {NoStop}%
\bibitem [{\citenamefont {Nelson}\ and\ \citenamefont
  {Kim}(2012)}]{nelson2012}%
  \BibitemOpen
  \bibfield  {author} {\bibinfo {author} {\bibfnamefont {W.~C.}\ \bibnamefont
  {Nelson}}\ and\ \bibinfo {author} {\bibfnamefont {C.-J.}\ \bibnamefont
  {Kim}},\ }\href {\doibase 10.1163/156856111X599562} {\bibfield  {journal}
  {\bibinfo  {journal} {\jast}\ }\textbf {\bibinfo {volume} {26}},\ \bibinfo
  {pages} {1747} (\bibinfo {year} {2012})}\BibitemShut {NoStop}%
\bibitem [{\citenamefont {Pollack}\ \emph {et~al.}(2002)\citenamefont
  {Pollack}, \citenamefont {Shenderov},\ and\ \citenamefont
  {Fair}}]{pollack2002}%
  \BibitemOpen
  \bibfield  {author} {\bibinfo {author} {\bibfnamefont {M.}~\bibnamefont
  {Pollack}}, \bibinfo {author} {\bibfnamefont {A.}~\bibnamefont {Shenderov}},
  \ and\ \bibinfo {author} {\bibfnamefont {R.}~\bibnamefont {Fair}},\ }\href
  {\doibase 10.1039/b110474h} {\bibfield  {journal} {\bibinfo  {journal}
  {\labchip}\ }\textbf {\bibinfo {volume} {2}},\ \bibinfo {pages} {96}
  (\bibinfo {year} {2002})}\BibitemShut {NoStop}%
\bibitem [{\citenamefont {Srinivasan}\ \emph {et~al.}(2004)\citenamefont
  {Srinivasan}, \citenamefont {Pamula},\ and\ \citenamefont
  {Fair}}]{srinivasan2004}%
  \BibitemOpen
  \bibfield  {author} {\bibinfo {author} {\bibfnamefont {V.}~\bibnamefont
  {Srinivasan}}, \bibinfo {author} {\bibfnamefont {V.~K.}\ \bibnamefont
  {Pamula}}, \ and\ \bibinfo {author} {\bibfnamefont {R.~B.}\ \bibnamefont
  {Fair}},\ }\href {\doibase 10.1039/b403341h} {\bibfield  {journal} {\bibinfo
  {journal} {\labchip}\ }\textbf {\bibinfo {volume} {4}},\ \bibinfo {pages}
  {310} (\bibinfo {year} {2004})}\BibitemShut {NoStop}%
\bibitem [{\citenamefont {Lee}\ and\ \citenamefont {Kim}(2000)}]{lee2000}%
  \BibitemOpen
  \bibfield  {author} {\bibinfo {author} {\bibfnamefont {J.}~\bibnamefont
  {Lee}}\ and\ \bibinfo {author} {\bibfnamefont {C.-J.}\ \bibnamefont {Kim}},\
  }\href {\doibase 10.1109/84.846697} {\bibfield  {journal} {\bibinfo
  {journal} {\jmems}\ }\textbf {\bibinfo {volume} {9}},\ \bibinfo {pages} {171}
  (\bibinfo {year} {2000})}\BibitemShut {NoStop}%
\bibitem [{\citenamefont {Siria}\ \emph {et~al.}(2017)\citenamefont {Siria},
  \citenamefont {Bocquet},\ and\ \citenamefont {Bocquet}}]{siria2017}%
  \BibitemOpen
  \bibfield  {author} {\bibinfo {author} {\bibfnamefont {A.}~\bibnamefont
  {Siria}}, \bibinfo {author} {\bibfnamefont {M.-L.}\ \bibnamefont {Bocquet}},
  \ and\ \bibinfo {author} {\bibfnamefont {L.}~\bibnamefont {Bocquet}},\ }\href
  {\doibase 10.1038/s41570-017-0091} {\bibfield  {journal} {\bibinfo  {journal}
  {\natrevchem}\ }\textbf {\bibinfo {volume} {1}},\ \bibinfo {pages} {0091}
  (\bibinfo {year} {2017})}\BibitemShut {NoStop}%
\bibitem [{\citenamefont {Beni}\ and\ \citenamefont
  {Hackwood}(1981)}]{beni1981}%
  \BibitemOpen
  \bibfield  {author} {\bibinfo {author} {\bibfnamefont {G.}~\bibnamefont
  {Beni}}\ and\ \bibinfo {author} {\bibfnamefont {S.}~\bibnamefont
  {Hackwood}},\ }\href {\doibase 10.1063/1.92322} {\bibfield  {journal}
  {\bibinfo  {journal} {\appphyslett}\ }\textbf {\bibinfo {volume} {38}},\
  \bibinfo {pages} {207} (\bibinfo {year} {1981})}\BibitemShut {NoStop}%
\bibitem [{\citenamefont {Beni}\ and\ \citenamefont {Tenan}(1981)}]{beni1981b}%
  \BibitemOpen
  \bibfield  {author} {\bibinfo {author} {\bibfnamefont {G.}~\bibnamefont
  {Beni}}\ and\ \bibinfo {author} {\bibfnamefont {M.}~\bibnamefont {Tenan}},\
  }\href {\doibase 10.1063/1.329822} {\bibfield  {journal} {\bibinfo  {journal}
  {\jap}\ }\textbf {\bibinfo {volume} {52}},\ \bibinfo {pages} {6011} (\bibinfo
  {year} {1981})}\BibitemShut {NoStop}%
\bibitem [{\citenamefont {Beni}\ \emph {et~al.}(1982)\citenamefont {Beni},
  \citenamefont {Hackwood},\ and\ \citenamefont {Jackel}}]{beni1982}%
  \BibitemOpen
  \bibfield  {author} {\bibinfo {author} {\bibfnamefont {G.}~\bibnamefont
  {Beni}}, \bibinfo {author} {\bibfnamefont {S.}~\bibnamefont {Hackwood}}, \
  and\ \bibinfo {author} {\bibfnamefont {J.}~\bibnamefont {Jackel}},\ }\href
  {\doibase 10.1063/1.92920} {\bibfield  {journal} {\bibinfo  {journal}
  {\appphyslett}\ }\textbf {\bibinfo {volume} {40}},\ \bibinfo {pages} {912}
  (\bibinfo {year} {1982})}\BibitemShut {NoStop}%
\bibitem [{\citenamefont {Hayes}\ and\ \citenamefont
  {Feenstra}(2003)}]{hayes2003}%
  \BibitemOpen
  \bibfield  {author} {\bibinfo {author} {\bibfnamefont {R.~A.}\ \bibnamefont
  {Hayes}}\ and\ \bibinfo {author} {\bibfnamefont {B.~J.}\ \bibnamefont
  {Feenstra}},\ }\href {\doibase 10.1038/nature01988} {\bibfield  {journal}
  {\bibinfo  {journal} {Nature}\ }\textbf {\bibinfo {volume} {425}},\ \bibinfo
  {pages} {383} (\bibinfo {year} {2003})}\BibitemShut {NoStop}%
\bibitem [{\citenamefont {Pride}\ and\ \citenamefont
  {Morgan}(1991)}]{pride1991}%
  \BibitemOpen
  \bibfield  {author} {\bibinfo {author} {\bibfnamefont {S.~R.}\ \bibnamefont
  {Pride}}\ and\ \bibinfo {author} {\bibfnamefont {F.}~\bibnamefont {Morgan}},\
  }\href {\doibase 10.1190/1.1443125} {\bibfield  {journal} {\bibinfo
  {journal} {Geophysics}\ }\textbf {\bibinfo {volume} {56}},\ \bibinfo {pages}
  {914} (\bibinfo {year} {1991})}\BibitemShut {NoStop}%
\bibitem [{\citenamefont {Fiorentino}\ \emph
  {et~al.}(2016{\natexlab{a}})\citenamefont {Fiorentino}, \citenamefont
  {Toussaint},\ and\ \citenamefont {Jouniaux}}]{fiorentino2016a}%
  \BibitemOpen
  \bibfield  {author} {\bibinfo {author} {\bibfnamefont {E.-A.}\ \bibnamefont
  {Fiorentino}}, \bibinfo {author} {\bibfnamefont {R.}~\bibnamefont
  {Toussaint}}, \ and\ \bibinfo {author} {\bibfnamefont {L.}~\bibnamefont
  {Jouniaux}},\ }\href {\doibase 10.1093/gji/ggw041} {\bibfield  {journal}
  {\bibinfo  {journal} {\gji}\ }\textbf {\bibinfo {volume} {205}},\ \bibinfo
  {pages} {648} (\bibinfo {year} {2016}{\natexlab{a}})}\BibitemShut {NoStop}%
\bibitem [{\citenamefont {Fiorentino}\ \emph
  {et~al.}(2016{\natexlab{b}})\citenamefont {Fiorentino}, \citenamefont
  {Toussaint},\ and\ \citenamefont {Jouniaux}}]{fiorentino2016b}%
  \BibitemOpen
  \bibfield  {author} {\bibinfo {author} {\bibfnamefont {E.-A.}\ \bibnamefont
  {Fiorentino}}, \bibinfo {author} {\bibfnamefont {R.}~\bibnamefont
  {Toussaint}}, \ and\ \bibinfo {author} {\bibfnamefont {L.}~\bibnamefont
  {Jouniaux}},\ }\href {\doibase 10.1093/gji/ggw417} {\bibfield  {journal}
  {\bibinfo  {journal} {\gji}\ }\textbf {\bibinfo {volume} {208}},\ \bibinfo
  {pages} {1139} (\bibinfo {year} {2016}{\natexlab{b}})}\BibitemShut {NoStop}%
\bibitem [{\citenamefont {Hassenkam}\ \emph {et~al.}(2011)\citenamefont
  {Hassenkam}, \citenamefont {Pedersen}, \citenamefont {Dalby}, \citenamefont
  {Austad},\ and\ \citenamefont {Stipp}}]{hassenkam2011}%
  \BibitemOpen
  \bibfield  {author} {\bibinfo {author} {\bibfnamefont {T.}~\bibnamefont
  {Hassenkam}}, \bibinfo {author} {\bibfnamefont {C.~S.}\ \bibnamefont
  {Pedersen}}, \bibinfo {author} {\bibfnamefont {K.}~\bibnamefont {Dalby}},
  \bibinfo {author} {\bibfnamefont {T.}~\bibnamefont {Austad}}, \ and\ \bibinfo
  {author} {\bibfnamefont {S.~L.~S.}\ \bibnamefont {Stipp}},\ }\href {\doibase
  10.1016/j.colsurfa.2011.09.025} {\bibfield  {journal} {\bibinfo  {journal}
  {\csurfa}\ }\textbf {\bibinfo {volume} {390}},\ \bibinfo {pages} {179}
  (\bibinfo {year} {2011})}\BibitemShut {NoStop}%
\bibitem [{\citenamefont {Hilner}\ \emph {et~al.}(2015)\citenamefont {Hilner},
  \citenamefont {Andersson}, \citenamefont {Hassenkam}, \citenamefont
  {Matthiesen}, \citenamefont {Salino},\ and\ \citenamefont
  {Stipp}}]{hilner2015}%
  \BibitemOpen
  \bibfield  {author} {\bibinfo {author} {\bibfnamefont {E.}~\bibnamefont
  {Hilner}}, \bibinfo {author} {\bibfnamefont {M.~P.}\ \bibnamefont
  {Andersson}}, \bibinfo {author} {\bibfnamefont {T.}~\bibnamefont
  {Hassenkam}}, \bibinfo {author} {\bibfnamefont {J.}~\bibnamefont
  {Matthiesen}}, \bibinfo {author} {\bibfnamefont {P.}~\bibnamefont {Salino}},
  \ and\ \bibinfo {author} {\bibfnamefont {S.~L.~S.}\ \bibnamefont {Stipp}},\
  }\href {\doibase 10.1038/srep09933} {\bibfield  {journal} {\bibinfo
  {journal} {Scientific Reports}\ }\textbf {\bibinfo {volume} {5}},\ \bibinfo
  {pages} {9933} (\bibinfo {year} {2015})}\BibitemShut {NoStop}%
\bibitem [{\citenamefont {RezaeiDoust}\ \emph {et~al.}(2009)\citenamefont
  {RezaeiDoust}, \citenamefont {Puntervold}, \citenamefont {Strand},\ and\
  \citenamefont {Austad}}]{rezaeidoust2009}%
  \BibitemOpen
  \bibfield  {author} {\bibinfo {author} {\bibfnamefont {A.}~\bibnamefont
  {RezaeiDoust}}, \bibinfo {author} {\bibfnamefont {T.}~\bibnamefont
  {Puntervold}}, \bibinfo {author} {\bibfnamefont {S.}~\bibnamefont {Strand}},
  \ and\ \bibinfo {author} {\bibfnamefont {T.}~\bibnamefont {Austad}},\ }\href
  {\doibase 10.1021/ef900185q} {\bibfield  {journal} {\bibinfo  {journal}
  {\energyfuels}\ }\textbf {\bibinfo {volume} {23}},\ \bibinfo {pages} {4479}
  (\bibinfo {year} {2009})}\BibitemShut {NoStop}%
\bibitem [{\citenamefont {Pedersen}\ \emph {et~al.}(2016)\citenamefont
  {Pedersen}, \citenamefont {Hassenkam}, \citenamefont {Ceccato}, \citenamefont
  {Dalby}, \citenamefont {Mogensen},\ and\ \citenamefont
  {Stipp}}]{pedersen2016}%
  \BibitemOpen
  \bibfield  {author} {\bibinfo {author} {\bibfnamefont {N.}~\bibnamefont
  {Pedersen}}, \bibinfo {author} {\bibfnamefont {T.}~\bibnamefont {Hassenkam}},
  \bibinfo {author} {\bibfnamefont {M.}~\bibnamefont {Ceccato}}, \bibinfo
  {author} {\bibfnamefont {K.~N.}\ \bibnamefont {Dalby}}, \bibinfo {author}
  {\bibfnamefont {K.}~\bibnamefont {Mogensen}}, \ and\ \bibinfo {author}
  {\bibfnamefont {S.~L.~S.}\ \bibnamefont {Stipp}},\ }\href {\doibase
  10.1021/acs.energyfuels.5b02562} {\bibfield  {journal} {\bibinfo  {journal}
  {\energyfuels}\ }\textbf {\bibinfo {volume} {30}},\ \bibinfo {pages} {3768}
  (\bibinfo {year} {2016})}\BibitemShut {NoStop}%
\bibitem [{\citenamefont {Pl{\"u}mper}\ \emph {et~al.}(2017)\citenamefont
  {Pl{\"u}mper}, \citenamefont {Botan}, \citenamefont {Los}, \citenamefont
  {Liu}, \citenamefont {Malthe-S{\o}renssen},\ and\ \citenamefont
  {Jamtveit}}]{plumper2017}%
  \BibitemOpen
  \bibfield  {author} {\bibinfo {author} {\bibfnamefont {O.}~\bibnamefont
  {Pl{\"u}mper}}, \bibinfo {author} {\bibfnamefont {A.}~\bibnamefont {Botan}},
  \bibinfo {author} {\bibfnamefont {C.}~\bibnamefont {Los}}, \bibinfo {author}
  {\bibfnamefont {Y.}~\bibnamefont {Liu}}, \bibinfo {author} {\bibfnamefont
  {A.}~\bibnamefont {Malthe-S{\o}renssen}}, \ and\ \bibinfo {author}
  {\bibfnamefont {B.}~\bibnamefont {Jamtveit}},\ }\href {\doibase
  10.1038/ngeo3009} {\bibfield  {journal} {\bibinfo  {journal} {\natgeo}\
  }\textbf {\bibinfo {volume} {10}},\ \bibinfo {pages} {685} (\bibinfo {year}
  {2017})}\BibitemShut {NoStop}%
\bibitem [{\citenamefont {De~Gennes}(1985)}]{degennes1985}%
  \BibitemOpen
  \bibfield  {author} {\bibinfo {author} {\bibfnamefont {P.-G.}\ \bibnamefont
  {De~Gennes}},\ }\href {\doibase 10.1103/RevModPhys.57.827} {\bibfield
  {journal} {\bibinfo  {journal} {\revmodphys}\ }\textbf {\bibinfo {volume}
  {57}},\ \bibinfo {pages} {827} (\bibinfo {year} {1985})}\BibitemShut
  {NoStop}%
\bibitem [{\citenamefont {Bonn}\ \emph {et~al.}(2009)\citenamefont {Bonn},
  \citenamefont {Eggers}, \citenamefont {Indekeu}, \citenamefont {Meunier},\
  and\ \citenamefont {Rolley}}]{bonn2009}%
  \BibitemOpen
  \bibfield  {author} {\bibinfo {author} {\bibfnamefont {D.}~\bibnamefont
  {Bonn}}, \bibinfo {author} {\bibfnamefont {J.}~\bibnamefont {Eggers}},
  \bibinfo {author} {\bibfnamefont {J.}~\bibnamefont {Indekeu}}, \bibinfo
  {author} {\bibfnamefont {J.}~\bibnamefont {Meunier}}, \ and\ \bibinfo
  {author} {\bibfnamefont {E.}~\bibnamefont {Rolley}},\ }\href {\doibase
  10.1103/RevModPhys.81.739} {\bibfield  {journal} {\bibinfo  {journal}
  {\revmodphys}\ }\textbf {\bibinfo {volume} {81}},\ \bibinfo {pages} {739}
  (\bibinfo {year} {2009})}\BibitemShut {NoStop}%
\bibitem [{\citenamefont {Snoeijer}\ and\ \citenamefont
  {Andreotti}(2013)}]{snoeijer2013}%
  \BibitemOpen
  \bibfield  {author} {\bibinfo {author} {\bibfnamefont {J.~H.}\ \bibnamefont
  {Snoeijer}}\ and\ \bibinfo {author} {\bibfnamefont {B.}~\bibnamefont
  {Andreotti}},\ }\href {\doibase 10.1146/annurev-fluid-011212-140734}
  {\bibfield  {journal} {\bibinfo  {journal} {\arfm}\ }\textbf {\bibinfo
  {volume} {45}},\ \bibinfo {pages} {269} (\bibinfo {year} {2013})}\BibitemShut
  {NoStop}%
\bibitem [{\citenamefont {Melcher}\ and\ \citenamefont
  {Taylor}(1969)}]{melcher1969}%
  \BibitemOpen
  \bibfield  {author} {\bibinfo {author} {\bibfnamefont {J.}~\bibnamefont
  {Melcher}}\ and\ \bibinfo {author} {\bibfnamefont {G.}~\bibnamefont
  {Taylor}},\ }\href {\doibase 10.1146/annurev.fl.01.010169.000551} {\bibfield
  {journal} {\bibinfo  {journal} {\arfm}\ }\textbf {\bibinfo {volume} {1}},\
  \bibinfo {pages} {111} (\bibinfo {year} {1969})}\BibitemShut {NoStop}%
\bibitem [{\citenamefont {Saville}(1997)}]{saville1997}%
  \BibitemOpen
  \bibfield  {author} {\bibinfo {author} {\bibfnamefont {D.~A.}\ \bibnamefont
  {Saville}},\ }\href {\doibase 10.1146/annurev.fluid.29.1.27} {\bibfield
  {journal} {\bibinfo  {journal} {\arfm}\ }\textbf {\bibinfo {volume} {29}},\
  \bibinfo {pages} {27} (\bibinfo {year} {1997})}\BibitemShut {NoStop}%
\bibitem [{\citenamefont {Fylladitakis}\ \emph {et~al.}(2014)\citenamefont
  {Fylladitakis}, \citenamefont {Theodoridis},\ and\ \citenamefont
  {Moronis}}]{fylladitakis2014}%
  \BibitemOpen
  \bibfield  {author} {\bibinfo {author} {\bibfnamefont {E.~D.}\ \bibnamefont
  {Fylladitakis}}, \bibinfo {author} {\bibfnamefont {M.~P.}\ \bibnamefont
  {Theodoridis}}, \ and\ \bibinfo {author} {\bibfnamefont {A.~X.}\ \bibnamefont
  {Moronis}},\ }\href {\doibase 10.1109/TPS.2013.2297173} {\bibfield  {journal}
  {\bibinfo  {journal} {\ieeetps}\ }\textbf {\bibinfo {volume} {42}},\ \bibinfo
  {pages} {358} (\bibinfo {year} {2014})}\BibitemShut {NoStop}%
\bibitem [{\citenamefont {Taylor}(1966)}]{taylor1966}%
  \BibitemOpen
  \bibfield  {author} {\bibinfo {author} {\bibfnamefont {G.}~\bibnamefont
  {Taylor}},\ }\href@noop {} {\bibfield  {journal} {\bibinfo  {journal}
  {Proceedings of the Royal Society of London. Series A. Mathematical and
  Physical Sciences}\ }\textbf {\bibinfo {volume} {291}},\ \bibinfo {pages}
  {159} (\bibinfo {year} {1966})}\BibitemShut {NoStop}%
\bibitem [{\citenamefont {Schnitzer}\ and\ \citenamefont
  {Yariv}(2015)}]{schnitzer2015}%
  \BibitemOpen
  \bibfield  {author} {\bibinfo {author} {\bibfnamefont {O.}~\bibnamefont
  {Schnitzer}}\ and\ \bibinfo {author} {\bibfnamefont {E.}~\bibnamefont
  {Yariv}},\ }\href {\doibase 10.1017/jfm.2015.242} {\bibfield  {journal}
  {\bibinfo  {journal} {\jfm}\ }\textbf {\bibinfo {volume} {773}},\ \bibinfo
  {pages} {1} (\bibinfo {year} {2015})}\BibitemShut {NoStop}%
\bibitem [{\citenamefont {Zholkovskij}\ \emph {et~al.}(2002)\citenamefont
  {Zholkovskij}, \citenamefont {Masliyah},\ and\ \citenamefont
  {Czarnecki}}]{zholkovskij2002}%
  \BibitemOpen
  \bibfield  {author} {\bibinfo {author} {\bibfnamefont {E.~K.}\ \bibnamefont
  {Zholkovskij}}, \bibinfo {author} {\bibfnamefont {J.~H.}\ \bibnamefont
  {Masliyah}}, \ and\ \bibinfo {author} {\bibfnamefont {J.}~\bibnamefont
  {Czarnecki}},\ }\href {\doibase 10.1017/S0022112002001441} {\bibfield
  {journal} {\bibinfo  {journal} {\jfm}\ }\textbf {\bibinfo {volume} {472}},\
  \bibinfo {pages} {1} (\bibinfo {year} {2002})}\BibitemShut {NoStop}%
\bibitem [{\citenamefont {Monroe}\ \emph
  {et~al.}(2006{\natexlab{a}})\citenamefont {Monroe}, \citenamefont {Daikhin},
  \citenamefont {Urbakh},\ and\ \citenamefont {Kornyshev}}]{monroe2006}%
  \BibitemOpen
  \bibfield  {author} {\bibinfo {author} {\bibfnamefont {C.~W.}\ \bibnamefont
  {Monroe}}, \bibinfo {author} {\bibfnamefont {L.~I.}\ \bibnamefont {Daikhin}},
  \bibinfo {author} {\bibfnamefont {M.}~\bibnamefont {Urbakh}}, \ and\ \bibinfo
  {author} {\bibfnamefont {A.~A.}\ \bibnamefont {Kornyshev}},\ }\href {\doibase
  10.1103/PhysRevLett.97.136102} {\bibfield  {journal} {\bibinfo  {journal}
  {\physrevlett}\ }\textbf {\bibinfo {volume} {97}},\ \bibinfo {pages} {136102}
  (\bibinfo {year} {2006}{\natexlab{a}})}\BibitemShut {NoStop}%
\bibitem [{\citenamefont {Monroe}\ \emph
  {et~al.}(2006{\natexlab{b}})\citenamefont {Monroe}, \citenamefont {Daikhin},
  \citenamefont {Urbakh},\ and\ \citenamefont {Kornyshev}}]{monroe2006b}%
  \BibitemOpen
  \bibfield  {author} {\bibinfo {author} {\bibfnamefont {C.~W.}\ \bibnamefont
  {Monroe}}, \bibinfo {author} {\bibfnamefont {L.~I.}\ \bibnamefont {Daikhin}},
  \bibinfo {author} {\bibfnamefont {M.}~\bibnamefont {Urbakh}}, \ and\ \bibinfo
  {author} {\bibfnamefont {A.~A.}\ \bibnamefont {Kornyshev}},\ }\href {\doibase
  10.1088/0953-8984/18/10/009} {\bibfield  {journal} {\bibinfo  {journal}
  {\jpcm}\ }\textbf {\bibinfo {volume} {18}},\ \bibinfo {pages} {2837}
  (\bibinfo {year} {2006}{\natexlab{b}})}\BibitemShut {NoStop}%
\bibitem [{\citenamefont {Yang}\ \emph {et~al.}(2013)\citenamefont {Yang},
  \citenamefont {Li},\ and\ \citenamefont {Ding}}]{yang2013}%
  \BibitemOpen
  \bibfield  {author} {\bibinfo {author} {\bibfnamefont {Q.}~\bibnamefont
  {Yang}}, \bibinfo {author} {\bibfnamefont {B.~Q.}\ \bibnamefont {Li}}, \ and\
  \bibinfo {author} {\bibfnamefont {Y.}~\bibnamefont {Ding}},\ }\href {\doibase
  j.ijmultiphaseflow.2013.06.006} {\bibfield  {journal} {\bibinfo  {journal}
  {\ijmf}\ }\textbf {\bibinfo {volume} {57}},\ \bibinfo {pages} {1} (\bibinfo
  {year} {2013})}\BibitemShut {NoStop}%
\bibitem [{\citenamefont {Yang}\ \emph {et~al.}(2014)\citenamefont {Yang},
  \citenamefont {Li}, \citenamefont {Shao},\ and\ \citenamefont
  {Ding}}]{yang2014}%
  \BibitemOpen
  \bibfield  {author} {\bibinfo {author} {\bibfnamefont {Q.}~\bibnamefont
  {Yang}}, \bibinfo {author} {\bibfnamefont {B.~Q.}\ \bibnamefont {Li}},
  \bibinfo {author} {\bibfnamefont {J.}~\bibnamefont {Shao}}, \ and\ \bibinfo
  {author} {\bibfnamefont {Y.}~\bibnamefont {Ding}},\ }\href {\doibase
  10.1016/j.ijheatmasstransfer.2014.07.039} {\bibfield  {journal} {\bibinfo
  {journal} {\ijhmt}\ }\textbf {\bibinfo {volume} {78}},\ \bibinfo {pages}
  {820} (\bibinfo {year} {2014})}\BibitemShut {NoStop}%
\bibitem [{\citenamefont {Berry}\ \emph {et~al.}(2013)\citenamefont {Berry},
  \citenamefont {Davidson},\ and\ \citenamefont {Harvie}}]{berry2013}%
  \BibitemOpen
  \bibfield  {author} {\bibinfo {author} {\bibfnamefont {J.}~\bibnamefont
  {Berry}}, \bibinfo {author} {\bibfnamefont {M.}~\bibnamefont {Davidson}}, \
  and\ \bibinfo {author} {\bibfnamefont {D.~J.}\ \bibnamefont {Harvie}},\
  }\href {\doibase 10.1016/j.jcp.2013.05.026} {\bibfield  {journal} {\bibinfo
  {journal} {\jcompp}\ }\textbf {\bibinfo {volume} {251}},\ \bibinfo {pages}
  {209} (\bibinfo {year} {2013})}\BibitemShut {NoStop}%
\bibitem [{\citenamefont {Zeng}(2011)}]{zeng2011}%
  \BibitemOpen
  \bibfield  {author} {\bibinfo {author} {\bibfnamefont {J.}~\bibnamefont
  {Zeng}},\ }\href {\doibase 10.3390/ijms12031633} {\bibfield  {journal}
  {\bibinfo  {journal} {\ijms}\ }\textbf {\bibinfo {volume} {12}},\ \bibinfo
  {pages} {1633} (\bibinfo {year} {2011})}\BibitemShut {NoStop}%
\bibitem [{\citenamefont {Lu}\ \emph {et~al.}(2007)\citenamefont {Lu},
  \citenamefont {Glasner}, \citenamefont {Bertozzi},\ and\ \citenamefont
  {Kim}}]{lu2007}%
  \BibitemOpen
  \bibfield  {author} {\bibinfo {author} {\bibfnamefont {H.-W.}\ \bibnamefont
  {Lu}}, \bibinfo {author} {\bibfnamefont {K.}~\bibnamefont {Glasner}},
  \bibinfo {author} {\bibfnamefont {A.}~\bibnamefont {Bertozzi}}, \ and\
  \bibinfo {author} {\bibfnamefont {C.-J.}\ \bibnamefont {Kim}},\ }\href
  {\doibase 10.1017/S0022112007008154} {\bibfield  {journal} {\bibinfo
  {journal} {\jfm}\ }\textbf {\bibinfo {volume} {590}},\ \bibinfo {pages} {411}
  (\bibinfo {year} {2007})}\BibitemShut {NoStop}%
\bibitem [{\citenamefont {Walker}\ \emph {et~al.}(2009)\citenamefont {Walker},
  \citenamefont {Shapiro},\ and\ \citenamefont {Nochetto}}]{walker2009}%
  \BibitemOpen
  \bibfield  {author} {\bibinfo {author} {\bibfnamefont {S.~W.}\ \bibnamefont
  {Walker}}, \bibinfo {author} {\bibfnamefont {B.}~\bibnamefont {Shapiro}}, \
  and\ \bibinfo {author} {\bibfnamefont {R.~H.}\ \bibnamefont {Nochetto}},\
  }\href {\doibase 10.1063/1.3254022} {\bibfield  {journal} {\bibinfo
  {journal} {\physfluids}\ }\textbf {\bibinfo {volume} {21}},\ \bibinfo {pages}
  {102103} (\bibinfo {year} {2009})}\BibitemShut {NoStop}%
\bibitem [{\citenamefont {Logg}\ \emph {et~al.}(2012)\citenamefont {Logg},
  \citenamefont {Mardal},\ and\ \citenamefont {Wells}}]{logg2012}%
  \BibitemOpen
  \bibfield  {author} {\bibinfo {author} {\bibfnamefont {A.}~\bibnamefont
  {Logg}}, \bibinfo {author} {\bibfnamefont {K.-A.}\ \bibnamefont {Mardal}}, \
  and\ \bibinfo {author} {\bibfnamefont {G.}~\bibnamefont {Wells}},\
  }\href@noop {} {\emph {\bibinfo {title} {Automated solution of differential
  equations by the finite element method: The FEniCS book}}},\ Vol.~\bibinfo
  {volume} {84}\ (\bibinfo  {publisher} {Springer Science \& Business Media},\
  \bibinfo {year} {2012})\BibitemShut {NoStop}%
\bibitem [{\citenamefont {Mortensen}\ and\ \citenamefont
  {Valen-Sendstad}(2015)}]{mortensen2015}%
  \BibitemOpen
  \bibfield  {author} {\bibinfo {author} {\bibfnamefont {M.}~\bibnamefont
  {Mortensen}}\ and\ \bibinfo {author} {\bibfnamefont {K.}~\bibnamefont
  {Valen-Sendstad}},\ }\href {\doibase 10.1016/j.cpc.2014.10.026} {\bibfield
  {journal} {\bibinfo  {journal} {\cpc}\ }\textbf {\bibinfo {volume} {188}},\
  \bibinfo {pages} {177} (\bibinfo {year} {2015})}\BibitemShut {NoStop}%
\bibitem [{\citenamefont {Mitscha-Baude}\ \emph {et~al.}(2017)\citenamefont
  {Mitscha-Baude}, \citenamefont {Buttinger-Kreuzhuber}, \citenamefont
  {Tulzer},\ and\ \citenamefont {Heitzinger}}]{mitscha-baude2017}%
  \BibitemOpen
  \bibfield  {author} {\bibinfo {author} {\bibfnamefont {G.}~\bibnamefont
  {Mitscha-Baude}}, \bibinfo {author} {\bibfnamefont {A.}~\bibnamefont
  {Buttinger-Kreuzhuber}}, \bibinfo {author} {\bibfnamefont {G.}~\bibnamefont
  {Tulzer}}, \ and\ \bibinfo {author} {\bibfnamefont {C.}~\bibnamefont
  {Heitzinger}},\ }\href {\doibase 10.1016/j.jcp.2017.02.072} {\bibfield
  {journal} {\bibinfo  {journal} {\jcompp}\ }\textbf {\bibinfo {volume}
  {338}},\ \bibinfo {pages} {452} (\bibinfo {year} {2017})}\BibitemShut
  {NoStop}%
\bibitem [{\citenamefont {Bolet}\ \emph {et~al.}(2018)\citenamefont {Bolet},
  \citenamefont {Linga},\ and\ \citenamefont {Mathiesen}}]{bolet2018}%
  \BibitemOpen
  \bibfield  {author} {\bibinfo {author} {\bibfnamefont {A.}~\bibnamefont
  {Bolet}}, \bibinfo {author} {\bibfnamefont {G.}~\bibnamefont {Linga}}, \ and\
  \bibinfo {author} {\bibfnamefont {J.}~\bibnamefont {Mathiesen}},\ }\href
  {\doibase 10.1103/PhysRevE.97.043114} {\bibfield  {journal} {\bibinfo
  {journal} {\physreve}\ }\textbf {\bibinfo {volume} {97}},\ \bibinfo {pages}
  {043114} (\bibinfo {year} {2018})}\BibitemShut {NoStop}%
\bibitem [{\citenamefont {Prosperetti}\ and\ \citenamefont
  {Tryggvason}(2009)}]{prosperetti2009}%
  \BibitemOpen
  \bibfield  {author} {\bibinfo {author} {\bibfnamefont {A.}~\bibnamefont
  {Prosperetti}}\ and\ \bibinfo {author} {\bibfnamefont {G.}~\bibnamefont
  {Tryggvason}},\ }\href@noop {} {\emph {\bibinfo {title} {Computational
  methods for multiphase flow}}}\ (\bibinfo  {publisher} {Cambridge University
  Press},\ \bibinfo {year} {2009})\BibitemShut {NoStop}%
\bibitem [{\citenamefont {Yang}\ and\ \citenamefont {Li}(1996)}]{yang1996}%
  \BibitemOpen
  \bibfield  {author} {\bibinfo {author} {\bibfnamefont {C.}~\bibnamefont
  {Yang}}\ and\ \bibinfo {author} {\bibfnamefont {D.}~\bibnamefont {Li}},\
  }\href {\doibase 10.1016/0927-7757(96)03544-3} {\bibfield  {journal}
  {\bibinfo  {journal} {\csurfa}\ }\textbf {\bibinfo {volume} {113}},\ \bibinfo
  {pages} {51} (\bibinfo {year} {1996})}\BibitemShut {NoStop}%
\bibitem [{\citenamefont {Qian}\ \emph {et~al.}(2006)\citenamefont {Qian},
  \citenamefont {Wang},\ and\ \citenamefont {Sheng}}]{qian2006}%
  \BibitemOpen
  \bibfield  {author} {\bibinfo {author} {\bibfnamefont {T.}~\bibnamefont
  {Qian}}, \bibinfo {author} {\bibfnamefont {X.-P.}\ \bibnamefont {Wang}}, \
  and\ \bibinfo {author} {\bibfnamefont {P.}~\bibnamefont {Sheng}},\ }\href
  {\doibase 10.1017/S0022112006001935} {\bibfield  {journal} {\bibinfo
  {journal} {\jfm}\ }\textbf {\bibinfo {volume} {564}},\ \bibinfo {pages} {333}
  (\bibinfo {year} {2006})}\BibitemShut {NoStop}%
\bibitem [{\citenamefont {Qian}\ \emph {et~al.}(2003)\citenamefont {Qian},
  \citenamefont {Wang},\ and\ \citenamefont {Sheng}}]{qian2003}%
  \BibitemOpen
  \bibfield  {author} {\bibinfo {author} {\bibfnamefont {T.}~\bibnamefont
  {Qian}}, \bibinfo {author} {\bibfnamefont {X.-P.}\ \bibnamefont {Wang}}, \
  and\ \bibinfo {author} {\bibfnamefont {P.}~\bibnamefont {Sheng}},\ }\href
  {\doibase 10.1103/PhysRevE.68.016306} {\bibfield  {journal} {\bibinfo
  {journal} {\physreve}\ }\textbf {\bibinfo {volume} {68}},\ \bibinfo {pages}
  {016306} (\bibinfo {year} {2003})}\BibitemShut {NoStop}%
\bibitem [{\citenamefont {Sui}\ \emph {et~al.}(2014)\citenamefont {Sui},
  \citenamefont {Ding},\ and\ \citenamefont {Spelt}}]{sui2014}%
  \BibitemOpen
  \bibfield  {author} {\bibinfo {author} {\bibfnamefont {Y.}~\bibnamefont
  {Sui}}, \bibinfo {author} {\bibfnamefont {H.}~\bibnamefont {Ding}}, \ and\
  \bibinfo {author} {\bibfnamefont {P.~D.~M.}\ \bibnamefont {Spelt}},\ }\href
  {\doibase 10.1146/annurev-fluid-010313-141338} {\bibfield  {journal}
  {\bibinfo  {journal} {\arfm}\ }\textbf {\bibinfo {volume} {46}},\ \bibinfo
  {pages} {97} (\bibinfo {year} {2014})}\BibitemShut {NoStop}%
\bibitem [{\citenamefont {Ervik}\ \emph {et~al.}(2016)\citenamefont {Ervik},
  \citenamefont {Lysgaard}, \citenamefont {Herdes}, \citenamefont
  {Jiménez-Serratos}, \citenamefont {M{\"u}ller}, \citenamefont {Munkejord},\
  and\ \citenamefont {M{\"u}ller}}]{ervik2016}%
  \BibitemOpen
  \bibfield  {author} {\bibinfo {author} {\bibfnamefont {{\AA}.}~\bibnamefont
  {Ervik}}, \bibinfo {author} {\bibfnamefont {M.~O.}\ \bibnamefont {Lysgaard}},
  \bibinfo {author} {\bibfnamefont {C.}~\bibnamefont {Herdes}}, \bibinfo
  {author} {\bibfnamefont {G.}~\bibnamefont {Jiménez-Serratos}}, \bibinfo
  {author} {\bibfnamefont {E.~A.}\ \bibnamefont {M{\"u}ller}}, \bibinfo
  {author} {\bibfnamefont {S.~T.}\ \bibnamefont {Munkejord}}, \ and\ \bibinfo
  {author} {\bibfnamefont {B.}~\bibnamefont {M{\"u}ller}},\ }\href {\doibase
  10.1016/j.jcp.2016.09.039} {\bibfield  {journal} {\bibinfo  {journal}
  {\jcompp}\ }\textbf {\bibinfo {volume} {327}},\ \bibinfo {pages} {576 }
  (\bibinfo {year} {2016})}\BibitemShut {NoStop}%
\bibitem [{\citenamefont {Walker}\ and\ \citenamefont
  {Shapiro}(2006)}]{walker2006}%
  \BibitemOpen
  \bibfield  {author} {\bibinfo {author} {\bibfnamefont {S.~W.}\ \bibnamefont
  {Walker}}\ and\ \bibinfo {author} {\bibfnamefont {B.}~\bibnamefont
  {Shapiro}},\ }\href {\doibase 10.1109/JMEMS.2006.878876} {\bibfield
  {journal} {\bibinfo  {journal} {\jmems}\ }\textbf {\bibinfo {volume} {15}},\
  \bibinfo {pages} {986} (\bibinfo {year} {2006})}\BibitemShut {NoStop}%
\bibitem [{\citenamefont {Teigen}\ and\ \citenamefont
  {Munkejord}(2010)}]{teigen2010}%
  \BibitemOpen
  \bibfield  {author} {\bibinfo {author} {\bibfnamefont {K.~E.}\ \bibnamefont
  {Teigen}}\ and\ \bibinfo {author} {\bibfnamefont {S.~T.}\ \bibnamefont
  {Munkejord}},\ }\href {\doibase 10.1063/1.3504271} {\bibfield  {journal}
  {\bibinfo  {journal} {\physfluids}\ }\textbf {\bibinfo {volume} {22}},\
  \bibinfo {pages} {112104} (\bibinfo {year} {2010})}\BibitemShut {NoStop}%
\bibitem [{\citenamefont {Tomar}\ \emph {et~al.}(2007)\citenamefont {Tomar},
  \citenamefont {Gerlach}, \citenamefont {Biswas}, \citenamefont {Alleborn},
  \citenamefont {Sharma}, \citenamefont {Durst}, \citenamefont {Welch},\ and\
  \citenamefont {Delgado}}]{tomar2007}%
  \BibitemOpen
  \bibfield  {author} {\bibinfo {author} {\bibfnamefont {G.}~\bibnamefont
  {Tomar}}, \bibinfo {author} {\bibfnamefont {D.}~\bibnamefont {Gerlach}},
  \bibinfo {author} {\bibfnamefont {G.}~\bibnamefont {Biswas}}, \bibinfo
  {author} {\bibfnamefont {N.}~\bibnamefont {Alleborn}}, \bibinfo {author}
  {\bibfnamefont {A.}~\bibnamefont {Sharma}}, \bibinfo {author} {\bibfnamefont
  {F.}~\bibnamefont {Durst}}, \bibinfo {author} {\bibfnamefont
  {S.}~\bibnamefont {Welch}}, \ and\ \bibinfo {author} {\bibfnamefont
  {A.}~\bibnamefont {Delgado}},\ }\href {\doibase 10.1016/j.jcp.2007.09.003}
  {\bibfield  {journal} {\bibinfo  {journal} {\jcompp}\ }\textbf {\bibinfo
  {volume} {227}},\ \bibinfo {pages} {1267 } (\bibinfo {year}
  {2007})}\BibitemShut {NoStop}%
\bibitem [{\citenamefont {L{\'o}pez-Herrera}\ \emph {et~al.}(2011)\citenamefont
  {L{\'o}pez-Herrera}, \citenamefont {Popinet},\ and\ \citenamefont
  {Herrada}}]{lopez-herrera2011}%
  \BibitemOpen
  \bibfield  {author} {\bibinfo {author} {\bibfnamefont {J.}~\bibnamefont
  {L{\'o}pez-Herrera}}, \bibinfo {author} {\bibfnamefont {S.}~\bibnamefont
  {Popinet}}, \ and\ \bibinfo {author} {\bibfnamefont {M.}~\bibnamefont
  {Herrada}},\ }\href {\doibase 10.1016/j.jcp.2010.11.042} {\bibfield
  {journal} {\bibinfo  {journal} {\jcompp}\ }\textbf {\bibinfo {volume}
  {230}},\ \bibinfo {pages} {1939} (\bibinfo {year} {2011})}\BibitemShut
  {NoStop}%
\bibitem [{\citenamefont {Anderson}\ \emph {et~al.}(1998)\citenamefont
  {Anderson}, \citenamefont {McFadden},\ and\ \citenamefont
  {Wheeler}}]{anderson1998}%
  \BibitemOpen
  \bibfield  {author} {\bibinfo {author} {\bibfnamefont {D.~M.}\ \bibnamefont
  {Anderson}}, \bibinfo {author} {\bibfnamefont {G.~B.}\ \bibnamefont
  {McFadden}}, \ and\ \bibinfo {author} {\bibfnamefont {A.~A.}\ \bibnamefont
  {Wheeler}},\ }\href {\doibase 10.1146/annurev.fluid.30.1.139} {\bibfield
  {journal} {\bibinfo  {journal} {\arfm}\ }\textbf {\bibinfo {volume} {30}},\
  \bibinfo {pages} {139} (\bibinfo {year} {1998})}\BibitemShut {NoStop}%
\bibitem [{\citenamefont {Hohenberg}\ and\ \citenamefont
  {Halperin}(1977)}]{hohenberg1977}%
  \BibitemOpen
  \bibfield  {author} {\bibinfo {author} {\bibfnamefont {P.~C.}\ \bibnamefont
  {Hohenberg}}\ and\ \bibinfo {author} {\bibfnamefont {B.~I.}\ \bibnamefont
  {Halperin}},\ }\href {\doibase 10.1103/RevModPhys.49.435} {\bibfield
  {journal} {\bibinfo  {journal} {\revmodphys}\ }\textbf {\bibinfo {volume}
  {49}},\ \bibinfo {pages} {435} (\bibinfo {year} {1977})}\BibitemShut
  {NoStop}%
\bibitem [{\citenamefont {Lowengrub}\ and\ \citenamefont
  {Truskinovsky}(1998)}]{lowengrub1998}%
  \BibitemOpen
  \bibfield  {author} {\bibinfo {author} {\bibfnamefont {J.}~\bibnamefont
  {Lowengrub}}\ and\ \bibinfo {author} {\bibfnamefont {L.}~\bibnamefont
  {Truskinovsky}},\ }\href {\doibase 10.1098/rspa.1998.0273} {\bibfield
  {journal} {\bibinfo  {journal} {\prsa}\ }\textbf {\bibinfo {volume} {454}},\
  \bibinfo {pages} {2617} (\bibinfo {year} {1998})}\BibitemShut {NoStop}%
\bibitem [{\citenamefont {Abels}\ \emph {et~al.}(2012)\citenamefont {Abels},
  \citenamefont {Garcke},\ and\ \citenamefont {Gr{\"u}n}}]{abels2012}%
  \BibitemOpen
  \bibfield  {author} {\bibinfo {author} {\bibfnamefont {H.}~\bibnamefont
  {Abels}}, \bibinfo {author} {\bibfnamefont {H.}~\bibnamefont {Garcke}}, \
  and\ \bibinfo {author} {\bibfnamefont {G.}~\bibnamefont {Gr{\"u}n}},\ }\href
  {\doibase 10.1142/S0218202511500138} {\bibfield  {journal} {\bibinfo
  {journal} {\mmmas}\ }\textbf {\bibinfo {volume} {22}},\ \bibinfo {pages}
  {1150013} (\bibinfo {year} {2012})}\BibitemShut {NoStop}%
\bibitem [{Note1()}]{Note1}%
  \BibitemOpen
  \bibinfo {note} {Instead of the full Navier--Stokes equations, which would be
  necessary in the presence of boundaries in the two in-plane
  dimensions.}\BibitemShut {Stop}%
\bibitem [{\citenamefont {Lin}\ \emph {et~al.}(2012)\citenamefont {Lin},
  \citenamefont {Skjetne},\ and\ \citenamefont {Carlson}}]{lin2012}%
  \BibitemOpen
  \bibfield  {author} {\bibinfo {author} {\bibfnamefont {Y.}~\bibnamefont
  {Lin}}, \bibinfo {author} {\bibfnamefont {P.}~\bibnamefont {Skjetne}}, \ and\
  \bibinfo {author} {\bibfnamefont {A.}~\bibnamefont {Carlson}},\ }\href
  {\doibase 10.1016/j.ijmultiphaseflow.2012.04.002} {\bibfield  {journal}
  {\bibinfo  {journal} {International Journal of Multiphase Flow}\ }\textbf
  {\bibinfo {volume} {45}},\ \bibinfo {pages} {1} (\bibinfo {year}
  {2012})}\BibitemShut {NoStop}%
\bibitem [{\citenamefont {Campillo-Funollet}\ \emph {et~al.}(2012)\citenamefont
  {Campillo-Funollet}, \citenamefont {Gr{\"u}n},\ and\ \citenamefont
  {Klingbeil}}]{campillo-funollet2012}%
  \BibitemOpen
  \bibfield  {author} {\bibinfo {author} {\bibfnamefont {E.}~\bibnamefont
  {Campillo-Funollet}}, \bibinfo {author} {\bibfnamefont {G.}~\bibnamefont
  {Gr{\"u}n}}, \ and\ \bibinfo {author} {\bibfnamefont {F.}~\bibnamefont
  {Klingbeil}},\ }\href {\doibase 10.1137/120861333} {\bibfield  {journal}
  {\bibinfo  {journal} {\siap}\ }\textbf {\bibinfo {volume} {72}},\ \bibinfo
  {pages} {1899} (\bibinfo {year} {2012})}\BibitemShut {NoStop}%
\bibitem [{\citenamefont {Eck}\ \emph {et~al.}(2009)\citenamefont {Eck},
  \citenamefont {Fontelos}, \citenamefont {Gr{\"u}n}, \citenamefont
  {Klingbeil},\ and\ \citenamefont {Vantzos}}]{eck2009}%
  \BibitemOpen
  \bibfield  {author} {\bibinfo {author} {\bibfnamefont {C.}~\bibnamefont
  {Eck}}, \bibinfo {author} {\bibfnamefont {M.}~\bibnamefont {Fontelos}},
  \bibinfo {author} {\bibfnamefont {G.}~\bibnamefont {Gr{\"u}n}}, \bibinfo
  {author} {\bibfnamefont {F.}~\bibnamefont {Klingbeil}}, \ and\ \bibinfo
  {author} {\bibfnamefont {O.}~\bibnamefont {Vantzos}},\ }\href {\doibase
  10.4171/IFB/211} {\bibfield  {journal} {\bibinfo  {journal} {Interfaces and
  free boundaries}\ }\textbf {\bibinfo {volume} {11}},\ \bibinfo {pages} {259}
  (\bibinfo {year} {2009})}\BibitemShut {NoStop}%
\bibitem [{\citenamefont {Nochetto}\ \emph {et~al.}(2014)\citenamefont
  {Nochetto}, \citenamefont {Salgado},\ and\ \citenamefont
  {Walker}}]{nochetto2014}%
  \BibitemOpen
  \bibfield  {author} {\bibinfo {author} {\bibfnamefont {R.~H.}\ \bibnamefont
  {Nochetto}}, \bibinfo {author} {\bibfnamefont {A.~J.}\ \bibnamefont
  {Salgado}}, \ and\ \bibinfo {author} {\bibfnamefont {S.~W.}\ \bibnamefont
  {Walker}},\ }\href {\doibase 10.1142/S0218202513500474} {\bibfield  {journal}
  {\bibinfo  {journal} {\mmmas}\ }\textbf {\bibinfo {volume} {24}},\ \bibinfo
  {pages} {67} (\bibinfo {year} {2014})}\BibitemShut {NoStop}%
\bibitem [{\citenamefont {Gr{\"u}n}\ and\ \citenamefont
  {Klingbeil}(2014)}]{grun2014}%
  \BibitemOpen
  \bibfield  {author} {\bibinfo {author} {\bibfnamefont {G.}~\bibnamefont
  {Gr{\"u}n}}\ and\ \bibinfo {author} {\bibfnamefont {F.}~\bibnamefont
  {Klingbeil}},\ }\href {\doibase 10.1016/j.jcp.2013.10.028} {\bibfield
  {journal} {\bibinfo  {journal} {\jcompp}\ }\textbf {\bibinfo {volume}
  {257}},\ \bibinfo {pages} {708} (\bibinfo {year} {2014})}\BibitemShut
  {NoStop}%
\bibitem [{\citenamefont {Hysing}\ \emph {et~al.}(2009)\citenamefont {Hysing},
  \citenamefont {Turek}, \citenamefont {Kuzmin}, \citenamefont {Parolini},
  \citenamefont {Burman}, \citenamefont {Ganesan},\ and\ \citenamefont
  {Tobiska}}]{hysing2009}%
  \BibitemOpen
  \bibfield  {author} {\bibinfo {author} {\bibfnamefont {S.-R.}\ \bibnamefont
  {Hysing}}, \bibinfo {author} {\bibfnamefont {S.}~\bibnamefont {Turek}},
  \bibinfo {author} {\bibfnamefont {D.}~\bibnamefont {Kuzmin}}, \bibinfo
  {author} {\bibfnamefont {N.}~\bibnamefont {Parolini}}, \bibinfo {author}
  {\bibfnamefont {E.}~\bibnamefont {Burman}}, \bibinfo {author} {\bibfnamefont
  {S.}~\bibnamefont {Ganesan}}, \ and\ \bibinfo {author} {\bibfnamefont
  {L.}~\bibnamefont {Tobiska}},\ }\href {\doibase 10.1002/fld.1934} {\bibfield
  {journal} {\bibinfo  {journal} {\ijnmf}\ }\textbf {\bibinfo {volume} {60}},\
  \bibinfo {pages} {1259} (\bibinfo {year} {2009})}\BibitemShut {NoStop}%
\bibitem [{\citenamefont {Aland}\ and\ \citenamefont
  {Voigt}(2012)}]{aland2012}%
  \BibitemOpen
  \bibfield  {author} {\bibinfo {author} {\bibfnamefont {S.}~\bibnamefont
  {Aland}}\ and\ \bibinfo {author} {\bibfnamefont {A.}~\bibnamefont {Voigt}},\
  }\href {\doibase 10.1002/fld.2611} {\bibfield  {journal} {\bibinfo  {journal}
  {\ijnmf}\ }\textbf {\bibinfo {volume} {69}},\ \bibinfo {pages} {747}
  (\bibinfo {year} {2012})}\BibitemShut {NoStop}%
\bibitem [{\citenamefont {Guill{\'e}n-Gonz{\'a}lez}\ and\ \citenamefont
  {Tierra}(2014)}]{guillen-gonzalez2014}%
  \BibitemOpen
  \bibfield  {author} {\bibinfo {author} {\bibfnamefont {F.}~\bibnamefont
  {Guill{\'e}n-Gonz{\'a}lez}}\ and\ \bibinfo {author} {\bibfnamefont
  {G.}~\bibnamefont {Tierra}},\ }\href {\doibase 10.4208/jcm.1405-m4410}
  {\bibfield  {journal} {\bibinfo  {journal} {\jcompm}\ }\textbf {\bibinfo
  {volume} {32}},\ \bibinfo {pages} {643} (\bibinfo {year} {2014})}\BibitemShut
  {NoStop}%
\bibitem [{\citenamefont {Gr{\"u}n}\ \emph {et~al.}(2016)\citenamefont
  {Gr{\"u}n}, \citenamefont {Guill{\'e}n-Gonz{\'a}lez},\ and\ \citenamefont
  {Metzger}}]{grun2016}%
  \BibitemOpen
  \bibfield  {author} {\bibinfo {author} {\bibfnamefont {G.}~\bibnamefont
  {Gr{\"u}n}}, \bibinfo {author} {\bibfnamefont {F.}~\bibnamefont
  {Guill{\'e}n-Gonz{\'a}lez}}, \ and\ \bibinfo {author} {\bibfnamefont
  {S.}~\bibnamefont {Metzger}},\ }\href {\doibase 10.4208/cicp.scpde14.39s}
  {\bibfield  {journal} {\bibinfo  {journal} {\commcompphys}\ }\textbf
  {\bibinfo {volume} {19}},\ \bibinfo {pages} {1473} (\bibinfo {year}
  {2016})}\BibitemShut {NoStop}%
\bibitem [{\citenamefont {Metzger}(2015)}]{metzger2015}%
  \BibitemOpen
  \bibfield  {author} {\bibinfo {author} {\bibfnamefont {S.}~\bibnamefont
  {Metzger}},\ }\href {\doibase 10.1002/pamm.201510346} {\bibfield  {journal}
  {\bibinfo  {journal} {Proc. Appl. Math. Mech.}\ }\textbf {\bibinfo {volume}
  {15}},\ \bibinfo {pages} {715} (\bibinfo {year} {2015})}\BibitemShut
  {NoStop}%
\bibitem [{\citenamefont {Metzger}(2018)}]{metzger2018}%
  \BibitemOpen
  \bibfield  {author} {\bibinfo {author} {\bibfnamefont {S.}~\bibnamefont
  {Metzger}},\ }\href {\doibase 10.1007/s11075-018-0530-2} {\bibfield
  {journal} {\bibinfo  {journal} {Numerical Algorithms}\ ,\ \bibinfo {pages}
  {1}} (\bibinfo {year} {2018})}\BibitemShut {NoStop}%
\bibitem [{\citenamefont {N{\"u}rnberg}\ and\ \citenamefont
  {Tucker}(2017)}]{nurnberg2017}%
  \BibitemOpen
  \bibfield  {author} {\bibinfo {author} {\bibfnamefont {R.}~\bibnamefont
  {N{\"u}rnberg}}\ and\ \bibinfo {author} {\bibfnamefont {E.~J.~W.}\
  \bibnamefont {Tucker}},\ }\href {\doibase 10.1017/S0956792516000395}
  {\bibfield  {journal} {\bibinfo  {journal} {\ejam}\ }\textbf {\bibinfo
  {volume} {28}},\ \bibinfo {pages} {470} (\bibinfo {year} {2017})}\BibitemShut
  {NoStop}%
\bibitem [{\citenamefont {Langtangen}\ and\ \citenamefont
  {Logg}(2017)}]{langtangen2017}%
  \BibitemOpen
  \bibfield  {author} {\bibinfo {author} {\bibfnamefont {H.~P.}\ \bibnamefont
  {Langtangen}}\ and\ \bibinfo {author} {\bibfnamefont {A.}~\bibnamefont
  {Logg}},\ }\href {\doibase 10.1007/978-3-319-52462-7} {\emph {\bibinfo
  {title} {Solving PDEs in Python}}}\ (\bibinfo  {publisher} {Springer},\
  \bibinfo {year} {2017})\BibitemShut {NoStop}%
\bibitem [{Note2()}]{Note2}%
  \BibitemOpen
  \bibinfo {note} {The interpretation of this pressure depends on the
  formulation of the force on the right hand side of Eq.~\protect \textup
  {\hbox {\mathsurround \z@ \protect \normalfont (\ignorespaces \ref
  {eq:sharp_NS2}\unskip \@@italiccorr )}}.}\BibitemShut {Stop}%
\bibitem [{\citenamefont {Nielsen}\ and\ \citenamefont
  {Bruus}(2014)}]{nielsen2014}%
  \BibitemOpen
  \bibfield  {author} {\bibinfo {author} {\bibfnamefont {C.~P.}\ \bibnamefont
  {Nielsen}}\ and\ \bibinfo {author} {\bibfnamefont {H.}~\bibnamefont
  {Bruus}},\ }\href {\doibase 10.1103/PhysRevE.90.043020} {\bibfield  {journal}
  {\bibinfo  {journal} {\physreve}\ }\textbf {\bibinfo {volume} {90}},\
  \bibinfo {pages} {043020} (\bibinfo {year} {2014})}\BibitemShut {NoStop}%
\bibitem [{\citenamefont {Linga}\ \emph
  {et~al.}(2018{\natexlab{a}})\citenamefont {Linga}, \citenamefont {Bolet},\
  and\ \citenamefont {Mathiesen}}]{linga2018decoupled}%
  \BibitemOpen
  \bibfield  {author} {\bibinfo {author} {\bibfnamefont {G.}~\bibnamefont
  {Linga}}, \bibinfo {author} {\bibfnamefont {A.}~\bibnamefont {Bolet}}, \ and\
  \bibinfo {author} {\bibfnamefont {J.}~\bibnamefont {Mathiesen}},\ }\href@noop
  {} {\enquote {\bibinfo {title} {Transient electrohydrodynamic flow with
  concentration dependent fluid properties: modelling and energy-stable
  schemes},}\ } (\bibinfo {year} {2018}{\natexlab{a}}),\ \bibinfo {note}
  {submitted}\BibitemShut {NoStop}%
\bibitem [{\citenamefont {Carlson}\ \emph {et~al.}(2012)\citenamefont
  {Carlson}, \citenamefont {Bellani},\ and\ \citenamefont
  {Amberg}}]{carlson2012}%
  \BibitemOpen
  \bibfield  {author} {\bibinfo {author} {\bibfnamefont {A.}~\bibnamefont
  {Carlson}}, \bibinfo {author} {\bibfnamefont {G.}~\bibnamefont {Bellani}}, \
  and\ \bibinfo {author} {\bibfnamefont {G.}~\bibnamefont {Amberg}},\ }\href
  {\doibase 10.1103/PhysRevE.85.045302} {\bibfield  {journal} {\bibinfo
  {journal} {\physreve}\ }\textbf {\bibinfo {volume} {85}},\ \bibinfo {pages}
  {045302} (\bibinfo {year} {2012})}\BibitemShut {NoStop}%
\bibitem [{\citenamefont {Shen}\ and\ \citenamefont {Yang}(2015)}]{shen2015}%
  \BibitemOpen
  \bibfield  {author} {\bibinfo {author} {\bibfnamefont {J.}~\bibnamefont
  {Shen}}\ and\ \bibinfo {author} {\bibfnamefont {X.}~\bibnamefont {Yang}},\
  }\href {\doibase 10.1137/140971154} {\bibfield  {journal} {\bibinfo
  {journal} {\sinum}\ }\textbf {\bibinfo {volume} {53}},\ \bibinfo {pages}
  {279} (\bibinfo {year} {2015})}\BibitemShut {NoStop}%
\bibitem [{\citenamefont {Brenner}\ and\ \citenamefont
  {Scott}(2007)}]{brenner2007}%
  \BibitemOpen
  \bibfield  {author} {\bibinfo {author} {\bibfnamefont {S.~C.}\ \bibnamefont
  {Brenner}}\ and\ \bibinfo {author} {\bibfnamefont {L.~R.}\ \bibnamefont
  {Scott}},\ }\href@noop {} {\emph {\bibinfo {title} {The mathematical theory
  of finite element methods}}},\ Vol.~\bibinfo {volume} {15}\ (\bibinfo
  {publisher} {Springer Science \& Business Media},\ \bibinfo {year}
  {2007})\BibitemShut {NoStop}%
\bibitem [{\citenamefont {Langtangen}\ \emph {et~al.}(2002)\citenamefont
  {Langtangen}, \citenamefont {Mardal},\ and\ \citenamefont
  {Winther}}]{langtangen2002}%
  \BibitemOpen
  \bibfield  {author} {\bibinfo {author} {\bibfnamefont {H.~P.}\ \bibnamefont
  {Langtangen}}, \bibinfo {author} {\bibfnamefont {K.-A.}\ \bibnamefont
  {Mardal}}, \ and\ \bibinfo {author} {\bibfnamefont {R.}~\bibnamefont
  {Winther}},\ }\href {\doibase 10.1016/S0309-1708(02)00052-0} {\bibfield
  {journal} {\bibinfo  {journal} {Advances in Water Resources}\ }\textbf
  {\bibinfo {volume} {25}},\ \bibinfo {pages} {1125} (\bibinfo {year}
  {2002})}\BibitemShut {NoStop}%
\bibitem [{\citenamefont {Chorin}(1967)}]{chorin1967}%
  \BibitemOpen
  \bibfield  {author} {\bibinfo {author} {\bibfnamefont {A.~J.}\ \bibnamefont
  {Chorin}},\ }\href {\doibase 10.1006/jcph.1997.5716} {\bibfield  {journal}
  {\bibinfo  {journal} {\jcompp}\ }\textbf {\bibinfo {volume} {2}},\ \bibinfo
  {pages} {12} (\bibinfo {year} {1967})}\BibitemShut {NoStop}%
\bibitem [{\citenamefont {Chorin}(1968)}]{chorin1968}%
  \BibitemOpen
  \bibfield  {author} {\bibinfo {author} {\bibfnamefont {A.~J.}\ \bibnamefont
  {Chorin}},\ }\href {\doibase 10.1090/S0025-5718-1968-0242392-2} {\bibfield
  {journal} {\bibinfo  {journal} {\mathcomput}\ }\textbf {\bibinfo {volume}
  {22}},\ \bibinfo {pages} {745} (\bibinfo {year} {1968})}\BibitemShut
  {NoStop}%
\bibitem [{\citenamefont {Guermond}\ \emph {et~al.}(2006)\citenamefont
  {Guermond}, \citenamefont {Minev},\ and\ \citenamefont
  {Shen}}]{guermond2006}%
  \BibitemOpen
  \bibfield  {author} {\bibinfo {author} {\bibfnamefont {J.-L.}\ \bibnamefont
  {Guermond}}, \bibinfo {author} {\bibfnamefont {P.}~\bibnamefont {Minev}}, \
  and\ \bibinfo {author} {\bibfnamefont {J.}~\bibnamefont {Shen}},\ }\href
  {\doibase 10.1016/j.cma.2005.10.010} {\bibfield  {journal} {\bibinfo
  {journal} {\cmame}\ }\textbf {\bibinfo {volume} {195}},\ \bibinfo {pages}
  {6011} (\bibinfo {year} {2006})}\BibitemShut {NoStop}%
\bibitem [{\citenamefont {Linga}\ and\ \citenamefont
  {Bolet}(2018)}]{bernaise2018}%
  \BibitemOpen
  \bibfield  {author} {\bibinfo {author} {\bibfnamefont {G.}~\bibnamefont
  {Linga}}\ and\ \bibinfo {author} {\bibfnamefont {A.}~\bibnamefont {Bolet}},\
  }\href@noop {} {\enquote {\bibinfo {title} {{Bernaise: Git repository}},}\
  }\bibinfo {howpublished} {\url{https://www.github.com/gautelinga/BERNAISE}}
  (\bibinfo {year} {2018})\BibitemShut {NoStop}%
\bibitem [{Note3()}]{Note3}%
  \BibitemOpen
  \bibinfo {note} {The latter also contains an equilibrium solver for the
  quiescent electrochemistry problem, mainly to be used for initialization
  purposes.}\BibitemShut {Stop}%
\bibitem [{Note4()}]{Note4}%
  \BibitemOpen
  \bibinfo {note} {Obviously, when the physical interface thickness may be
  resolved by the phase field, the sharp-interface assumption might be less
  sensible than the diffuse. Hence, in such cases this point might be too
  crude.}\BibitemShut {Stop}%
\bibitem [{\citenamefont {Pillai}\ \emph {et~al.}(2015)\citenamefont {Pillai},
  \citenamefont {Berry}, \citenamefont {Harvie},\ and\ \citenamefont
  {Davidson}}]{pillai2015}%
  \BibitemOpen
  \bibfield  {author} {\bibinfo {author} {\bibfnamefont {R.}~\bibnamefont
  {Pillai}}, \bibinfo {author} {\bibfnamefont {J.}~\bibnamefont {Berry}},
  \bibinfo {author} {\bibfnamefont {D.}~\bibnamefont {Harvie}}, \ and\ \bibinfo
  {author} {\bibfnamefont {M.}~\bibnamefont {Davidson}},\ }\href {\doibase
  10.1103/PhysRevE.92.013007} {\bibfield  {journal} {\bibinfo  {journal}
  {Physical Review E}\ }\textbf {\bibinfo {volume} {92}},\ \bibinfo {pages}
  {013007} (\bibinfo {year} {2015})}\BibitemShut {NoStop}%
\bibitem [{\citenamefont {Logg}\ and\ \citenamefont {Wells}(2010)}]{logg2010}%
  \BibitemOpen
  \bibfield  {author} {\bibinfo {author} {\bibfnamefont {A.}~\bibnamefont
  {Logg}}\ and\ \bibinfo {author} {\bibfnamefont {G.~N.}\ \bibnamefont
  {Wells}},\ }\href {\doibase 10.1145/1731022.1731030} {\bibfield  {journal}
  {\bibinfo  {journal} {ACM Trans. Math. Softw.}\ }\textbf {\bibinfo {volume}
  {37}},\ \bibinfo {pages} {20:1} (\bibinfo {year} {2010})}\BibitemShut
  {NoStop}%
\bibitem [{\citenamefont {Balay}\ \emph {et~al.}(2017)\citenamefont {Balay},
  \citenamefont {Abhyankar}, \citenamefont {Adams}, \citenamefont {Brown},
  \citenamefont {Brune}, \citenamefont {Buschelman}, \citenamefont {Dalcin},
  \citenamefont {Eijkhout}, \citenamefont {Gropp}, \citenamefont {Kaushik},
  \citenamefont {Knepley}, \citenamefont {May}, \citenamefont {McInnes},
  \citenamefont {Rupp}, \citenamefont {Smith}, \citenamefont {Zampini},
  \citenamefont {Zhang},\ and\ \citenamefont {Zhang}}]{petsc2017}%
  \BibitemOpen
  \bibfield  {author} {\bibinfo {author} {\bibfnamefont {S.}~\bibnamefont
  {Balay}}, \bibinfo {author} {\bibfnamefont {S.}~\bibnamefont {Abhyankar}},
  \bibinfo {author} {\bibfnamefont {M.~F.}\ \bibnamefont {Adams}}, \bibinfo
  {author} {\bibfnamefont {J.}~\bibnamefont {Brown}}, \bibinfo {author}
  {\bibfnamefont {P.}~\bibnamefont {Brune}}, \bibinfo {author} {\bibfnamefont
  {K.}~\bibnamefont {Buschelman}}, \bibinfo {author} {\bibfnamefont
  {L.}~\bibnamefont {Dalcin}}, \bibinfo {author} {\bibfnamefont
  {V.}~\bibnamefont {Eijkhout}}, \bibinfo {author} {\bibfnamefont {W.~D.}\
  \bibnamefont {Gropp}}, \bibinfo {author} {\bibfnamefont {D.}~\bibnamefont
  {Kaushik}}, \bibinfo {author} {\bibfnamefont {M.~G.}\ \bibnamefont
  {Knepley}}, \bibinfo {author} {\bibfnamefont {D.~A.}\ \bibnamefont {May}},
  \bibinfo {author} {\bibfnamefont {L.~C.}\ \bibnamefont {McInnes}}, \bibinfo
  {author} {\bibfnamefont {K.}~\bibnamefont {Rupp}}, \bibinfo {author}
  {\bibfnamefont {B.~F.}\ \bibnamefont {Smith}}, \bibinfo {author}
  {\bibfnamefont {S.}~\bibnamefont {Zampini}}, \bibinfo {author} {\bibfnamefont
  {H.}~\bibnamefont {Zhang}}, \ and\ \bibinfo {author} {\bibfnamefont
  {H.}~\bibnamefont {Zhang}},\ }\href {http://www.mcs.anl.gov/petsc} {\enquote
  {\bibinfo {title} {{PETS}c web page},}\ }\bibinfo {howpublished}
  {\url{http://www.mcs.anl.gov/petsc}} (\bibinfo {year} {2017})\BibitemShut
  {NoStop}%
\bibitem [{\citenamefont {Mugele}\ and\ \citenamefont
  {Buehrle}(2007)}]{mugele2007}%
  \BibitemOpen
  \bibfield  {author} {\bibinfo {author} {\bibfnamefont {F.}~\bibnamefont
  {Mugele}}\ and\ \bibinfo {author} {\bibfnamefont {J.}~\bibnamefont
  {Buehrle}},\ }\href {\doibase 10.1088/0953-8984/19/37/375112} {\bibfield
  {journal} {\bibinfo  {journal} {\jpcm}\ }\textbf {\bibinfo {volume} {19}},\
  \bibinfo {pages} {375112} (\bibinfo {year} {2007})}\BibitemShut {NoStop}%
\bibitem [{\citenamefont {Linga}\ \emph
  {et~al.}(2018{\natexlab{b}})\citenamefont {Linga}, \citenamefont {Bolet},\
  and\ \citenamefont {Mathiesen}}]{linga2018controlling}%
  \BibitemOpen
  \bibfield  {author} {\bibinfo {author} {\bibfnamefont {G.}~\bibnamefont
  {Linga}}, \bibinfo {author} {\bibfnamefont {A.}~\bibnamefont {Bolet}}, \ and\
  \bibinfo {author} {\bibfnamefont {J.}~\bibnamefont {Mathiesen}},\ }\href@noop
  {} {\enquote {\bibinfo {title} {Controlling wetting with electrolytic
  solutions: phase-field simulations of a droplet-conductor system},}\ }
  (\bibinfo {year} {2018}{\natexlab{b}}),\ \bibinfo {note}
  {submitted}\BibitemShut {NoStop}%
\bibitem [{\citenamefont {Huang}\ \emph {et~al.}(2015)\citenamefont {Huang},
  \citenamefont {Huang},\ and\ \citenamefont {Wang}}]{huang2015}%
  \BibitemOpen
  \bibfield  {author} {\bibinfo {author} {\bibfnamefont {J.-J.}\ \bibnamefont
  {Huang}}, \bibinfo {author} {\bibfnamefont {H.}~\bibnamefont {Huang}}, \ and\
  \bibinfo {author} {\bibfnamefont {X.}~\bibnamefont {Wang}},\ }\href {\doibase
  10.1002/fld.3975} {\bibfield  {journal} {\bibinfo  {journal} {\ijnmf}\
  }\textbf {\bibinfo {volume} {77}},\ \bibinfo {pages} {123} (\bibinfo {year}
  {2015})}\BibitemShut {NoStop}%
\bibitem [{\citenamefont {Teigen}\ \emph {et~al.}(2011)\citenamefont {Teigen},
  \citenamefont {Song}, \citenamefont {Lowengrub},\ and\ \citenamefont
  {Voigt}}]{teigen2011}%
  \BibitemOpen
  \bibfield  {author} {\bibinfo {author} {\bibfnamefont {K.~E.}\ \bibnamefont
  {Teigen}}, \bibinfo {author} {\bibfnamefont {P.}~\bibnamefont {Song}},
  \bibinfo {author} {\bibfnamefont {J.}~\bibnamefont {Lowengrub}}, \ and\
  \bibinfo {author} {\bibfnamefont {A.}~\bibnamefont {Voigt}},\ }\href
  {\doibase 10.1016/j.jcp.2010.09.020} {\bibfield  {journal} {\bibinfo
  {journal} {\jcompp}\ }\textbf {\bibinfo {volume} {230}},\ \bibinfo {pages}
  {375 } (\bibinfo {year} {2011})}\BibitemShut {NoStop}%
\end{thebibliography}%

\end{document}